# Advancing Heliophysics and Space Weather Modeling through Open Science

**C. Corti[1,2,3], M.M. Kuznetsova[1], M.A. Reiss[1,3], J. Yue[1,4], J. Karpen[1], C.N. Arge[1], F. Bacchini[5], C. Bard[1], S. Bruinsma[6], R.M. Caplan[7], L.K.S. Daldorff[1,4], P.J. Deka[5], C.R. DeVore[*1,4], S. Elvidge[8], N. Ganushkina[9], J.D. Huba[10], B.V. Jackson[11], V. Jordanova[12], J.A. Linker[7], H. Liu[13], J.G. Luhmann[14], S. Markidis[15], P. Mayank[13], V. Merkin[16], N. Moens[5], D. Odstrcil[1,17], Y.A. Omelchenko[18], M. Palmroth[19,20], S. Poedts[5,21], A.J. Ridley[9], Y. Shou[9], V. Tenishev[22], D.R. Themens[8], G. Toth[9], W. Wang[13], R.-P. Wilhelm[5], M.A. Young[23], B. Cecconi[24], M.-Y. Chou[1], D. De Zeeuw[1,4], G.L. Delzanno[12], C. Didigu[1], M. El Alaoui[†1,4], S. Fung[‡1], J. Green[25], Z. Huang[9,26], L.K. Jian[1], L.J. Landwer[27,28], M. Lesko[1], P. MacNeice[1], A. Masson[29], M.L. Mays[1], P.M. Mehta[30], M.S. Miesch[27], E. Palmerio[7], M. Petrenko[1], E. Provornikova[16], L. Rastätter[1], L. Rusaitis[1,4], N. Sachdeva[9], E. Samara[1,4], D. Sur[1,4], A. Taktakishvili[1,4], J. Topper[1], T. Tsui[1], C. Verbeke[1,3,5], J. Wang[1,4], C. Wiegand[1], M. Wiltberger[13], Y. Zheng[1], M.M. Bisi[31], M.K. Georgoulis[32,33], T. Kodikara[34], T. Pulkkinen[9], A. Chartier[16], D. da Silva[1], A. Faturahman[35], K. Garcia-Sage[1], D. Kondrashov[36], V.E. Ledvina[37], W. Liu[9], C. Pandey[38], E. Resnick[1,39], C. Shi[40], R.S. Weigel[17], K. Whitman[41,42], I. Zakharenkova[13], and K. Zhang[43,44]**

[1]NASA Goddard Space Flight Center, Heliophysics Science Division, Greenbelt, MD, USA

[2]University of Hawaii at Manoa, Physics and Astronomy Department, Honolulu, HI, USA

[3]Universities Space Research Association, Washington, DC, USA

[4]The Catholic University of America, Washington, DC, USA

[5]Centre for Mathematical Plasma Astrophysics, Department of Mathematics, KU Leuven, Leuven, Belgium

[6]CNES, Space Geodesy Office, Toulouse, France

[7]Predictive Science Inc., San Diego, CA, USA

[8]School of Engineering, University of Birmingham, Birmingham, UK

[9]Department of Climate and Space Sciences and Engineering, University of Michigan, Ann Arbor, MI, USA

[10]Syntek Technologies, Fairfax, VA

[11]Department of Astronomy and Astrophysics, University of California San Diego, San Diego, CA, USA

[12]Los Alamos National Laboratory, Los Alamos, NM, USA

[13]University Corporation for Atmospheric Research (UCAR), Boulder, CO, USA

[14]Space Sciences Laboratory, University of California Berkeley, Berkeley, CA, USA

[15]KTH Royal Institute of Technology, Stockholm, Sweden

[16]Johns Hopkins University Applied Physics Laboratory, Laurel, MD, USA

[17]Department of Physics and Astronomy, George Mason University, Fairfax, VA, USA

[18]Space Science Institute, Boulder, CO, USA

[19]Department of Physics, University of Helsinki, Helsinki, Finland

[20]Space and Earth Observation Centre, Finnish Meteorological Institute, Helsinki, Finland

[21]Maria Curie-Skłodowska University, Lublin, Poland

[22]NASA Marshall Space Flight Center, Huntsville, AL, USA

[23]University of New Hampshire, Durham, NH, USA

[24]LIRA, Observatoire de Paris, Université PSL, Sorbonne Université, Université Paris Cité, CY Cergy Paris Université, CNRS, Meudon, France

[25]Space Hazards Applications, Golden, CO, USA

[26]School of Earth and Space Sciences, University of Science and Technology of China

[27]CIRES, University of Colorado, Boulder, CO, USA

[28]NOAA, Space Weather Prediction Center, Boulder, CO, USA

[29]European Space Agency, European Space Astronomy Centre, Madrid, Spain

[30]West Virginia University, Morgantown, WV, USA

[31]RAL Space, United Kingdom Research and Innovation, Science & Technology Facilities Council, Rutherford Appleton Laboratory, Harwell Campus Oxfordshire, UK

[32]Johns Hopkins Applied Physics Laboratory, Laurel, MD, USA

[33]RCAAM of the Academy of Athens, Athens, Greece (on leave)

[34]German Aerospace Center (DLR), Institute for Solar-Terrestrial Physics, Neustrelitz, Germany

[35]Research Center for Space, National Research and Innovation Agency (BRIN), Jakarta, Indonesia

[36]Department of Atmospheric and Oceanic Sciences, University of California Los Angeles, Los Angeles, CA, USA

[37]Geophysical Institute, University of Alaska Fairbanks, Fairbanks, AK, USA

[38]Texas Christian University, Fort Worth, TX, USA

[39]ASRC Federal, Beltsville, MD, USA

[40]Department of Physics, Auburn University, Auburn, AL, USA

[41]KBR, Houston, TX, USA

[42]NASA Johnson Space Flight Center, Houston, TX, USA

[43]Department of Earth, Planetary, and Space Sciences, University of California Los Angeles, Los Angeles, CA, USA

[44]Department of Physical Sciences, Embry-Riddle Aeronautical University, Daytona Beach, FL, USA

Corresponding authors: Claudio Corti (corti@hawaii.edu), Maria M. Kuznetsova (maria.m.kuznetsova@nasa.gov)

[*] Retired independent consultant

[†] Mostafa El Alaoui sadly passed away during the preparation of this paper.

[‡] Affiliation at the time of submission. Shing Fung retired from NASA in May 2026.

**Key Points:**

- Open science in heliophysics and space weather modeling poses unique challenges with respect to experimental data and analysis software.
- We present recommendations to address the challenges and gaps in current open science practices and guidelines.
- We establish a platform for the modeling community to guide the future of open science in heliophysics and space weather modeling.

**Abstract**

We present a community-wide effort to develop a strategy and action plan to advance heliophysics and space weather modeling through open science. While open science has the potential to enhance the quality and pace of scientific discovery, its application to scientific modeling requires more careful consideration regarding open data and open software guidelines, as scientific models differ significantly from data analysis software. We gathered feedback from modeling teams worldwide through a living survey and discussion sessions at the 2024 Open Science Workshop in College Park, USA, and at the 2025 COSPAR ISWAT Working Meeting in Cape Canaveral, USA. We complement these findings with lessons learned from almost 25 years of experience at the Community Coordinated Modeling Center in enabling open use of models. We identify key roadblocks in current open science practices and guidelines and offer recommendations for future progress across four overlapping themes: open use of models and simulation results, open validation, open development, and open collaboration. An essential outcome of the discussion is the need for model developers and model users to speak with a united voice and promote the role of models in future open science efforts. We introduce a new cross-domain community initiative called the Heliophysics Open Modeling Environment (HOME), which will be integrated as an overarching activity within COSPAR ISWAT. HOME will serve as a platform for modelers and model users to work together, facilitate community modeling, improve the scientific return on modeling investment, and advance understanding, modeling, and forecasting in heliophysics and space weather.

**Plain Language Summary**

Scientific models are crucial for advancing research and forecasting in heliophysics and space weather. To enable efficient community science and build trust in scientific results, models should be openly accessible to everyone with minimal barriers. At present, many advanced heliophysics and space weather models are available for open use through simulation service providers such as the CCMC and the VSWMC. Nevertheless, the complexity and computational requirements of advanced scientific models pose challenges for creating open and replicable workflows. After collecting feedback from the modeling community worldwide on current challenges, we present a path forward around four overlapping themes: making models broadly accessible (open use), increasing trust and transparency in simulation results (open validation), enabling the community to develop models collaboratively (open development), and building trusting working relationships (open collaboration). Our work furthermore highlights the need for modelers to speak with a united voice to promote the unique role of models in open science, as well as to foster a work culture that rewards an open science approach in heliophysics and space weather modeling.

## 1 Introduction

### 1.1 Background

The overarching objective of open science is to enable our society to make efficient use of past scientific achievements and thereby enhance both the quality and pace of new scientific breakthroughs. The term *open science* refers to specific principles and practices that are geared towards making scientific knowledge accessible to everyone and ensuring transparency and collaboration in how new discoveries are made. Because of its transformative potential, open science has received elevated attention worldwide. Nevertheless, the scope, focus, and the pace of implementation of open science policies vary by institution, agency, and/or organization.

The United Nations Educational, Scientific, and Cultural Organization (UNESCO) has promoted open science globally. In 2021, it published an international definition of open science that was endorsed by 193 countries. UNESCO defines open science as a framework “that combines different movements and practices to make scientific knowledge openly available, accessible, and usable for everyone”. They highlight the importance of “making the entire process of scientific knowledge creation and evaluation open”, and of “strengthening scientific collaborations and sharing of information”, all for the benefit of science and society (UNESCO, 2021).

Well before the UNESCO definition of open science was published, multiple entities had adopted open science through their own policies and programs. In the United States, the National Aeronautics and Space Administration (NASA) Science Missions Directorate (SMD) has been pioneering open science, particularly regarding the open access to mission data, since the early 1990s. NASA's Heliophysics Division formally adopted its Science Data Management Policy in 2007, which laid the groundwork for open data practices within the heliophysics community. NASA, in its Scientific Information Policy for the SMD (also known as SPD-41a, NASA, 2022), describes guidelines for sharing scientific information, broadly categorized as publications, data, and software. Additionally, the U.S. National Science Foundation (NSF), through its Public Access Initiative (https://www.nsf.gov/public-access), prioritizes that outputs from NSF research grants, including publications and scientific data, are made openly available with the fewest constraints possible. In Europe, the European Union has an open science policy, embraced by various scientific organizations such as the European Geosciences Union. For more than a decade, the European Space Agency (ESA) Earth Observation (EO) directorate has promoted open science by making its data, tools, and research openly accessible and reusable. Key initiatives include the Earth Science Collaborative Open Development Environment (https://earthcode.esa.int/), the Open Science Catalog (https://opensciencedata.esa.int/), and participation in initiatives such as the European Open Science Cloud (https://eosc.eu/), supported by the European Union. These efforts foster a collaborative and transparent scientific ecosystem. The EO directorate hosts heliophysics data from, e.g., the ionospheric Swarm and CHAMP magnetometry missions. The ESA science directorate is also promoting open science by making all its data publicly and freely accessible. It plays a key role in data interoperability by working hand in hand with other space agencies and main data providers through international data alliances, in particular the International Heliophysics Data Environment Alliance (IHDEA, https://ihdea.net). IHDEA is composed of various working groups focused at adhering to and promoting the use of a set of governing data standards, data exchange protocols, and data visualization and analysis tools, actively pushing open science goals and reporting at yearly hybrid meetings together with the Data, Analysis, and Software in Heliophysics forum (DASH, https://dash.heliophysics.net/).

### 1.2 Open Science Policies, Principles, and Resources

Examples of open science policies commonly adopted by funding agencies include Science Data Management Policy and Open-Source Policy. Throughout different policies, the FAIR principles (https://www.go-fair.org/fair-principles/, Wilkinson et al., 2016) are common technical guidelines for the implementation of open science. These principles aim to ensure that research outputs are Findable, Accessible, Interoperable, and Reusable. Here, *Findable* means that data and research output can be discovered easily, for instance, through unique identifiers and extensive metadata descriptions. *Accessible* means that, once the data and research output have been found, everyone can get access to them through clear pathways, which may include guidance on communication and authentication protocols. *Interoperable* is about research products being described with standard formats so that they can be used across different platforms and scientific disciplines. *Reusable* is the final element, which means that once data or research output have been identified, accessed, and understood they can be reused for the same purpose or for new scientific investigations.

The FAIR principles are mentioned repeatedly in recommendations on open science because they provide the technical specificity that is needed to transition high-level definitions to practical usage. The usage of FAIR as a practical guideline is not restricted to publications and data but is also applicable to software. Scientific software has been identified as a core component of open science. UNESCO lists open-source software among its key components under “open scientific knowledge”, and NASA’s SPD-41a policy identifies software along with data and publication as essential elements.

The application of the FAIR principles on scientific software has so far heavily relied on the open-source software infrastructure and collaborative environments originating from the industrial sector. Some of the common practices and resources include:

- *code repositories*, used for tracking code changes (i.e., version control), sharing code under active development, and managing the development process among collaborators, hosted on personal, institutional, or commercial servers, such as GitHub (https://github.com/), GitLab (https://about.gitlab.com/), BitBucket (https://bitbucket.org/), and Codeberg (https://codeberg.org/);
- *code archives*, used for long-term preservation of specific software versions, usually hosted on institutional or non-governmental organization servers, such as Software Heritage (https://www.softwareheritage.org/), Zenodo (https://zenodo.org/), Open Science Framework (https://osf.io/), and Dataverse (https://dataverse.org/), but also on commercial servers, such as figshare (https://figshare.com/);

- *code licenses*, used to specify the terms under which code can be used, modified, and shared (https://opensource.org/licenses, https://spdx.org/licenses/);
- *code documentation*, preferably itself under version control, hosted on platforms such as Read the Docs (https://about.readthedocs.com/) and GitHub Pages (https://pages.github.com/) or peer-reviewed in dedicated journals, such as the Journal of Open Source Software (https://joss.theoj.org/).

### 1.3 Misconceptions in Analogies and Ambiguity in Terminology

Expanding open science policies, principles, and resources from observations to modeling frequently relies on analogies. While analogies can be useful in making a point, they are not always accurate representations of the issue. By relying on analogies, one may overlook key differences between things being compared, leading to flawed conclusions and inefficient policies.

Misleading or weak analogies between modeling and observations include:

- equating open-source code requirements with open experimental data policies,
- conflating scientific models with analysis software,
- mistaking simulation runs for observational datasets, and confusing simulation service providers with observational data repositories.

Ambiguity in terminology contributes to misconceptions. The terms *model* and *data* allow multiple interpretations. For example, in data science the term *data model* is used in reference to a framework that organizes how data elements are structured, stored, and related within a database or information system. The term *data*, typically understood as observational data, is also used sometimes in reference to simulation output. The frequently used term *model-data comparison* implies comparison of simulated products derived from outputs of simulation runs with observational data products. A discussion on disambiguating terminology differences initiated by Edmund Henley (UK Metoffice) and Eric Adamson (NOAA Space Weather Prediction Center) is included in Kuznetsova et al. (2026b, Appendix B) with focus on terms most relevant to space weather capabilities assessment and transition to operations. Disambiguating terminology and resolving tricky analogies in heliophysics and space weather modeling are important and should be addressed in a dedicated community paper.

To better frame the discussion on how to advance modeling through open science, we propose more appropriate analogies:

- scientific models are tools for numerical experiments, the same way that instruments are used to perform observational experiments;
- model source code is akin to an instrument blueprint;
- simulation runs resemble experimental campaigns for a specific mission, with simulation output having the role of raw data collected from the mission's multiple instruments;
- timelines and space weather products derived from simulation outputs are comparable to observational datasets;

- post-processing and visualization software corresponds to data analysis software.

### 1.4 Community Engagement and Open Co-Authorship

To gather community feedback on how to best advance heliophysics and space weather modeling through open science, the Community Coordinated Modeling Center (CCMC, https://ccmc.gsfc.nasa.gov/) organized an Open Science Workshop in June 2024 (https://ccmc.gsfc.nasa.gov/ccmc-workshops/ccmc-2024-workshop/). This workshop was part of the 11th CCMC community biennial workshop, which, since 2001, brings together model developers, users, and agency representatives for an in-depth exchange of experiences, opinions, needs, and pathways forward in heliophysics modeling. The 2024 Open Science Workshop was sponsored by the NSF Innovations in Open Science Planning Workshops program. The workshop featured sessions focused on various aspects of open science in modeling including infrastructure, use of simulation results, use of models, model validation, model development, and the research-to-operations-to-research (R2O2R) transitions. Preparation for the workshop by the CCMC was conducted collaboratively, with each discussion session organized by modeling community representatives and dedicated CCMC facilitators. All the planned discussion topics were made available to the community for comments and suggestions two weeks before the workshop. Summaries from all sessions are publicly accessible (https://drive.google.com/drive/folders/1ChHWrCkLrWeOHlQU_jeDSofL3YTeUonO).

Questions addressed during the Workshop include: How can we make simulation runs reproducible? How can we capture all the information needed to describe a complex modeling framework consisting of multiple models exchanging information across different physical domains? Is access to the full source code really needed for the open use of models and their simulation outputs? Is open source alone sufficient for developing improved or new modeling capabilities?

One important outcome of the 2024 Open Science Workshop was that the community identified the need to create a bottom-up Heliophysics Open Modeling Environment (HOME) initiative. The major objective of HOME is to unite and give voice to all researchers across all physical domains in heliophysics with primary expertise in modeling and model-related software, as well as those using models and simulation results in their research. It will furthermore serve as a platform to continue the exchange on open science in heliophysics modeling. The HOME initiative, recently set as a Committee on Space Research (COSPAR) International Space Weather Action Teams (ISWAT, https://iswat-cospar.org) overarching activity, is discussed in more detail in Section 8.

In this paper, we present the outcomes of the discussions from the 2024 Open Science Workshop, along with opinions, feedback, and inputs collected after the Workshop, including those from the COSPAR ISWAT Initiative Working Meeting held in February 2025 in Florida. We complement these discussions with almost 25 years of experience in enabling open use of models at the CCMC and consider lessons from more recently established simulation service providers such as the Virtual Space Weather Modeling Center (VSWMC, https://spaceweather.hpc.kuleuven.be).

Preparation of this manuscript was carried out in an open manner. All workshop participants were encouraged to contribute to the paper. An open invitation to contribute was also advertised at the COSPAR ISWAT website and at the COSPAR ISWAT Working Meeting.

### 1.5 Open Science Themes Focused on Actions and Outcomes

Discussions in this paper are grouped by overlapping themes (see Figure 1) focused on actions (use, development, validation, and collaboration) aiming at goals (e.g., advancing space weather modeling and forecasting) rather than on policies and principles:

- *Open Use*: how to maximize science return on investments in model development and to enable broader utilization of state-of-the art numerical models in research without the need to modify (and even access) source codes; how to visualize and analyze simulation results; and how to enhance discovery and reusability of past simulation outputs.
- *Open Validation*: how to improve transparency of code quality, model robustness, and model output uncertainty; how to facilitate model evaluations by peers; how to prepare both experimental and simulated data for model validation and publicly communicate results of validations to interested stakeholders.
- *Open Development*: how to increase momentum of model development by building upon existing capabilities; how to enable model development with the fewest impediments to potential contributors; how to foster and grow the modeler community and create an environment for collaborative model improvements.
- *Open Collaboration*: how to facilitate tighter interactions between model developers and users; how to improve openness of collaborations and encourage interdisciplinary and international partnerships.

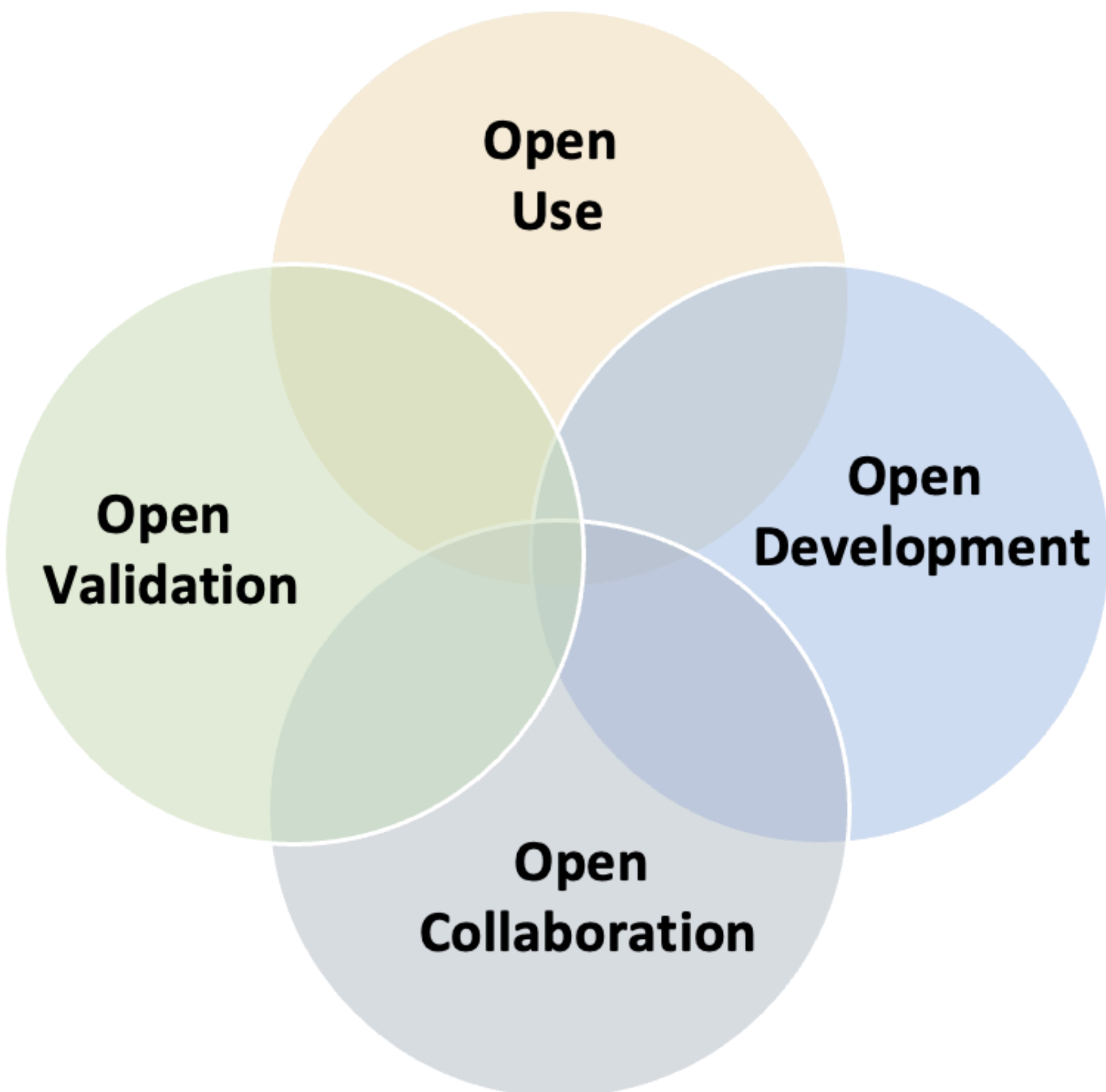

**Figure 1.** Overlapping themes of open science in modeling, aiming to push frontiers in heliophysics research and space weather forecasting.

1.6 Paper Organization

This paper is structured as follows. Section 2 discusses the importance and challenges of modeling in heliophysics and space weather. Section 3 presents the outcomes of a survey we conducted on the perspectives of modeling teams regarding successes and challenges in open science. Sections 4 to 7 focus on the four open science themes illustrated in Figure 1. Section 4 describes current approaches for the open use of models and simulation outputs, highlighting the role of simulation service providers (such as the CCMC and the Virtual Space Weather Modelling Centre, VSWMC) in enabling the global heliophysics and space weather communities to take advantage of state-of-the-art numerical models. Limitations and difficulties, both technical (e.g., how to access very large simulation outputs) and organizational (e.g., how to define metadata standards accepted by all actors involved), are discussed. In addition to suggestions for simulation service providers and model developers, the responsibilities of users in following best practices are also emphasized. Section 5 reviews current efforts in validating models, both for scientific and operational use, focusing on community validation campaigns and the infrastructure necessary for carrying out a truly open validation process. Section 6 presents various topics related to open development, including: how to incorporate best practices from the software industry in the model development process; how to increase the coordination among modelers, simulation service providers, and funding and operational agencies to develop the next generation of scientific and operational models; and how to address some challenges specific to the intersection of scientific and operational models and open source. Section 7 discusses current (e.g., ISWAT) and future ways in which various actors can collaborate to solve specific problems, such as how to integrate simulation outputs and observational data products, how to allow users to enhance reusability of simulation outputs, and how to better coordinate forecasting activities among international operational centers. Section 8 describes the HOME initiative and its role in advancing modeling through open science. Section 9 summarizes key recommendations and pathways for advancing heliophysics modeling through open science. Appendix A includes the full perspectives on open science from the modeling teams that participated in the survey summarized in Section 3, along with description of their numerical models. This serves as a comprehensive, yet incomplete, view of the status of the heliophysics and space weather modeling community, illustrating the variety of physics phenomena simulated as well as the technical approaches and organizational structures adopted by the various developing teams. Appendix B clarifies the advantages and disadvantages of the use of cloud file storage solutions (such as Amazon S3) in scientific computing, while Appendix C lists acronyms and definitions used throughout the paper, organized by topic.

An executive summary of this paper can be found in Reiss et al. (2026), published in AGU Commentaries on Space Weather and Space Physics.

## 2 The Role of Models in Heliophysics and Space Weather

Successful implementation of open science practices requires recognition of the role models play in heliophysics and space weather. Numerical models are crucial in advancing fundamental research in heliophysics and operational forecasting of space weather events. Broadly speaking, models enable executable science, where theories that explain physical processes are captured in simulation codes, developed continually, and verified against empirical experiments. Models are essential for understanding the connections between the Sun and Earth, particularly in unraveling the links between solar eruptive phenomena and their effects in interplanetary space and planetary environments. Additionally, model solutions can link observations from vastly separated locations in the heliosphere to create a coherent physical picture. Many heliophysics models result from years of collaborative work among domain specialists/researchers, algorithm experts, software engineers, system engineers, students, and other dedicated individuals. Their further development, improvement, upkeep, and knowledge transfer to the next generation of researchers deserves special attention within the community. Models and coupled modeling systems (also called modeling frameworks) can be seen as *instruments for numerical experiments* that complement space missions and ground-based facilities.

The U.S. Decadal Survey for Solar and Space Physics (U.S. Decadal Survey in the rest of the paper) correctly recognizes the importance of models to advance our knowledge of the heliosphere, putting them on equal footing with observations. Large-scale models are a critical element of an envisioned integrated HelioSystem Laboratory, along with ground- and space-based assets, since they generate simulated data that can be used to verify theories and interpret experimental data. The U.S. Decadal Survey also recognizes how developing large-scale models is a complex endeavor, akin to the design and construction of space- and ground-based instruments, missions, networks, and infrastructures: "Model development has become as complex as space missions or ground-based facilities." (NASEM, 2025, p. 28). In addition, the U.S. Decadal Survey points out that "In the past decade, physics-based models have made great strides in delivering scientific breakthroughs, explaining data from existing assets, motivating new missions, and furthering the national space weather enterprise. These advances were achieved with relatively modest investment by the funding agencies via a hierarchy of programs [...]." (NASEM, 2025, p. 157). For these reasons, the U.S. Decadal Survey recommended the creation of a flagship community modeling program, with expanded funding support for model development and workforce training and retention, capable of addressing large-scale heliophysics problems by exploiting the evolving supercomputing resources, and meeting the open science requirements: "A significant expansion of the funding hierarchy for theory and modeling is needed to meet the challenges of ever-increasing physical complexity of the models, rapid development of the high-performance computing landscape, and adherence to open science standards." (NASEM, 2025, p. 158). Similarly, as highlighted in Kepko et al. (2024, Fig. 7), the vision for the next generation of the International Solar Terrestrial Physics (ISTP) program, often referred to as ISTPNext, includes theory and modeling as tools for comprehensive, coordinated system-of-systems science together with space- and ground-based observations.

Although implemented through programming languages, models are much more than just software — they are elaborate mathematical representations of physical systems. This is also recognized in other disciplines. For example, numerical weather prediction models have never been simply considered software. Like many space physics models, numerical weather prediction models have improved with growing computing power, new observations, and the development of more accurate and efficient numerical methods (Lynch, 2008). While heliophysics and space weather models focus on physical phenomena (with scientific and operational objectives, respectively), model-related software supports and facilitates the use of these models in research and forecasting. Examples include compilation and installation scripts, supporting infrastructure software (such as data pipelines, queue management systems, and containers), as well as post-processing, visualization, analysis, and validation software. This software is akin to data analysis tools used to study observational data, where models act as experiments generating simulated data. It also resembles mission-related software (like calibration and monitoring), where models function as instruments and infrastructures to be managed. The complexity of the computing ecosystems needed for both models and experiments is illustrated in Figure 2, which shows relationships between models, experiments, data, and theory. For these reasons, models should be recognized as crucial research tools and distinguished from model-related software when discussing open science guidelines.

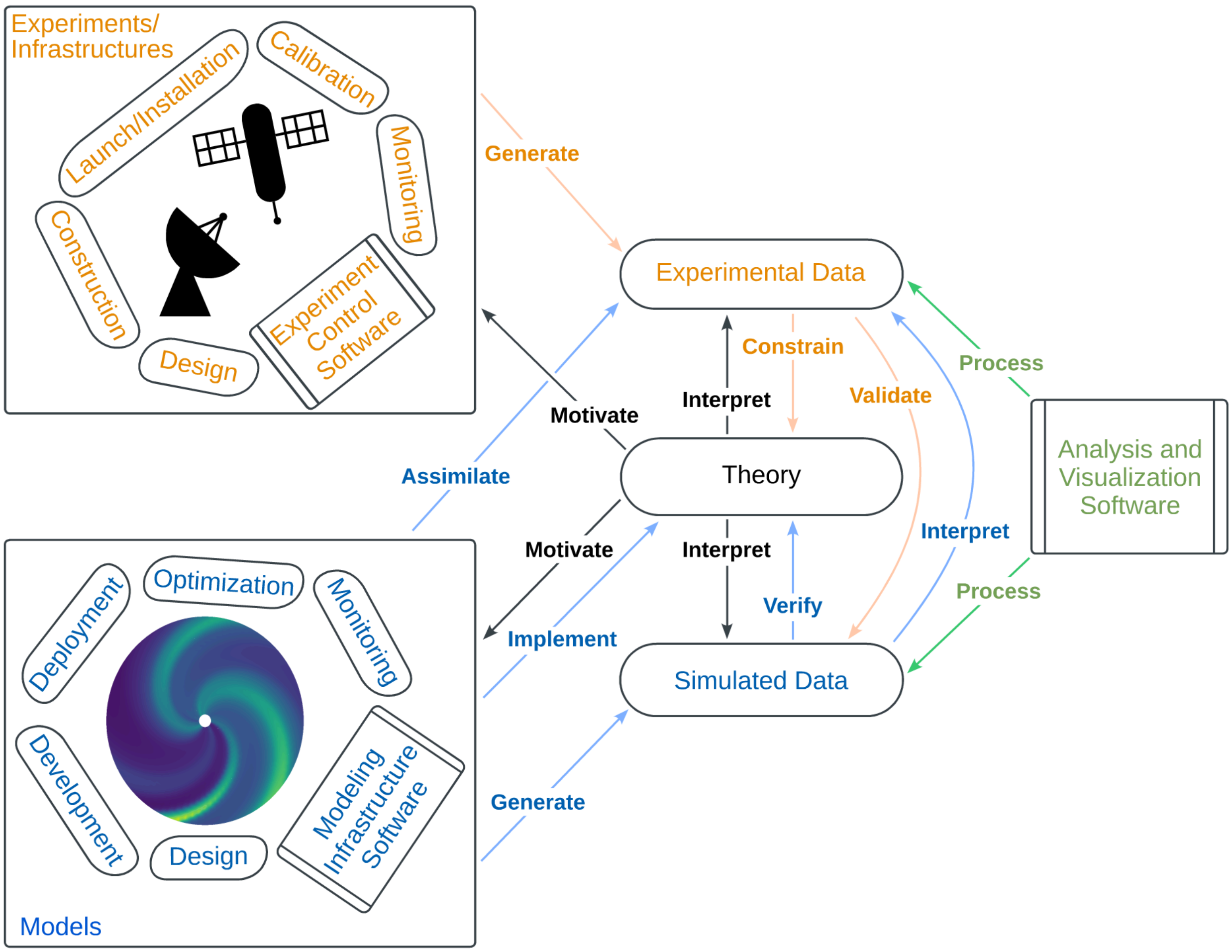

**Figure 2.** Relationships between experiments and models, illustrating their roles with different computing ecosystems. Arrows are directed from the components that perform the specified action to the components on which that action is performed. To simplify the diagram, the Assimilate arrow between Models and Experimental Data includes both proper data assimilation (i.e., continuous update of model predictions based on latest available data) and data being used only once as input to drive a model as boundary/initial conditions.

## 3 Successes and Challenges in Open Science in Heliophysics Modeling: Modelers' Perspectives

Following the 2024 Open Science Workshop, the Community Coordinated Modeling Center (CCMC, https://ccmc.gsfc.nasa.gov/) created a living online survey to gather modelers' perspectives on open science in model development and community usage. The survey was administered via Google Form to all modelers that agreed to contribute to the paper and consists of 39 questions organized in 6 sections: model general information (e.g., development history, computational requirements), open development information (e.g., code accessibility, licensing, software development practices), open use information (e.g., model accessibility, pre- and post-processing tools, output format and metadata), questions on open science successes and challenges, and additional comments. The full survey questions are available on Zenodo (Corti et al., 2026). Table 1 lists the 27 participating models in alphabetical order, along with details about development history and requirements for running their models and storing the outputs. A comprehensive description of all the models, along with some developers' experiences with open science, is provided in Appendix A. The contributions range from single developers to full teams of scientists and software engineers, from models created decades ago to recent modeling efforts, spanning a wide spectrum of codebase size and various levels of openness, from models available only as source code to models available to be run at the CCMC or at the Virtual Space Weather Modelling Centre (VSWMC, https://spaceweather.hpc.kuleuven.be).

**Table 1**. *List of Models Participating in Modelers' Perspectives and Survey. Expanded acronyms can be found in Appendix C. Links are listed in full in each model's subsection in Appendix A.*

| Model name | Model type and domain | Developers[a]; Codebase size (LOC[b]); Programming language | Computational requirements | Storage requirements; Output format | Available at |
|---|---|---|---|---|---|
| AMPS | Particle tracer in space and planetary environments | 20, over the course of 24 years; 360k; Fortran | 500–1000 CPU hours for planetary applications | 10–15 GB; Tecplot | CCMC |
| COCONUT | MHD solver for the corona | 40+ (5), over the course of 20 years; 1.1M; C++ | From a few hundred CPU hours, 0.2 GB RAM, up to 70k CPU hours, 5+ GB RAM, per simulated Carrington rotation, depending on grid resolution and on steady-state or time-evolving | A few GB per snapshot, up to 200 GB for a high-resolution time-evolving Carrington rotation simulation; Tecplot, VTK, Text | GitHub, VSWMC (planned) |

| Model name | Model type and domain | Developers[a]; Codebase size (LOC[b]); Programming language | Computational requirements | Storage requirements; Output format | Available at |
|---|---|---|---|---|---|
| | | | simulation | | |
| DTM | Semi-empirical atmospheric drag | 2/year, over the course of 50 years; 10k; Fortran | Single CPU core | 20 kB per orbital location; Text | GitHub, CCMC |
| E-CHAIM | Empirical high-frequency radio propagation in ionosphere | 3 (1), over the course of 10 years; thousands; IDL, Matlab, C | Single CPU core | < 1 MB; Text | Model website |
| ENLIL | MHD solver for the heliosphere | 1, over the course of 25 years; 220k; Fortran, IDL, scripts | 10 min on 16 cores for 7 simulated days at low resolution | From hundreds of MB up to a few GB per simulated hour, depending on resolution; NetCDF, Text | CCMC |
| EPREM | Numerical solver for solar energetic particle acceleration and transport in heliosphere | 10, over the course of 20 years; 10k; C | From 1 CPU hour, 1–2 GB RAM, up to 100 CPU hours, 10 GB RAM | From hundreds of MB up to hundreds of GB; NetCDF, Text | GitLab, CCMC |
| EUHFORIA | MHD solver for the heliosphere | 15 (1), over the course of 9 years; 400k; C++, Python | 20 min on 224 cores per simulated day | 50-70 GB, depending on frequency of update; NPZ, Text | VSWMC, CCMC |
| GITM/Aether | Numerical solver for neutrals and ions in ionosphere and thermosphere | 4, over the course of 22 years; 90k; Fortran, C++ | 500 CPU hours per simulated hour at medium resolution | 180 MB every 5 simulated minutes; NetCDF, Binary | GitHub, CCMC |

| Model name | Model type and domain | Developers[a]; Codebase size (LOC[b]); Programming language | Computational requirements | Storage requirements; Output format | Available at |
|---|---|---|---|---|---|
| Heliotomo | Time-dependent 3D reconstruction of solar wind density and velocity, including transients across the inner heliosphere | 25 to 30 (10), over the course of 40 years; 200k; Fortran, IDL | From 1 GB to 1 TB RAM | From 1 GB to 2 TB, depending on resolution; Text | CCMC |
| HYPERS | Hybrid (particle-in-cell + fluid) solver for planetary environments | 2, over the course of 14 years; tens of thousands; C++, Fortran, Perl, IDL, Python | 20 hours on 20k cores, 2–10 GB RAM | 100–300 GB; Text, Binary | CCMC |
| Icarus | MHD solver for the heliosphere | 5, over the course of 5 years; 4.8M; Fortran | 150 CPU hours for standard resolution | 1.3 GB every 3 simulated hours; Text, VTK | GitHub, VSWMC (planned), CCMC (planned) |
| IMPTAM | Particle tracer and numerical solver for magnetosphere | 1, over the course of 15 years; 50k; C++, Fortran, Python | From 12 CPUs, 4 GB RAM to 48 CPUs, 24 GB RAM per simulated day, depending on resolution | 40 GB per simulated day; Text | Model website, CCMC |
| iPIC3D | Particle-in-cell solver for space plasma and planetary environments | 20–25 (2), over the course of 20 years; 100k; C++, Matlab, Python, Shell | Up to 40k CPU hours, 1k GPU hours, 512 GB RAM | From 100 GB up to tens of TB; HDF5, VTK, BP5 | GitHub |
| MAGE | MHD solver for | 25 (3), over the | From 400 up to | From 6 up to | GitHub, |

| Model name | Model type and domain | Developers[a]; Codebase size (LOC[b]); Programming language | Computational requirements | Storage requirements; Output format | Available at |
|---|---|---|---|---|---|
| | magnetosphere and ionosphere | course of 8 years; 100k; Fortran | 100k CPU hours per simulated hour, depending on resolution | 400 GB per simulated hour, depending on resolution; HDF5 | CCMC |
| MARBLE/ AGATE | MHD solver for planetary environments | 5, over the course of 2.5 years; 100k; Python | From single CPU core up to tens of GPUs (or thousands of CPUs) for several days/weeks, depending on problem size | From a few GB up to a few TB, depending on problem size; HDF5 | GitHub |
| MAS/CORHEL/ CORHEL-CME | MHD solver for solar corona and heliosphere | 2-7 over the course of 30 years; several hundred thousand; Fortran, Shell, Python, C/C++ | From single CPU core up to tens of GPUs (or thousands of CPU cores) for under an hour to several days depending on problem type and size | From less than 1 GB to multiple TB depending on problem type and size; HDF4, HDF5 | GitHub, CCMC |
| PARADISE | Numerical solver for solar energetic particles in the heliosphere | 5, over the course of 9 years; 23k; C++ | 150 CPU hours per simulated day, 1.5 GB per core, for 25M particles | 600 GB per simulated day; HDF5 | VSWMC (planned) |
| RAM-SCB | Kinetic solver for particles and fields in the magnetosphere and atmosphere | 10, over the course of 30 years; 100k; Fortran | 30 CPU minutes, 2.5 GB RAM per simulated hour at medium resolution | 900 MB per simulated hour at medium resolution; NetCDF, Text | GitHub, CCMC |
| SAMI3 | Numerical solver for ions in | 3, over the course of 27 | From 128 CPU hours, 12 GB | From 24 GB up to 360 GB per | CCMC |

| Model name | Model type and domain | Developers[a]; Codebase size (LOC[b]); Programming language | Computational requirements | Storage requirements; Output format | Available at |
|---|---|---|---|---|---|
| | ionosphere and plasmasphere | years;<br>20k;<br>Fortran | RAM up to 3k CPU hours, 128 GB RAM per simulated day | simulated day;<br>NetCDF, Binary | |
| SEPMOD | Numerical solver for solar energetic particles in the heliosphere | 1, over the course of 15 years;<br>1k;<br>Fortran | Single CPU core | < 1 MB;<br>Text | CCMC |
| SPS | Numerical solver for plasma in planetary magnetospheres | 2–5, over the course of 14 years;<br>15k;<br>C | From 5k up to 20M CPU hours, from 8 GB to 3.2 TB RAM, depending on resolution | From 3 GB to 1.2 TB (reduced data), from 60 GB to 24 TB (full data snapshots), depending on resolution;<br>HDF5 | – |
| SWASTi | Semi-empirical solar corona + MHD solver for heliosphere | 6, over the course of 5 years;<br>1M;<br>C/C++, Python | 200 CPU hours per Carrington rotation at medium resolution | 300 MB at medium resolution;<br>VTK | – |
| SWMF | MHD solvers for corona, heliosphere, and magnetosphere, coupled with numerical solvers for solar energetic particle in heliosphere/ magnetosphere and other atmosphere/ ionosphere models | 50+ (4), over the course of 22 years;<br>1M;<br>Fortran, C++, IDL | From 1 up to 1M CPU hours, from 1 GB up to 1 TB RAM | From 1 GB up to 10 TB;<br>HDF5, FITS, Text, Binary, Tecplot, VTK | GitHub, CCMC |

| Model name | Model type and domain | Developers[a]; Codebase size (LOC[b]); Programming language | Computational requirements | Storage requirements; Output format | Available at |
|---|---|---|---|---|---|
| TIE-GCM | Numerical solver for thermosphere and ionosphere | 50 (10), over the course of 50 years; 70K; Fortran | 400 CPU hours, 235 GB RAM per simulated day at default resolution | 300 GB per simulated day at default resolution; NetCDF | GitHub, CCMC |
| Vlasiator | Numerical solver for plasma in geospace | 20 (5), over the course of 15 years; 220k; C++ | 3–30 million CPU hours | 50–100 TB; VTK-compatible binary | GitHub |
| WACCM-X | Hybrid solver for thermosphere and ionosphere | 20 (3), over the course of 18 years; 3M; Fortran | From tens up to tens of thousands of CPUs, from hundreds of GB to tens of TB RAM, depending on resolution | From 8 MB up to 1 GB per 3D field per frame, depending on resolution; NetCDF | CCMC |
| WSA | Combined empirical and physics-based model of the corona and solar wind | 2–5/year (2), over the course of 30 years; 50k; Fortran, Python | 6 CPU hours, 16 GB per simulated day (4x for ADAPT ensemble maps) | <100 MB per simulated day; HDF5, FITS, Text | CCMC |

[a] In parenthesis, the number of software engineers in the team, if any.
[b] LOC: lines of code.

Our living survey reveals that not all modelers face the same challenges and have similar experiences, e.g., different levels of sustained support for model development. It also highlights diverse user bases (students, researchers, operational entities, and private sector), each with distinct behaviors regarding code modification and model acknowledgment. This leads to differing opinions on how easy or difficult it is to embrace open science principles. For instance, in terms of openness, all but ten models in Table 1 make their source code available for download from a web page, a code archive, or a code repository. For the rest, either the source code is not available (six models) or users need to contact the developers via email or through registration on the model website. The source code licenses vary significantly among the teams.

Eleven models do not adopt any license, two models use a custom or modified open-source-like license, one model uses an academic license, while the rest are divided among Apache (https://spdx.org/licenses/Apache-2.0.html), GPL (https://spdx.org/licenses/GPL-2.0-or-later.html), MIT (https://spdx.org/licenses/MIT.html), and BSD (https://spdx.org/licenses/BSD-3-Clause.html) licenses. Only one model is released under the NASA Open Source Agreement license (NOSA, https://spdx.org/licenses/NASA-1.3.html). Most teams accept external code contributions, but only four teams have established guidelines for contributions in their code repository. Among the other models, eight groups do not allow external contributions, while two others allow trusted collaborators only. Most groups adopt one or more software development best practices, with a preference for testing and frequent commits (25 and 22 modeling groups, respectively). Build automation is used by fourteen groups, while only five groups adopt a continuous integration workflow.

All the participating models are currently available for open use, except for three. They either allow users to install locally (19 models) or run at the CCMC and/or VSWMC (18 models). To support the community with model usage, they primarily connect via email (26 modeling groups) and code repository platforms like GitHub or GitLab Issues (13 groups). Some groups provide support on social platforms like Slack and Discord (6 teams). The format of the simulation run outputs varies as well. The spectrum includes text files (14 models), HDF5 (8), NetCDF (7), VTK-compatible formats (6), custom binary formats (4), FITS (3), Tecplot files (3), and NPZ (1). Only three models produce output files that conform to a community-recognized metadata standard (CF for NetCDF files, Eaton et al., 2025; and WCS for FITS files, FITS Working group, 2018). Regarding pre- and post-processing tools needed to run the model or interpret the results, thirteen modeling groups provide both pre- and post-processing tools, six groups provide post-processing tools only, while nine other groups do not provide any processing tool.

While the open science approach varies for each team, common views on best practices do emerge. Table 2 shows the perspectives on open science from participating modelers, while Sections 4 and 6 discuss agreements and disagreements among modelers in open model use and development. There is a general agreement that making the source code public can improve its robustness. However, modelers highlight that unclear licensing and poor documentation are obstacles. Additionally, there is consensus that funding should specifically target documentation, maintenance, user support, and community engagement. Regarding code documentation, there is agreement that extensive documentation, including user training on the model's capabilities and limitations, is vital for model use. There is also consensus that open source broadens the user base, thereby opening collaboration possibilities. Another advantage mentioned is that open models can ideally be tailored to user needs. Some teams noted that acknowledgments to developers and funding agencies are sometimes missing. They also highlight that in some cases users modify models and publish results based on these modifications without contacting the original developers. Therefore, they recommend a more thorough peer review process to catch instances of such situations.

**Table 2**. *Summary of Recurring Modelers' Views on Open Science.*

| Theme | Recurring views | Modeling group |
|---|---|---|
| Verification and Robustness | Open source improves code robustness.<br>Regression and unit tests are needed to verify code modifications. | AMPS, MAGE, SWMF, TIE-GCM |
| Licensing | Unclear licensing and poor documentation hinder open-source code release, especially for legacy codes. | EPREM, MAGE, MARBLE/AGATE, WACCM-X, WSA |
| Funding | Funding and time are needed for documentation, maintenance, user support, and community engagement.<br>Limited HPC access is a bottleneck. | EUHFORIA, GITM/Aether, Heliotomo, HYPERS, Icarus, IMPTAM, iPIC3D, MAS/CORHEL, PARADISE, RAM-SCB, SEPMOD, SPS, TIE-GCM, Vlasiator, WACCM-X |
| Documentation and User Training | Extensive documentation and user training (e.g., summer schools, workshops) are important. | DTM, E-CHAIM, HYPERS, MAGE, RAM-SCB, SWMF |
| Tools and Visualization | Open pre-/post-processing tools (including standardized input data) and advanced visualization are needed to enhance usability. | DTM, E-CHAIM, EPREM, HYPERS, MAGE, MAS/CORHEL |
| User Base | Open source/use broadens user base and enables collaborations.<br>Models can be tailored to user needs.<br>Graphical user interfaces improve accessibility. | COCONUT, EUHFORIA, Icarus, MAS/CORHEL, PARADISE, RAM-SCB, SAMI3, SWMF, TIE-GCM |
| Credit | Users often fail to acknowledge developers and funding agencies.<br>Original developers might not be notified of custom code changes before publication.<br>Peer review should enforce acknowledgments. | E-CHAIM, SAMI3, Vlasiator |

To understand the challenges and successes of current open science principles in the community, we asked the teams to rank statements on open science success and challenges from 1 (strongly disagree) to 5 (strongly agree). Figures 3 and 4 show the modelers' responses to these statements. There is a general agreement that open use is valuable for broadening the usage of the models. However, there is less agreement that making the source code available alone has the same effect. Disagreements stem mainly from worries about misusing models (e.g., simulating conditions and scenarios beyond model tested capabilities) and/or introducing physics mistakes and performance degradation when non-experts modify the model's code, which then translate to reputational damage to the original model. A few modelers are also concerned about potential software piracy, where a model code is modified and repackaged with a different name without recognition for the original developers, and less innovation in model development due to the current funding structure (see also Sections 6.1 and 6.3). In addition to making the model available for open use and/or releasing the source code, a robust ecosystem (e.g., documentation, pre- and post-processing tools, user support) and advertising model capabilities via conferences and papers are also recognized as crucial factors in increasing the user base. Regarding user support, while open model use via web decreases the amount of technical requests about model installation and execution, this advantage can be offset by a broader adoption which leads to more requests about how to configure a model for a specific use case and how to properly interpret model results. This can become a problem if such support is not funded.

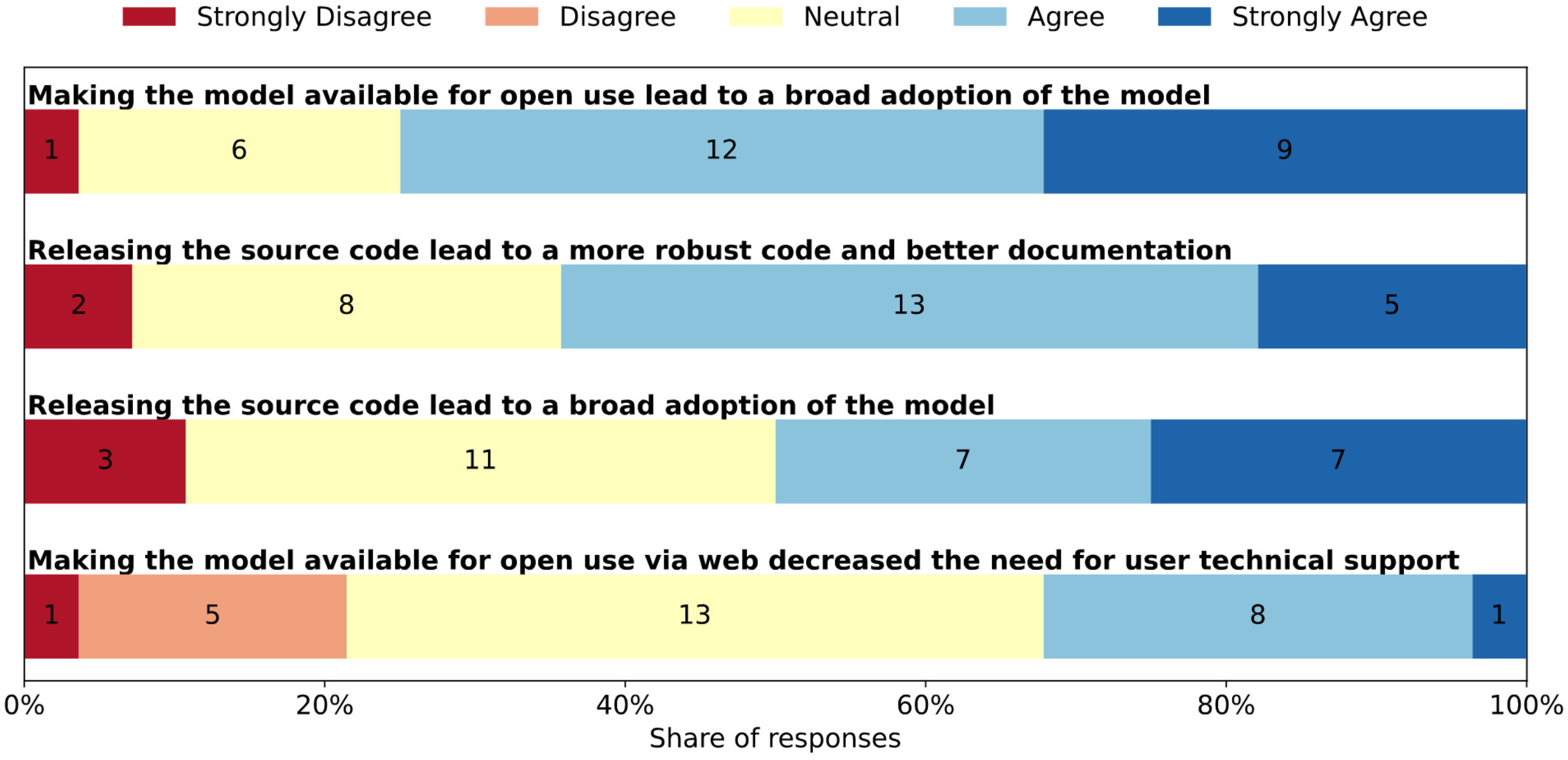

**Figure 3.** Survey results: number of modeling groups agreeing/disagreeing with specific statements on open science successes.

In general, most modeling teams have positive opinions on open source, open use, and their advantages, including for very computationally intensive models. Indeed, the majority of modelers disagree that open use should be reserved only for small to medium models. While it is recognized that typical users don't have the resources to run large/complex codes and it is unreasonable to expect simulation service providers to allow everybody to run such models on demand, the option to have access to the code and/or execution of large models should be preserved. Only two modelers prefer prioritizing the computational resources of simulation service providers and modelers to maintain a curated set of large models outputs for selected events and specific physics scenarios. Almost all modelers agree that releasing the source code is just the first step towards open use of models, and that pre-/post-processing and visualization tools are required for fully taking advantage of open use models (although the level of agreement on this point depends on the complexity of the model output). Similarly, almost all modelers point to a lack of financial and legal support from funding agencies to properly sustain code maintenance, documentation, development, and transition from closed/legacy codes to open-source codes. None of the modelers disagree with the potential risk that code modifications might reflect badly on models' capabilities and performance when the original model developers are not involved in the development process.

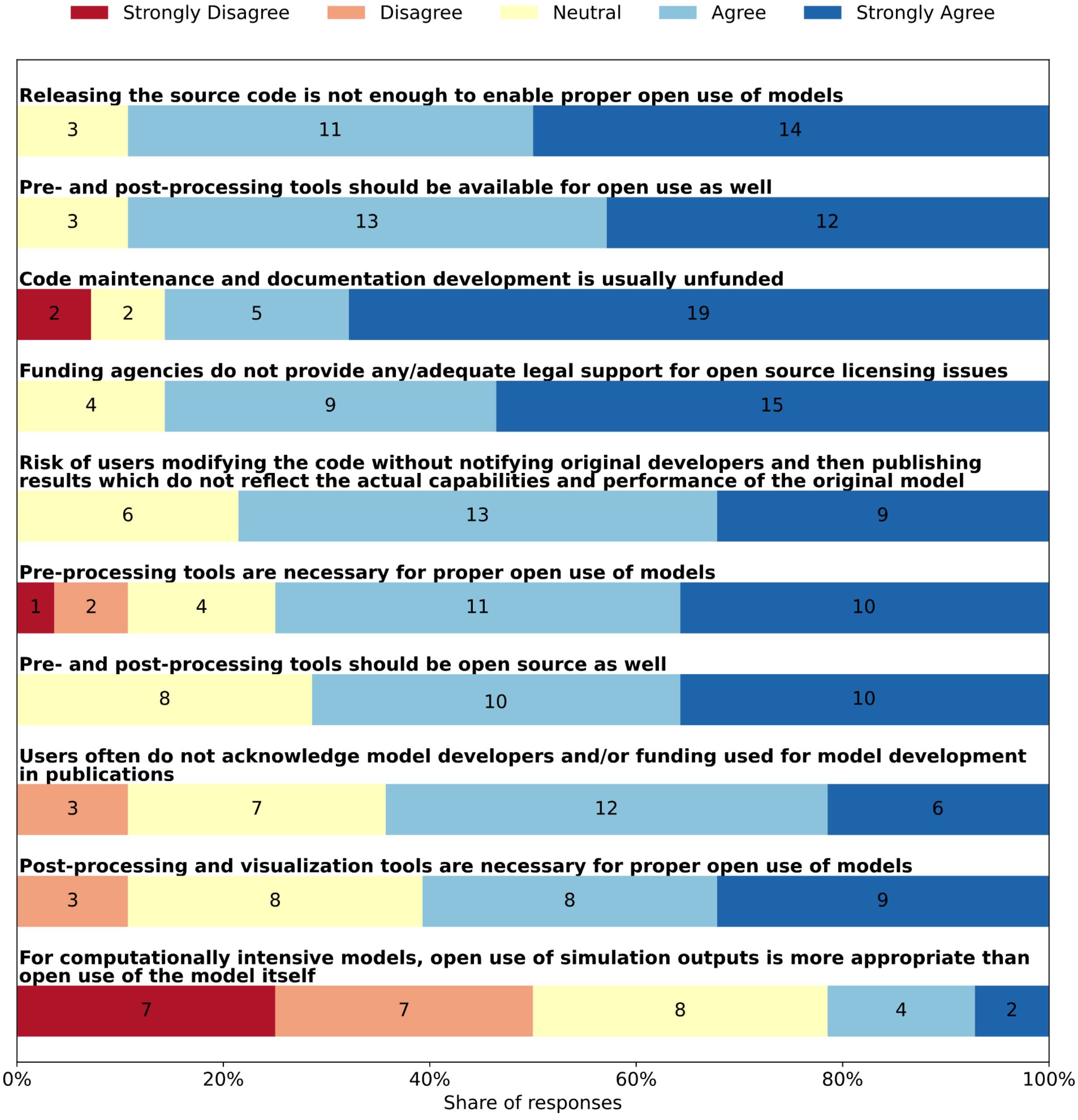

**Figure 4.** Survey results: number of modeling groups agreeing/disagreeing with specific statements on open science challenges.

We note that this open science survey conducted by the CCMC is an ongoing effort. Its objective is not only to capture the current state of open science in the modeling community, but also to serve as an ongoing collaborative activity in the HOME initiative. By maintaining timely questions and inviting more model developers to join, we aim to provide the community with up-to-date insights into the successes and challenges of open science in heliophysics and space weather.

## 4 Open Use

The open use theme is frequently overlooked in open science guidelines and program funding guidelines and is overshadowed by an open-source requirement. Open use of a model, however, even without access to its source code, enables model employment by a broad group of researchers who are not model developers themselves. This maximizes science return on investment in model development and advances innovation.

As pointed out by many modelers in Section 3 and Appendix A, the availability of the open-source code alone is not sufficient to ensure that models can be used without barriers. Proper use of a model requires installing the model, understanding its workflow, gathering and pre-processing necessary input data and/or parameters, configuring simulation settings tailored for science questions of interest, running it, and finally analyzing and visualizing its output in a useful way. Pre- and post-processing tools are crucial and need to be made available, whether as open source, via web interface, or via an application programming interface (API). Depending on the case, development of these tools can represent a sizable fraction of the resources needed to make the model openly usable, a fact that is not always recognized. Similarly important for users is the availability of clear and comprehensive documentation, even better if complemented by test cases, examples, and tutorials. Finally, dedicated user training (e.g., at summer schools and working meetings) is another way of ensuring that users know model capabilities and limitations and can maximize the return on model use, both in terms of extracted physics and usage efficiency. Since most code development today is done as part of science objective-driven research grants, all the above — which requires considerable time and effort from modelers — usually is not covered by grant funding. This is why it is imperative to recognize the need to have dedicated personnel and funding for such tasks.

The increase in computational resources of the last decade has enabled the advent of advanced models capable of simulating physics problems with higher spatial and time resolution. However, these new models sometimes require a large amount of storage and computing resources not affordable by most users, even with time allocations on high performance computing (HPC) clusters or when using cloud computing. For these models, open use is less about access to run the code, but rather about access to very large simulation outputs. In this case, modeling centers, such as the Community Coordinated Modeling Center (CCMC, https://ccmc.gsfc.nasa.gov/) and the Virtual Space Weather Modelling Centre (VSWMC, https://spaceweather.hpc.kuleuven.be, Poedts et al., 2020), could serve as hosts for simulation outputs, created either on premise or on HPC clusters, in the form of databases of common cases for specific physics domains, which users can then analyze as they wish. The essential requirement for open use of simulation outputs is a comprehensive, accessible, and searchable metadata description of each simulation run. The availability of free and/or open-source advanced 3D visualization tools as an alternative to very expensive commercial software is important to ensure the usability of these large simulation outputs.

Section 4.1 describes current approaches for the open use of models. Section 4.2 addresses different topics related to open use of simulation outputs including: challenges of open use of large outputs (4.2.1); output file formats (4.2.2); complexity of simulation run metadata (4.2.3); and peculiarity of metadata for simulations with artificial conditions (4.2.4). Sections 4.3 and 4.4 highlight simulation services at the CCMC and VSWMC. Section 4.5 underscores efforts on the query interface interoperability. Section 4.6 emphasizes the importance of enabling open use of simulation outputs produced by developers and users. Section 4.7 discusses implementation of ML techniques for analysis of open simulation output archives. Section 4.8 outlines open use best practices.

### 4.1 Open Use of Models

The open use of models does not require access to and modification of model source code. Modifying even a single line of source code requires following open development best practices, discussed in Section 6 of this paper. Instead, simulation runs for open use can be set through configuration/input files. Cutting-edge models are often experimental and rapidly evolve, with complex inputs and outputs and complicated setup and running requirements. For complex modeling systems, selecting parameters in a configuration file requires experience with specific models and understanding which simulation settings are the most relevant for reproducing specific physical phenomena or addressing specific science questions.

Enabling open use of models by a wider heliophysics and space weather community requires solutions that could close this complexity gap to a point where the model could be operated by an individual with limited domain knowledge and no knowledge of the internal organization of the modeling system. At the same time, such solutions should remain flexible and transparent enough to provide sufficient and traceable information on the underlying physics, assumptions, constants, data sources, inputs, formats, and other parameterizations used in the model.

There are several existing approaches to the open use of models, each with their advantages and challenges depending on the goals of model use and user experience (see Table 3). Simulation services at the CCMC and VSWMC enable users to execute models remotely on the provider's hardware, without having to install the models nor their supporting software on the user's computing systems.

The ability to access the source code and to configure, compile, and execute models on a user's choice of systems provides maximum flexibility and enables reproducibility tests of simulation runs on different systems. However, it requires access to computational resources, installation of the model and its supporting software, as well as understanding all simulation inputs and settings. For some models, installation requires the use of specific versions of libraries and compilers which might conflict with the default system environment. This hurdle can be overcome by clear documentation and by using build automation tools, such as CMake (https://cmake.org/), Autotools (https://www.gnu.org/software/automake/), or Meson (https://mesonbuild.com/). When feasible, the whole build and deployment process can be automated by using containers and/or dependency and environment management systems, such as Conda (https://docs.conda.io/), Module (https://envmodules.io/), or Spack (https://spack.io/). See Sections 4.3 and 4.4 for discussions on the advantages and challenges of this approach. With a local installation, users have more opportunities to experiment with simulation settings by editing configuration files than by using interfaces provided by the CCMC and the VSWMC. However, there are also more opportunities for misuse, including going beyond recommended parameter ranges. This approach is mostly utilized by advanced users interested in performing non-standard simulations and/or model development. The ability to tailor simulation settings through configuration files without the need to modify the source code is an important capability to facilitate broader use of this approach, especially for models that do not require significant computational resources.

**Table 3**. *Advantages and Challenges of Different Approaches to the Open Use of Models.*

| Approaches to the open use of models | Advantages | Challenges |
|---|---|---|
| *Simulation services through interactive web interfaces* | Free and easy access to advanced simulations by non-experts.<br><br>Does not require access to computational resources and experience in code installation, compilation, and run configuration.<br><br>Removes some of the technical support burden from developers.<br><br>Provides expert guidance in selecting simulation settings for modeling specific physical phenomena or addressing specific science questions.<br><br>Creating configuration files from information provided by users through web submission interfaces reduces the probability of misuse.<br><br>Enables interactive generation of input parameters and files.<br><br>Ensures preservation of simulation settings (including code versions, configuration files, and input data) in simulation service provider archives to enable reproducibility, if requested. | Less flexibility for advanced users to configure simulation settings.<br><br>Less optimal for on-the-fly model coupling, run submissions from Jupyter notebooks, and automatic pipelines. |

| Approaches to the open use of models | Advantages | Challenges |
|---|---|---|
| *Simulation services through command line via Application Programming Interface (API)* | Permits a model supported in one system to be readily used as a building block or a step in a model chain in another system or tool.<br><br>Makes it possible to run a simulation and analyze results in a Jupyter Notebook without having to navigate outside the notebook.<br><br>Allows for a more straightforward interconnection between model infrastructure systems.<br><br>Can be used as a back end for web-based services. | Less straightforward approach for non-expert users, requiring learning specific commands and syntax.<br><br>Lack of interactivity and user-friendliness during a run submission process, especially for complex API interactions that a web interface could simplify.<br><br>Typographical errors or misconfigured commands can lead to unintended actions.<br><br>Responses can be obscure, which may lead to misinterpretation.<br><br>Simulation service providers face potential security drawbacks that could lead to service disruptions for their intended users. |
| *Configuring and executing models by users on system of their choice* | Most flexibility for advanced users to select options in configuration files not available through web interfaces.<br><br>Encourage users to browse source code and to provide feedback to model developers.<br><br>Facilitates reproducibility tests of simulation runs on different systems.<br><br>A pathway for users interested in getting involved in model development. | Require access to computation resources and experience in code installation and system environment configuration.<br><br>Need comprehensive configuration files and documentation, including instructions for code installation.<br><br>Installation at the click of a button may not be a realistic |

| Approaches to the open use of models | Advantages | Challenges |
|---|---|---|
| | | expectation for rapidly developing modeling systems.<br><br>More opportunities for code misuse, since users are not obligated to read documentations and follow best practices. |

4.2 Open Use of Simulation Outputs

The open use of simulation outputs is closely connected to the open use of models. It is critical for open validation and open development, and it brings many opportunities for open collaborations.

The availability of open post-processing, visualization, and analysis tools is more important for open use of simulation outputs than access to model source codes. Simulation outputs can be of different types, e.g., 3D or 2D data on regular or irregular grids, 6D particle data in phase space, outputs aligned with magnetic field lines or flow lines. Simulation outputs also include derived products tailored for model-data comparisons and multi-model ensemble displays for space weather analysis and forecasting. Multi-component modeling systems produce different outputs for each component on different simulation grids. State-of-the-art multiscale codes (see, e.g., HYPERS described in Appendix A.11) take advantage of adaptive numerical algorithms in both space and time and incorporate elements of discrete-event simulation and artificial intelligence (AI). Simulations utilizing adaptive mesh refinements can have different grids for different time steps.

The characteristics that make open use of simulation outputs possible naturally map to the FAIR principles:

- *Findability, Accessibility*: ability to find and access simulation outputs used in specific publications;
- *Findability, Accessibility*: ability to find and access simulation outputs relevant to specific physical phenomena, research topics, science questions, and time periods of interest;
- *Accessibility*: open access to available archives of simulation runs;
- *Interoperability, Reuse*: ability to read, visualize, and analyze model outputs;
- *Interoperability, Reuse*: availability of well-documented, open source, easy-to-use tools for simulation output post-processing, interpolation, visualization, and analysis;

- *Interoperability, Reuse*: availability of metadata for each simulation output, describing how it was produced and how it can be used, including metadata/header for each output file describing what information is in the file.

Issues related to open use of simulation outputs, discussed during the 2024 Open Science Workshop and post-workshop conversations, include: strategies and technologies that can facilitate transparent collaborative analysis, output formats, information to be stored in metadata, approaches for storing and using large simulation outputs, implementation of machine learning (ML) or AI for output analysis, best practices, and codes of conduct. Some of these topics are addressed in Table 4 and following subsections.

**Table 4**. *Advantages and Challenges of Different Approaches to the Open Use of Simulation Outputs.*

| Approaches to the open use of simulation outputs | Advantages | Challenges |
|---|---|---|
| *Quick-look graphic (images and movies), derived time series* | Fast review of run output.<br><br>Ensemble displays in a similar format for multiple models.<br><br>Space weather products tailored for user-defined specifications.<br><br>Timelines ready for model-data and model-model comparisons.<br><br>Starting point for novice users. | No flexibility to generate different visualizations for different quantities, different plot modes, or different color scales.<br><br>Complete model output data might not be available for additional visualization and/or analysis. |
| *Interactive visualization through unified web interface similar for all models* | Free and easy access to complex simulation outputs by non-experts.<br><br>Does not require download of simulation output files and installation of visualization and analysis software.<br><br>Can enable generation of all material from the user's workstation, including plots, images, and movies needed for publications and presentations. | Less flexibility for advanced users to add new visualization features.<br><br>Less optimal for on-the-fly model coupling, run submissions from Jupyter notebooks, and automatic pipelines.<br><br>Visualization features added over time complicate the interface. Most users end up using defaults, if they make visualizations at all. |

| Approaches to the open use of simulation outputs | Advantages | Challenges |
|---|---|---|
| *Community developed software with a set of unified built-in capabilities, organized in a single, well-supported, and documented package* | Enables science users to work efficiently with simulation data of standardized types.<br><br>Includes best available post-processing and visualization.<br><br>Can be used as a backend of web visualization. | Requires format translation.<br><br>Requires compatibility with software/language used by the unified software. |
| *Jupyter Notebooks and executable papers* | Increased transparency and reproducibility for open science practices.<br><br>Can access derived outputs and reduced datasets from public servers.<br><br>A wider use of Jupyter notebooks and executable papers would facilitate greater collaboration and would influence promoting high-quality analysis and value of produced results.<br><br>Jupyter Notebooks provide a unique platform for integrating code, visualizations, and narrative text in a single document. | Not applicable to computationally intensive calculations.<br><br>Jupyter Notebooks may break if dependencies change or libraries are deprecated.<br><br>Requires tools like containers (e.g., Docker), Conda environments, or Python package requirement documentation to ensure long-term reproducibility. |
| *Downloading output files and using visualization tools of user's choice* | Most flexibility for advanced users to use the visualization tool of their choice.<br><br>Enable experimenting with using output of model A as an input to model B (one-way model coupling). | Requires access to computation resources and experience in visualization tools installation and implementation.<br><br>Need storage for large output volumes. |

| Approaches to the open use of simulation outputs | Advantages | Challenges |
| --- | --- | --- |
| | | Downloads through the public internet can take a long time for large output volumes. |

4.2.1 Challenges and Approaches for Open Use of Large Outputs

Modern simulations, in particular kinetic simulations, include both fields and particles and generate enormous amounts of data. It is not unusual for kinetic model outputs to be many tens of terabytes in size. To align with the open science vision, these outputs should be made available in some form upon request. For larger-scale simulations, open use of their outputs becomes even more important than using their open-source codes. Efficient computational algorithms for physical compression of large unstructured (e.g., particle) datasets should also be developed, implemented, and broadly disseminated as open source.

Where to store these data, how to store them, and how to enable fast data transfer are major issues that must be addressed. One of the ways to tackle these challenges could be investing in modern data-reduction techniques to reduce simulation data burden. Other techniques, such as feature extraction, would be valuable as well. ML tools might be very useful for these aspects. It is important to encourage model users to work with supercomputing facilities to enable efficient data sharing.

Because the computational resources needed to run some of the state-of-the-art models are not available to everyone, embracing open use of simulation outputs in the community would potentially allow for more scientific insight and maximize the value of the datasets. In this way, it is tempting to think of bigger simulation runs as space missions. A mission may be costly to develop and launch, but it provides data that multiple teams, often across different countries, find valuable. One can envision community-run models with agreed-upon parameters for large-scale simulations that might be too prohibitive for an individual team. This approach has been successfully adopted in the astrophysics community, where large-scale cosmological simulations are prepared and run by multi-institution collaborations and their outputs are released to the public (see, e.g., Illustris: https://www.illustris-project.org/, Vogelsberger et al., 2014; CAMELS, https://www.camel-simulations.org/, Villaescusa-Navarro et al., 2021; MillenniumTNG: https://www.mtng-project.org/, Pakmor et al., 2023; FLAMINGO: https://flamingo.strw.leidenuniv.nl/, Schaye et al., 2023).

Although open use of large outputs could be realized in different ways, depending on the context of models used and the types of outputs created, it is useful to imagine how a solution to display and analyze large (tens to hundreds of terabytes) datasets from a kinetic geospace model could work. Since it is inconvenient to move such large datasets, a future infrastructure should grant direct public read access to some simulation data residing on the HPC cluster. A data-retrieval queueing system could be activated when these datasets are in archival storage or need minimal post-processing/data reduction. We can imagine such a publicly available dataset being used in two ways.

First, a web interface could facilitate inspection of data products requested from the server (Figure 5, bottom left panel). With existing 3D web technologies, field (mesh-based) data could be manipulated in real time, since these data are relatively small in size compared to raw particle data. A user could plot different fields, identify areas of interest, and request particle distribution functions, rather than accessing particle data directly. Such data reduction or processing requests can be processed by data servers immediately or soon after a request is placed. With well-standardized datasets, even large datasets can be searched quickly and subsets retrieved and processed.

Second, some users will want to download datasets for local processing. In most cases, certain data-reduction techniques could be utilized to reduce the volume of data. The user should have an option to download only a particular quantity over some domain, e.g., particle data within a limited volume, or interpolate data on an adaptive grid, where high resolution may be preserved only in the region of interest. Reducing data size from terabytes to gigabytes is essential for practical purposes and will minimize the load on the network. Importantly, downloadable datasets must be well annotated and supplemented with usage examples, possibly provided in multiple programming languages, to expedite their local post-processing and visualization by end users.

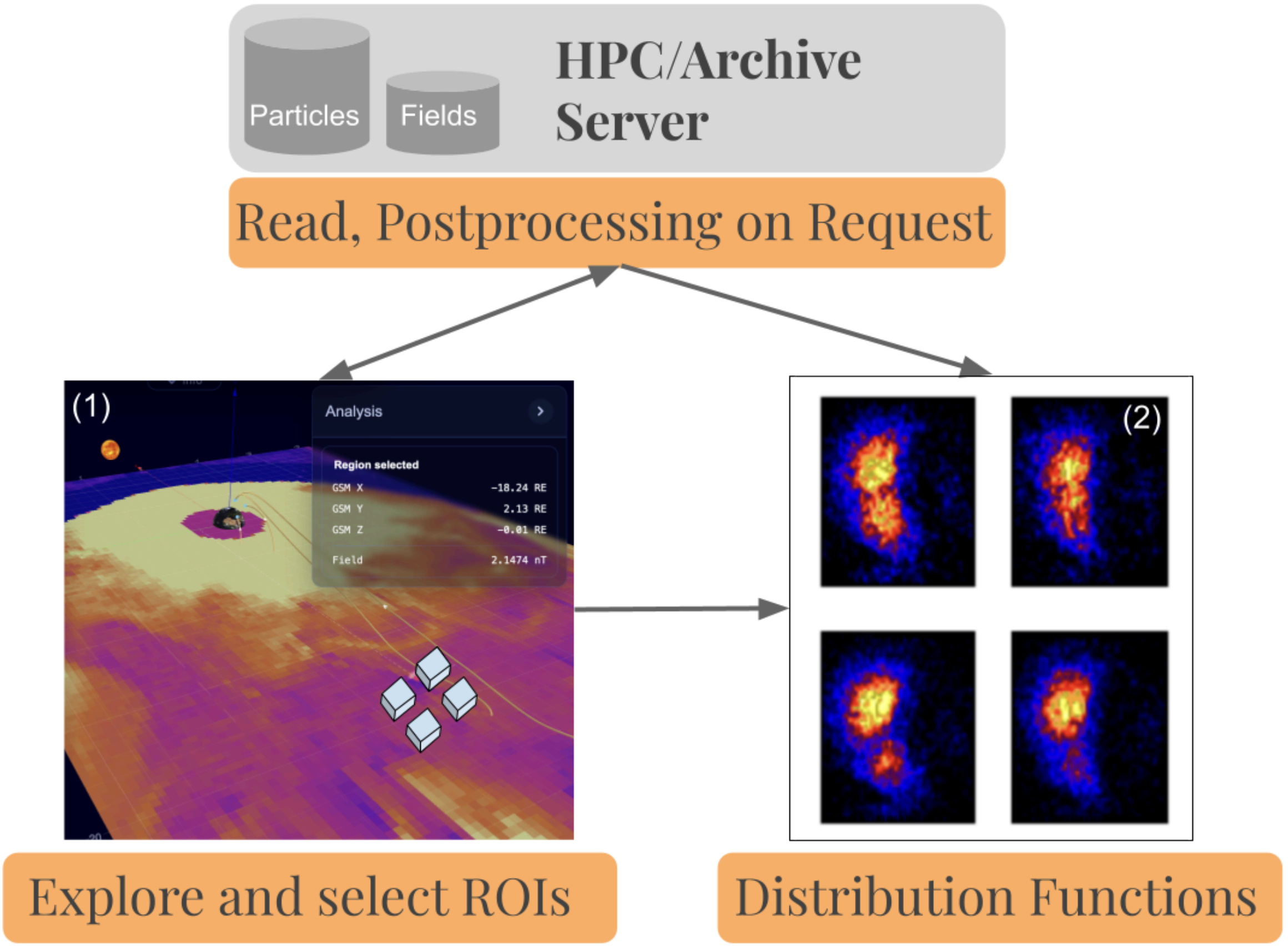

Figure 5. Conceptual vision for interacting with a large simulation dataset stored on an HPC/archive server. Through a web interface, the global fields data can be explored, and regions of interest (ROIs) can be selected for post-processing (1). Resulting from the post-processing request, particle distribution functions extracted from the particle data are made available via another web visualization or provided for download into a local analysis tool (2).

Note that, for models requiring very large output storage but relatively modest computational resources, an alternative solution could be to store only simulation run metadata (see Section 4.2.4) necessary to reproduce the simulation output, thereby saving the storage cost needed for archiving such a large output. The right tradeoff between the cost of storage and the amount of computational resources depends on the specific use case of a given simulation output and on the project priorities and available HPC resources of the simulation service/archive provider. Simulation run metadata might include information necessary to estimate storage vs rerunning costs as an aid to decide on the best strategy (see also Section 4.2.4.1).

#### 4.2.2 Output File Formats

The idea of using a single standard format for multi-dimensional model outputs has been discussed as an efficient approach to facilitate interoperability. One of the advantages of some standard science formats is that they allow direct access to certain portions of the data without the need to load the entire data file into the memory. This is very useful for generating time series of specific parameters at specific locations, e.g., while simulating a satellite flythrough over the entire simulation, with access to multiple files. Several multi-dimensional science standard file formats support direct access, each with its own strengths and weaknesses. The most popular options include NetCDF (Network Common Data Form, Rew et al., 1989), HDF5 (Hierarchical Data Format, version 5, The HDF Group), and NASA's CDF (Common Data Format, https://cdf.gsfc.nasa.gov/). Some models utilize outputs customized for specific commercial or open-source visualization software such as Tecplot (https://tecplot.com/) and VTK (https://vtk.org/). Some modelers designed their own text/binary formats that fit specific modeling approaches. Custom formats usually include headers/metadata with all information needed to read outputs, although sometimes metadata are not embedded in the output file, but rather hardcoded in the reading program/script. With the broader utilization of cloud environments, cloud-native data formats like Zarr (https://zarr.dev/, Jamison & Sommerville, 2023) are attracting attention and are already used in some scientific communities (e.g., Earth science and climate). Several comparative analyses of different formats for different purposes (see, e.g., Martin & Fisher, 2024) demonstrated that there is seldom a one-size-fits-all solution for model output management. While scientific data hosted on object storage (see Appendix B) is becoming more common, the bulk of scientific data used for data analysis, machine learning, and historic data archival is still stored in traditional computing ecosystems.

The climate and forecast metadata convention (CF, Hassell et al., 2017), incorporated into NetCDF format, includes naming standards for variables (physical quantities) adopted in Earth science and terrestrial weather. At present time, there are no naming standards for physical quantities in heliophysics commonly adopted neither for mission data, nor for model outputs. This complicates both data-data and model-data comparisons and reuse of data analysis software developed by the terrestrial weather community and by various experimental and modeling groups. On the mission data side, a partial solution is to use the International Solar-Terrestrial Physics (ISTP) dictionary keyword guidelines for CDF files (https://spdf.gsfc.nasa.gov/istp_guide/data_dictionary.html): this approach sidesteps the naming standard issue by augmenting each variable with one or more standard keyword describing the underlying physical quantity. However, the adoption of CDF/ISTP is usually limited to some NASA datasets, so the issue of inconsistent variable naming across multiple missions (including those funded by other agencies, e.g., NSF, NOAA, ESA) and multiple models remains. On the other hand, many tools (e.g., Python libraries such as SunPy, https://sunpy.org/, The SunPy community, 2020, or Astropy, https://www.astropy.org/, The Astropy Collaboration, 2022), databases (such as the ESA archives), and IHDEA (International Heliophysics Data Environment Alliance, https://ihdea.net) teams (e.g., Weigel et al., 2025) also use and recommend standards from the astronomy community (International Virtual Observatory

Alliance, IVOA, https://www.ivoa.net/), such as TAP (Dowler et al., 2019), VOTable (Ochsenbein et al., 2019), VOUnits (Gray et al., 2023).

Standard file formats require specific libraries to be properly created, read, and manipulated. Evolving library versions introduce new features and enhanced security but result in challenges such as lack of backward compatibility. For model developers, adding support for a given output metadata standard to an existing code base (including model output reading tools) or migrating from one standard to another (including from an older to newer standard version) can be time consuming and requires care to avoid breaking established workflows and pipelines.

These issues have been addressed in other disciplines using concepts defined in ontologies. For this purpose, the Research Data Alliance (RDA, https://www.rd-alliance.org/) defined the I-ADOPT (InteroperAble Descriptions of Observable Property Terminology, https://i-adopt.github.io/, Magagna et al., 2022) schema (see, e.g., Coussot et al., 2024, for an example implementation in the domain of environmental variables for continental surfaces). Thanks to this framework, communities can define concept terms, with contextual metadata (e.g., physical property, matrix, constraints). Coordination between data producers, repository curators, code developers, and modeling centers is necessary for setting this up, using, e.g., tools like Ontoportal-Astro (https://ontoportal-astro.eu/, Cecconi et al., 2025).

Generation of derived products like flythrough time series, field lines, cut planes, and isosurfaces, requires spatial and temporal interpolation. Standard file formats are not sufficient for this purpose without additional tools for interpolating output quantities at any requested location. Moreover, expecting all models to use the same file format for raw simulation outputs is not practical and could constrain innovations, as some models require unique technologies for file formats. On the other hand, some models offer a selection of file formats and output quantities during run configuration from a list of options tailored for specific applications. The interoperability issues arising from the diversity of file formats can be addressed by providing comprehensive metadata and scripts to convert data from one format to another. Metadata for output files should include normalization units for the physical quantities and other important information for reading and interpolation.

The ultimate goal is to enable users to analyze outputs from different models with commonly used software. The necessary requirements for a modeling system should include tools for reading simulation data into standard analysis software (like Python/Pandas/Matplotlib or ParaView/VisIt) and tools for interpolating output quantities at any requested location.

#### 4.2.3 Run Metadata

The Space Physics Archive Search and Extract (SPASE, Roberts et al., 2018) metadata registry (https://github.com/hpde/) is an example of a centralized collection of metadata on heliophysics and space weather resources. As discussed in Fung et al. (2023), SPASE-based middleware services, such as the NASA heliophysics data portal

(https://heliophysicsdata.gsfc.nasa.gov) or the heliophysics digital observatory (https://msqs.gsfc.nasa.gov/hdo/public, Fung et al., 2024), can then be developed and used to uniformly search for and obtain the resources needed for research using the data links provided in the SPASE metadata. The SPASE metadata model (SPASE Group, 2021) is a living model and currently provides the most mature description of heliophysics resources. While originally developed for observational datasets, SPASE was later extended to model artifacts, so both types of resources can be searched and accessed by the same middleware services. The extension has been working well as a resource description for standalone models and derived products, e.g., simulated timelines or synthetic images. However, applying the SPASE metadata schema to complex multi-component simulation runs is challenging and may not be practical.

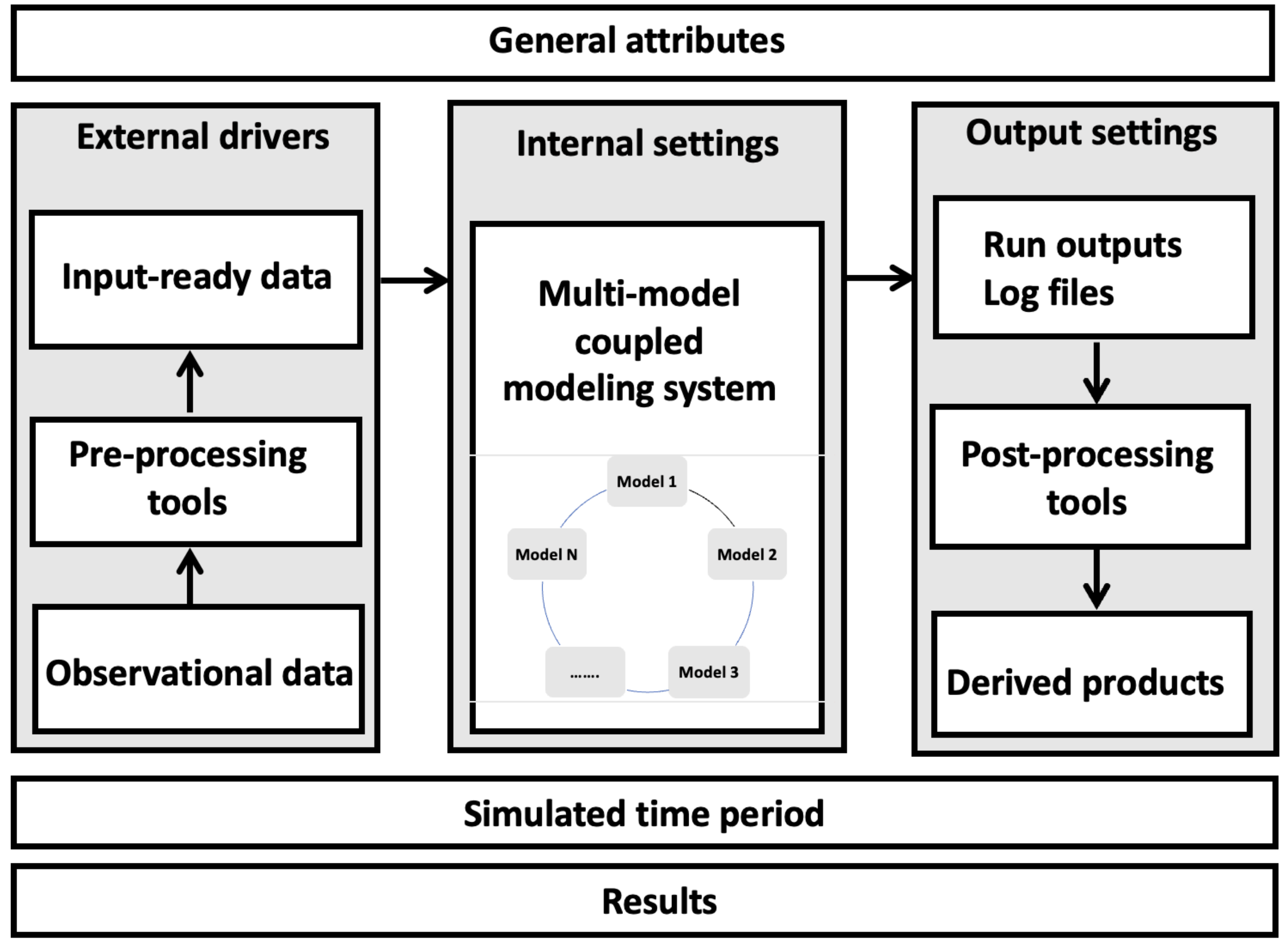

**Figure 6.** Simulation run metadata. Each run metadata component (general attributes, external drivers, internal settings, output settings, simulated time period, and results) has its own metadata. Simulated time period metadata can be applied to different runs. The same internal settings can be applied to different time periods to generate run series. Simulated time period and results are important components of the run metadata for search and discovery.

#### 4.2.3.1 Run metadata components

The scheme in Figure 6 illustrates components of a typical multi-model run. Each boxed component has metadata that can be described using SPASE or another schema. It is a challenge to produce a single SPASE hierarchical text string for an entire run. Metadata for simulation runs are grouped into general attributes, external drivers, internal settings, output settings, simulated time period, and results.

*General attributes* include run ID, information on who produced the run, where the run has been executed, guidelines for attribution (including credits to sponsors), the location of the *run directory* where all files (including inputs, outputs, configuration files, and logs) are stored and organized, and how it can be accessed. Other information, useful for archive management, could include output size and computation costs/complexity for reproducing a run (see also Section 4.2.1). General attributes can also include s*ystem attributes* that capture environment parameters such as operating system, hardware specifications, compiler types, supporting software, libraries and versions, etc.

*External drivers* include raw observational data and/or raw outputs from other runs, pre-processing tools, and input-ready data. Many raw observational data likely already have SPASE Resource ID and metadata entries. Pre-processing tools can be viewed as standalone software requiring their own configuration files and metadata. Input-ready data can be digested by models and should be preserved for possible run replay. Input-ready data include data used for assimilation. Run outputs stored in simulation archives and used as external drivers for other runs have their own run IDs and metadata. An example of run outputs used as external drivers are ionosphere electrodynamics outputs from global magnetosphere simulations that can be used to drive global ionosphere runs.

*Internal settings* include simulation settings for each model in a coupled modeling system (including one-way modeling chain) as well as settings on how information is passed between modeling components, and a version of the entire coupled modeling system. Models in coupled modeling systems can be of different types, cover different physical domains, and have different simulation grids. Assigning a concrete type (e.g., physics-based, empirical) to a multi-model run can be tricky. Many physics-based models include empirical elements, while empirical models are constrained by physical principles. For some models, simulation grids can change over time. Some physical parameters (e.g., Earth's dipole tilt) can be different for different time steps and not always consistent with actual physical parameters corresponding to specific real time due to different modeling approximations. Simulation settings may have hundreds of parameters. Some simulation settings are defined in configuration/input files. Simulation settings also include information on what and how frequently information is exchanged between modeling components. If settings are not defined in configuration/input files, default values are typically used. It is not feasible to include all default values into metadata. Default values for a specific model version can be retrieved from code documentation or included into supplementary configuration files even if those are not used as run inputs.

*Output settings* include output quantities, frequency of outputs, and file formats for each modeling component. For some models, they are hard coded, while for others they can be specified in configuration files.

A single simulation run can produce multiple output file sets in different dimensions and on different grids, generated by different modeling components. *Metadata for output file*s (sometimes called output file headers) should include a run ID linking to the parent run metadata, information on what is stored in each output file, quantities (parameters/variables), coordinate system, normalization units for physical quantities, and other important information for reading and interpolation. Readers and interpolators should be included in model output packages. There is no need for strict guidelines for output file naming convention. However, it is desirable that large multi-dimensional outputs are saved to separate files for different output times, with time stamps included in filenames. This would allow easy access to specific time intervals without needing to access the whole simulation output. Description of how output files are organized inside the run directory is also an important part of the run metadata.

*Post-processing tools* can be viewed as standalone software requiring their own configuration files and metadata. Post-processing tools produce derived products including movies, images, and timelines of quantities at specified locations (e.g., orbits or ground stations).

*Derived quantities* are quantities/parameters that are not directly provided in simulation outputs. Examples include: temperature, for magnetohydrodynamic (MHD) simulations using pressure and density as internal variables; ground magnetic perturbations; integrated heat flux into ionosphere; magnetic flux tube volume; and geomagnetic indices. Derived quantities include simulated in-situ observations used for model-data comparisons and surfaces such as magnetopause, bow-shock, magnetic topology separators, magnetic polarity separators, etc. Derived quantities are used as inputs for quick-look graphics (movies, images), space weather products, and applications, including displays, timelines, and dashboards used by forecasters and decision makers. Metadata for derived quantities for model-data comparisons and space weather products can use similar metadata schema as observational datasets with run ID indicating a resource.

*Naming standards for physical quantities* would facilitate the reuse of post-processing tools. Naming standards are especially beneficial for derived quantities used for timeline viewers and for model-model and model-data comparisons. However, no agreement has yet been reached on variable naming conventions in heliophysics. Variable metadata introduced in the SPASE scheme can be used as a starting point. It might also make sense to take advantage of CF convention guidelines used in terrestrial weather and IVOA vocabularies used in astronomy (Cecconi et al., 2025), or even use ontologies with concepts defined using, e.g., the I-ADOPT schema.

*Simulated time period* metadata include start and end time, as well as characteristics of the time period, e.g., minimum Dst, maximum Kp, shocks, flares, coronal mass ejections (CMEs), solar energetic particles (SEPs), and other observed features. Time period metadata can be reused for different runs. Simulated time period characteristics provide basic information for search and discovery. Note that extracting characteristics of simulated time periods from observational data for runs with artificial conditions (see Section 4.2.4) will likely have incorrect outcomes. Runs with artificial drivers should be clearly tagged to avoid this issue.

*Results metadata* is the most important component of run metadata for search and discovery. This type of metadata should include references to publications and presentations based on run outputs, simulated physical phenomena, special features, and new findings. Updating results metadata is the responsibility of run users and run producers. Users can rely on interfaces and guidance from the CCMC or other simulation services providers.

4.2.3.2 Use cases for run metadata

Use cases for run metadata include simulation archives management, run reproducibility, reuse of simulation settings already tailored for modeling specific physical phenomena, tracking differences between various simulation settings, and advanced search and discovery.

Most typical searches for simulation archives are based on metadata fields grouped under simulated time period, external driver characteristics, general attributes, and results. Details for internal settings are rarely used for searching.

Most of the run metadata information can be represented as a simple collection of key-value pairs that make these metadata easy to store in conventional databases and build client applications. Still, certain crucial parameters might have a significantly more complex structure encoded in model-specific input and output files. Such files might be impractical to capture in general-purpose metadata databases and could either be summarized as key-value pairs with a significant loss of information or be treated as independent supplementary units of metadata, supported by a tailored search and management applications.

One example of such model metadata is a simulation grid, one of the key parameters required for correct interpretation of global simulations. Uniform regular grids can be characterized by the number of grid cells, spatial resolution, and grid type (e.g., cartesian, spherical). Characterizing irregular grids with a few parameters and describing a grid as low/medium/high resolution can be misleading. The number of grid cells sends a message on how computationally expensive the run is. Increased resolution, however, can be set in different areas depending on physical phenomena under study. For instance, for geospace runs, increased resolution can be set near an inner boundary, the magnetosheath and magnetopause, or a tail plasma sheet. A grid with many cells and fine resolution at the dayside magnetosphere, described as high resolution in metadata, instead could be a low-resolution option for magnetotail studies. Irregular grids are typically defined through comprehensive schemes (e.g., SWMF, see Appendix A.24) or formulas (e.g., OpenGGCM) in configuration files. Thus, links to sample runs for specific simulation settings and a collection of annotated configuration/input files can be a useful part of run metadata, requiring special handling and supporting tools. AI tools could be used in the future to search through a more complete set of heterogeneous metadata to unearth complex information currently hidden in the inputs and outputs of the simulations.

Minimum requirements for reproducing a run are: a unique identifier of the modeling system (version number), access to source code or compiled executables for all modeling components, documentation and dependencies for all components, input-ready data used for the original run, and configuration files with all internal setting that are not defaults for a specific version. In many cases, it is not practical to reconstruct configuration files from key-value pairs stored in a metadata database (including SPASE schema). Entire configuration files with embedded explanations should be stored in run metadata. Including a supplementary file defining all defaults is also useful.

#### 4.2.4 Simulations with Artificial/Modeled Conditions

Not all runs simulate real time periods or use realistic settings for a simulated system (e.g., Earth's magnetosphere driven by solar wind). Outputs for runs with artificial conditions are not analogous to observational data. These runs are not intended for point-by-point comparisons with in-situ observations. Such runs, however, are broadly used for education, research, replicability experiments, and statistical data-model comparisons (see Section 4.3 in Kuznetsova et al., 2026b). Runs with artificial/modeled conditions include local physics runs, two-dimensional simulations of a three-dimensional system, runs with artificial drivers, and runs with artificial internal settings.

Examples of runs with artificial drivers include global magnetosphere simulations with constant solar wind conditions with different magnitudes of solar wind parameters (density, velocity), interplanetary magnetic field (IMF) strength, and orientation. Such runs are used to illustrate differences in global magnetosphere configuration for different drivers. In addition to constant solar wind conditions, runs with artificial drivers include different scenarios of changing inflow boundary conditions, e.g., north-south, south-north, or east-west IMF flip, pressure pulses, gradually increasing solar wind dynamic pressure or IMF magnitude. Such settings are used for replicability of experiments that compare different models using the same modeling approximations (e.g., ideal MHD). For example, the GEM 2009 Baseline Model Comparison project was designed to compare responses of different global MHD models to gradually increasingly negative solar wind Bz levels (https://ccmc.gsfc.nasa.gov/challenges/gem-2009-baseline-model-comparison/). Other examples of runs with artificial/modeled drivers are what-if runs that simulate potential extreme conditions, such as modeling Carrington-type space weather events (Ngwira et al., 2014) or using solar wind conditions observed at STEREO to drive Earth's magnetosphere (Ngwira et al., 2013).

Other examples include runs for an artificial Earth with different dipole tilts and strengths (see, e.g., https://ccmc.gsfc.nasa.gov/research-community-support/EXO/). These runs can also be used to illustrate simplistic examples of exoplanet environments or different scenarios of potential Earth's magnetic field evolution.

### 4.3 Simulation Services at the CCMC

The CCMC has been pioneering open use of advanced space environment models and coupled modeling systems since its establishment at the end of the last millennium, long before open science became a widespread concept. The CCMC simulation services, including signature Runs-on-Request (ROR, https://ccmc.gsfc.nasa.gov/tools/runs-on-request/) and Instant Run (https://ccmc.gsfc.nasa.gov/tools/instant-run/), are freely available worldwide, without need for registration. They enable users to configure simulation runs through an interactive self-explanatory web interface and to execute runs remotely on the CCMC computing infrastructure. The CCMC staff provides expert guidance to research users in selecting custom settings for simulating specific phenomena and other expert customer support. Some CCMC submission interfaces include advanced graphical tools that guide users through multiple steps to make high-fidelity parameter choices for modeling physical phenomena. After users complete the parameter selection, model input files are created, and runs are automatically initiated. Users can view generated inputs during the run submission process and restore them afterward. This highly popular system is heavily utilized by researchers who are not model developers themselves, with over 6000 requests by over 1000 distinct users processed in 2024. For complex modeling systems, the submission interface has an option for novice users with simulation settings tailored for specific phenomena, and options for advanced users that enable custom settings for a broad range of parameters (some of which have never been tested). Additionally, special requests can be submitted, where users ask for specific model configurations not available through the web interface: these require manual intervention from the CCMC personnel and sometimes from model developers themselves. Several models with outputs relevant for space weather operations are continuously running in batch mode using near real-time observation data as inputs (https://ccmc.gsfc.nasa.gov/tools/continuous-run/)

The CCMC hybrid computing infrastructure is composed of three components: on-premises; at a NASA High-End Computing Capability (HECC); and on the Amazon Web Services (AWS) cloud, frequently referred to as CCMC AWS Cloud or CCMC Cloud (see Appendix B for details). For each task, the CCMC infrastructure components are selected based on cost efficiency and performance. Seamless data flow between components is enabled by Apache Airflow (https://airflow.apache.org/). The CCMC utilizes a dedicated network for its on-premises systems, and benefits from a 10Gb connection that links other essential CCMC infrastructure components.

The CCMC maintains collaborative shared environments at NASA HECC and CCMC AWS Cloud to facilitate the involvement of model developers in model onboarding and upgrades. Modelers can build, install, and run their models in the collaborative shared environment, in preparation for final installation and execution on CCMC production servers. CCMC points of contact (scientists, software developers, and system engineers) are available to assist wherever necessary. In 2021-2023, for several complex onboarding tasks, modelers were funded through single source grants (i.e., CCMC surge funding for accelerated implementation). This approach significantly accelerated the onboarding process and enabled early open use of the latest modeling capabilities. Encouragement from model development programs and dedicated funding of model developers for their support of onboarding and implementation at the CCMC are essential. NASA's Living With a Star (LWS) Strategic Capabilities Program is an example of a successful implementation of this approach where model developers have been working with CCMC points of contact and utilizing the CCMC shared collaborative environment from early stages of projects.

The archive of simulation outputs at the CCMC is composed from a heterogenous collection of simulations produced in different ways:

- outputs from ROR service (currently more than 40K runs);
- outputs produced outside the CCMC and delivered to the CCMC through the simulation output onboarding process;
- run series produced by the CCMC for validation projects and specific physical phenomena with tailored simulation settings, e.g., geospace model runs for 200+ storm time periods https://kauai.ccmc.gsfc.nasa.gov/CMR/TimeInterval/viewAllTI
- outputs from continuous real-time runs;
- derived space weather products and time series for model-data comparisons.

Users can explore archived runs irrespective of who requested or produced them. Each run has a dedicated landing page, accessible through the CCMC website. Run landing pages, tagged by unique run IDs, contain all information relevant to a run, including input parameters selected by users, run metadata, all configuration files, and input data.

Outputs are offered in multiple ways, to address different use cases. Quick-look graphics (images and animations), developed in conjunction with model developers, provide a fast way of assessing simulation results. Customized online visualization tools allow users to explore complex simulation outputs without needing to download potentially large files, and, finally, users can access raw output files for more sophisticated data analyses.

The CCMC has a diverse set of tools in many languages that support utilization of simulation outputs accessible through CCMC interactive archives, including the stand-alone Kamodo open-source Python toolkit (see Section 3.2 and Annex 2 in Masson et al., 2024) and an online interactive visualization and analysis capability (CCMC-Vis). The CCMC-Vis, highly utilized for over two decades, was developed on a single software platform (IDL) to maintain portability and keep a uniform appearance across all simulation domains, models, and types of output. The CCMC-Vis framework enables rapid inclusion of new model output formats as soon as the CCMC is provided with a sample simulation output and source code (written in IDL, Perl, Fortran, C, Python, and other IDL-compatible languages) describing how the data are read. CCMC-Vis enables the user to visualize and interact with simulation results in multiple dimensions, and to perform scientific research without the need to download run outputs. CCMC-Vis produces publication-quality plots, images, and listings for basic model outputs, as well as derived quantities such as user-ordered custom variables. The system can generate Movies-on-Request and time series at a specified set of locations, grids, and flythroughs along requested orbits. All plotted data (including traced flow lines) and time series can be saved in a simple and unified text format.

The CCMC is also onboarding and implementing visualization and analysis software developed by the community. For example, Kaipy, a software developed by the CGS DRIVE center, is now being implemented at the CCMC as an option for interactive online visualization.

The CCMC is expanding implementation of APIs for starting and monitoring runs, downloading results, and enabling broader use of Jupyter notebooks. A new CCMC ROR Dashboard (https://ccmc.gsfc.nasa.gov/ror/requests/requests_dashboard.php) introduces transparency on what (and where) is currently running on all CCMC systems and on the status of each run. This includes links to run output landing pages with run metadata, and other details. Recently, a *My CCMC Items* service (https://ccmc.gsfc.nasa.gov/ror/search/my.php) was introduced to identify CCMC requests and their results, including ROR runs, output requests, movies, and other items based on one or more email addresses. Users can also compare different runs based on input files (https://ccmc.gsfc.nasa.gov/ror/results/run_diff.php), to better understand where differences in simulation output originate. A revised advanced search page, allowing ROR simulations to be filtered based on metadata, is in progress. All these tools enhance the findability, accessibility, and reusability of the CCMC simulation archive.

For simulation data to be easily findable and accessible, it is important that any published research work should contain clear references to corresponding archives that house details of the material used in the publication. Therefore, the CCMC is linking runs in CCMC archives with publications, abstracts, presentations/posters, and educational material (https://ccmc.gsfc.nasa.gov/publications/). Any user who reads a paper utilizing simulation outputs in CCMC archives can look at details of interest and even replicate results referenced in the archive of runs and CCMC resources.

Runs linked to publications, presentations, educational material, and utilized in community validation campaigns, together with run series generated by the CCMC or delivered to the CCMC, are tagged as high usability runs. Other runs, older than 2 years, are tagged as low usability runs. High-usability runs are being copied to high-performance redundant Dell PowerScale storage directly connected to CCMC visualization servers. Low-usability runs are moved to a modern tape archive, BlackPearl. The CCMC also uses AWS cloud storage capabilities, including object storage through AWS Simple Storage Service (S3), specifically for temporary storage of runs produced on the cloud or outputs copied to S3 on user request. For example, ENLIL (see Appendix A.6) run outputs are provided to the SWx TREC H3lioViz cloud-based web application via a S3 bucket. Disadvantages of object storage include: the marginal costs of access (bandwidth costs for data transferred), weak performance for computationally intensive analyses, and the possibility of software incompatibility. S3 object storage is the lowest cost per stored gigabyte in comparison with other storage solutions at AWS. However, it is still less cost effective than the CCMC on-premises high performance storage solution directly attached to CCMC visualization servers. For computationally intensive tasks, including interactive visualization and ML/AI training (see Section 4.6), cloud users can download run outputs directly from the CCMC to local or network storage on their cloud instances, using the same HTTPS protocol as for downloading from S3. Therefore, to avoid unnecessary costs, there is no need to transfer large run series stored at the CCMC on-premises storage system proactively to the cloud without explicit requests from users. See also Appendix B for more details on AWS S3, storage solutions costs, and transfer speed comparisons.

To accommodate the substantial computing and storage needs of large-scale computational runs, most heavy outputs are generated at NASA HECC. Ideally, visualization and analysis would occur directly at the source where heavy outputs are produced. While CCMC-modelers collaborative environments exist at NASA HECC, they offer less flexibility than the CCMC AWS Cloud. The CCMC is exploring opportunities for long term storage of large outputs, enabling on-demand post-processing, and for installation of visualization services at NASA HECC to enable fast access to simulation output for CCMC users.

Several limitations can impact the type and volume of models and simulation settings the CCMC can make available for open use, e.g.:

- Some model implementations are not organized in such a way that clearly separates different file types, such as inputs, outputs, and logs.
- Insufficient computational resources is a concern. Some large models can take weeks or even months to execute, making them less ideal for open use by individuals (see also the discussion in Section 4.2.1).
- Some input data sources are not completely openly available to the public, with some data providers requiring user registration or IP address whitelisting.
- Under NASA's cybersecurity policy, containerized models are subject to the same requirements as software installed directly on CCMC machines. This necessitates either complete network isolation or periodic container rebuilds to incorporate the latest operating system and library updates. However, the requirement for frequent rebuilds undermines the goal of scientific reproducibility, as the computational environment would constantly change.
- Some licensed software might treat each containerized model instance as a separate machine, even if all instances are running on the same physical machine, resulting in license shortages.

The CCMC has devised ways to mitigate some of these limitations, by working with modelers to make their models more flexible and portable, streamlining processes, providing monitoring and tracking, and supplying infrastructure that can better serve the needs of the community.

Collaborations between service providers, such as CCMC and VSWMC, are valuable for advancing open science. For example, developing common interoperability standards offers a fruitful avenue for collaboration, enabling the exchange of outputs, related or derived products, data, and metadata between their archives. These standards can significantly improve search, discovery, and compatibility across tools.

### 4.4 Simulation Services at the VSWMC

The VSWMC is a modular, end-to-end modeling framework designed to simulate the coupled Sun–Earth system for space weather applications. It enables the interactive execution and dynamic coupling of heterogeneous space weather models, which may be deployed locally or across geographically distributed computing environments. The current implementation integrates physics-based models covering key domains of the space weather system — from solar event initiation through the heliosphere, and into near-Earth regions, including the magnetosphere, radiation belts, ionosphere, and thermosphere. This architecture supports predictive capabilities analogous to those used in terrestrial weather forecasting, which facilitates assessment of space weather impacts on technological systems and human activities.

The VSWMC is designed to accommodate both advanced and general users. Expert users can construct customized model workflows based on specific scientific objectives or operational constraints, while standard users can access routine simulations involving pre-configured coupled model chains. Simulation outputs are compatible with downstream effects tools, enabling the forecasting of potentially deleterious space weather impacts on ground-based infrastructure, satellite systems, and human physiology in space environments.

The VSWMC system handling software is running on a server at KU Leuven (Belgium). ESA Space Weather Portal users can access the system via an interface that is controlled by ESA's Single Sign-On authentication and authorization mechanism.

In addition to interactive parameter setting and run submission through a graphical user interface, the VSWMC models and model chains can be run from the command line via API. The VSWMC Command-Line Interface (vswmc-cli: https://pypi.org/project/vswmc-cli/) allows the user to run a VSWMC model chain, to monitor the run logs, and to fetch run results via the VSWMC RESTful API (REST: representational state transfer). This mode of interaction also allows scripting and enables requests for EUHFORIA runs (see Appendix A.8) through the CCMC open web-interface for remote execution at the VSWMC servers at KU Leuven. This approach is valuable for configuring and executing space weather modeling chains and running submissions through Jupyter Notebooks.

To address reproducibility, the VSWMC team has initiated full containerization of the system, encapsulating all core components—including the vswmc-runner and vswmc-node—into portable, self-contained Docker containers. This approach ensures that each component includes all necessary dependencies, libraries, and runtime configurations, thereby enabling consistent deployment across heterogeneous environments with minimal setup effort.

As a result of this containerized architecture, the VSWMC system can now dynamically discover and interact with multiple distributed vswmc-runner instances, each potentially being hosted at a different site and managing a distinct set of models. When a model workflow (or chain) spans multiple locations, the coordinating vswmc-runner will delegate execution tasks to the appropriate remote vswmc-runner instances. These runners, in turn, instantiate vswmc-node containers locally to execute the assigned models within isolated and reproducible environments.

The VSWMC team created a development environment as a stripped-down version of the containerized VSWMC system that can be run locally by anyone on their workstation. The purpose of this environment is to allow the user to locally configure and integrate new models, i.e., configuration files with user parameters, configuration files for model coupling interfaces, generation of the graphical user interface (GUI), choice of what parameters the users have access to, including a definition of the allowed parameter space. Furthermore, they can test parameter transformations from the GUI to model requirements and configure and test the GUI for new simulation chains. It is currently impossible for developers to develop the model coupling interfaces (C++, vswmc-node) by themselves, and communication with the VSWMC development team is still needed for this purpose. This may change in the future. In theory, it is even possible to run your model locally, on your own computer, and couple it to VSWMC models or model chains running on the distributed VSWMC network.

### 4.5 Runs-on-request Query Interoperability

Efforts on the query interface interoperability have been studied for many years. We mention here two initiatives in this domain. First, the IMPEx project (Integrated Medium for Planetary Exploration, Modolo et al., 2018) proposed a unified interface for modeling space physics planetary environments with runs-on-request as well as slicing of existing runs. This work defined the Simulation extensions of the SPASE model (https://spase-group.org/data/simulation/). The IMPEx framework reused IVOA standards such as the VOTable format for the tabular output, which allows, e.g., rich metadata to be appended to the slicing process output. Second, in the frame of the IVOA, a standard interface has been defined to query a digital infrastructure for asynchronous run on request. This interface is called UWS (Universal Worker Service, Harrison & Rixon, 2016) and is used by many astronomy data centers and projects. It permits a smooth interface to an internal job management system (e.g., Slurm, https://slurm.schedmd.com/), including user authentication. Servillat et al. (2022) developed a web graphical user interface complementing the scriptable UWS web-service to provide a flexible runs-on-request interface (Servillat, 2024).

### 4.6 Simulation Outputs Produced by Developers and Users

Other than the CCMC and the VSWMC, a number of model developing groups maintain their own archives, e.g., outputs from the National Center for Atmospheric Research (NCAR) Whole Atmosphere Community Climate Model with thermosphere and ionosphere extension (WACCM-X, see Appendix A.27) on the NCAR Climate Data Gateway (Gasperini, 2019; Solomon et al., 2019), and outputs from the Thermosphere-Ionosphere-Electrodynamics Global Circulation Model (TIE-GCM, see Appendix A.25) on the World Data Center for Climate (WDCC) archive (https://www.wdc-climate.de/ui/project?acronym=OTHITACS). Upon agreement with NCAR and WDCC, the CCMC is enabling download, post-processing, visualization, and analysis through CCMC systems on requests from users.

Users who perform simulations on their own (outside of the CCMC, the VSWMC, or other simulation service providers) become run producers. Run producers (users or developers) should plan to maintain open archives of runs they produced with metadata following the same guidelines as simulation service providers. Run producers may also register simulation outputs in CCMC metadata registry with or without delivering actual outputs to the CCMC.

### 4.7 Artificial Intelligence and Machine Learning Approaches

In addition to solving physics equations, modeling approaches include ML and AI solutions. The 2024 Nobel Prizes in physics (Nobel Foundation, 2025b) and chemistry (Nobel Foundation, 2025a), awarded for computational designs and innovative numerical algorithms, attest to the growing recognition of modeling in research.

With modern simulation models producing increasingly large amounts of output data, using the data effectively becomes a greater priority. Elaborate algorithms are usually needed to reduce data size and to extract scientifically meaningful conclusions. To this end, the community has already started implementing ML techniques to compress datasets. A statistical representation of model outputs, much easier to share and use, is a particularly promising approach. However, such approaches are at a very early stage, and it is important not to place black boxes on top of each other, and instead to focus our efforts on ensuring that models and their outputs are properly documented, archived, and accessible first. A statistical representation of a model dataset that does not have community-adopted standards for metadata and model configuration will offer reduced benefits to science.

Utilizing ML/AI to emulate and speed up complex and expensive physics-based models is an ongoing effort in the geospace community. For example, Hu and Dong (2025) introduced an AI model (CAM-NET) to capture global-scale atmospheric dynamics by training on a decade of datasets from WACCM-X. An acceleration of over 1000 times in inference time was achieved. Similarly, Licata and Mehta (2023) employed long-short term memory neutral networks to emulate TIE-GCM, called TIE-GCM ROPE (reduced-order probabilistic emulator) and achieved <5 km bias of satellite propagation comparing to the original TIE-GCM density. As the CCMC stores decades of archived runs from multiple models, those output files become the natural training datasets to be used for creating model emulators and speeding up simulations. However, caution must be taken when training is conducted without consulting/collaborating with the original model developers or without proper knowledge of the domain physics. Model settings and input files are the necessary prerequisites to perform such emulations. Blindly applying ML techniques to physics-based models could lead to wrong interpretation of fundamental physics.

### 4.8 Open Use Best Practices

Modelers and simulation services providers should ensure that models can be used with as few barriers as possible and to facilitate user training. It is important for modelers and simulation services providers to follow the best practices discussed above. Equally important are best practices for users, aimed at improving the efficiency and effectiveness of the resources offered by simulation service providers. This section is based on experiences accumulated by the CCMC over the course of two decades. Although many best practices suggested here can be generalized, implementation details might vary depending on specific use cases, priorities, and resources available to other simulation service providers.

Novice users are advised to use web-based services to reduce the workload on model developers for support on model installation and execution. Users of CCMC simulation services should check, using the provided tools, if any run they are planning to request already exists in the archive of simulation outputs. This helps to avoid executing the same run multiple times. It is generally accepted to request a run for the same model and the same time period, but with different simulation settings. Users experimenting with custom untested simulation settings available at the CCMC are requested to inform the CCMC and model developers on results of their analysis to ensure that there are no numerical artifacts or other unphysical results. This helps to avoid misuse and misinterpretation.

The CCMC is not generating digital object identifiers (DOIs) for all requested runs, as not all runs are planned to be cited. A DOI implies a guarantee that the resource it points to will remain available forever. This means that, as the number of runs with DOI increases, so does the storage needed to keep them permanently publicly available. This would impose a high cost on the CCMC (or any other simulation service providers), which calls for a limit on how many runs can receive a DOI. For this reason, users are advised to inform the CCMC on the intended use of runs and request DOIs.

Users are requested to inform the CCMC on run output use in publications, presentations, and other activities. This can be done through the Results Submission Form https://ccmc.gsfc.nasa.gov/publication-submissions/ or by contacting CCMC staff. References to publications/presentations are now included in run metadata. Authors are requested to include run IDs and DOIs in publications to enable readers to interactively analyze model outputs used in the paper. Users are encouraged to insert new findings from simulation output analysis into the Results component of run metadata, and to link runs and simulation settings with physical phenomena and special features. This will facilitate search and discovery and help other users interested in studies of the same phenomena to reuse appropriate simulation settings. This also helps in deciding which runs need to be maintained in performance storage instead of being copied to tape storage (see discussion of low- and high-usability runs in Section 4.3).

Users interested in installing, configuring, and executing models on a system of their choice should read the documentation prior to contacting developers for help. This is especially important for users with limited modeling experience.

Users that become run producers need to make simulation outputs and metadata, including all input and configuration files, openly available. This could be achieved through user-maintained archives or through onboarding of run results at the CCMC.

## 5 Open Validation

Open validation intersects with the concepts of open use and open collaboration (Figure 1). When users engage with open use models, they compare the results to observations, other models, or simply their own expectations. In this way, every user run enabled by open use implicitly tests the model. The more widely and openly a model is used, the more varied the scenarios under which it is tested. Different scenarios (data inputs, boundary conditions, model settings, etc.) investigated by a research community lead to stress-testing models beyond what a single research team could achieve. Equally important for open validation is its overlap with open collaboration. Open collaboration can inform model developers about inconsistencies, bugs, and limitations, and help enhance the reliability and robustness of a model. Open development also inevitably links to open validation: clear documentation of a model's applicability and limitations is needed to correctly assess model performance and capabilities for both scientific and operational uses. At the same time, reproducibility in the context of open development (ensuring code modifications do not introduce unintended changes in simulation outputs) allows developers to verify external contributions to the model source code in a transparent and systematic way.

These intersections of open validation with open use and open collaboration set the stage for the rest of this section, where we first focus on openness about model science quality (Section 5.1), then on community validation campaigns (Section 5.2), and finally on the infrastructure and tools that enable open validation (Section 5.3). While in Kuznetsova et al. (2026b) we review the role of validation in developing new capabilities based on science and transitioning them to operations, in this section we narrow the focus to the open-science aspects of validation.

To avoid confusion about the terminology, we follow Halford et al. (2019), using the standard definitions of validation and verification typical in NASA and ESA publications. *Validation* refers to comparing a forecasted and observed physical quantity with metrics that quantitatively assess the differences. In contrast, the process of *verification* ensures that the model meets its requirements (e.g., numerical accuracy, convergence behavior, stability, etc.). More details are provided in Appendix B of Kuznetsova et al. (2026b).

### 5.1 Model Science Quality

Before model-data comparisons can be meaningful, model developers must be open about the scientific quality of their models. The minimum requirement for any model used in a scientific context is model verification. In many heliophysics modeling domains, papers introducing new numerical algorithms include verification tests to show that the implementation can reproduce the analytical solution or benchmark tests with the expected accuracy. However, the depth of such verification exercises varies across communities and domains. Having a standardized set of verification benchmark tests could be valuable to better assess the merits of different codes.

Without verification, the model is a mixture of mathematical solutions, numerical errors, and unknown errors due to potentially incorrect code. Such model output is below standards even if it seemingly agrees with observations. Model verification is the minimum baseline requirement for meaningful validation efforts. Reference problems and their expected outputs during the verification process should be made available so that users can verify for themselves the degree to which they can reproduce the expected model output.

However, verification is only the first step. After establishing correctness, it is important to quantify the model sensitivity. Such sensitivity assessments can include comparing model outputs under different inputs and parameter settings, which help users understand how these factors influence the model outputs. Ensembles of runs that systematically explore these factors should become standard practice and be made available. This knowledge is valuable for validation efforts because it helps distinguish between uncertainty arising from the model itself and that introduced by parameter choices.

Even after verification and uncertainty assessments are established, open communication about the capabilities and limitations of models is needed. Modelers need to state what their model can simulate and what it cannot reliably simulate. Quantitative indicators of model behavior, such as numerical diffusivity, should be provided so users understand which results are realistic and whether a discrepancy with observations is due to physical assumptions or from numerical effects and limitations of the model. Providing users with example runs of physical phenomena that can be modeled is crucial. It enables users to make informed decisions and avoid misuse when selecting models for their specific scientific questions. At the same time, users have the responsibility to carefully consider how they apply a model and ensure they use it in a manner that is consistent with the explanatory materials.

An efficient way to communicate the strengths and limitations of models is through explanatory materials. Explanatory materials are helpful for users to understand which physical phenomena a specific model can reproduce. For instance, not all ionospheric models can reproduce localized structures, such as storm-enhanced density and the tongue of ionization, and their dynamics. From the user's perspective, having a tutorial or documentation summarizing the major features of different models using specific space weather events is beneficial. This would help users understand what to expect when requesting a model for studying other similar events. Furthermore, it is important that this explanatory material offers insights into model limitations. For example, model-specific settings, such as upper altitudinal limits (e.g., 500, 1000, 2000 km), can affect absolute total electron content values in Earth's ionosphere and therefore comparison with data. While these kinds of subtleties may be obvious to developers, they are not apparent to users if not properly explained. An example of a YouTube tutorial (Downs, 2023) where developers from Predictive Science Inc. demonstrate usage of their model is available at the Community Coordinated Modeling Center (CCMC, https://ccmc.gsfc.nasa.gov/) from the CORHEL-CME landing page (https://ccmc.gsfc.nasa.gov/models/CORHEL-CME~1/, see Appendix A.17).

Open use not only makes models available, but also implicitly validates them, providing feedback to developers to guide future developments. Community campaigns have been instrumental in creating this feedback and should be further expanded upon.

### 5.2 Community Validation Campaigns

Open validation means that individual modeling teams are as transparent as possible about how they validate their models. Community-wide validation efforts go a step further by establishing a shared vision of model validation, ensuring that models are assessed using the same physical quantities of interest, metrics, and events or time periods. Validating models under the same conditions enables systematic, comparable, and above all fair assessments of their capabilities. To create this shared vision, community-wide validation projects bring the modeling community together to agree on shared goals, identify what signals model success, and define the metrics used to quantify this success.

Such efforts have been actively pursued through programs like NSF's Geospace Environment Modeling (GEM; see Liemohn et al., 2025, for an overview), Coupling Energetics and Dynamics of Atmospheric Regions (CEDAR), and Solar Heliospheric and Interplanetary Environment (SHINE). These programs often take the form of validation campaigns where solutions from participating models are collected and compared against observations and each other. Recently, several teams have emerged within the COSPAR ISWAT Initiative (Kuznetsova et al., 2026a) focused on unifying validation practices in space weather modeling (see, e.g., Verbeke et al., 2019; Reiss et al., 2023). An overview of past and recent community-wide validation efforts is provided in Section 4 in Kuznetsova et al. (2026b).

A major strength of community validation efforts is that they create an open competition among models within their respective modeling domains. By providing a level playing field for all participating models, modeling teams can trust the fairness of the validation process. A fair environment for model validation can become a powerful driver of innovation at all stages of the research-to-operation-to-research (R2O2R) chain. An example is the study by Pulkkinen et al. (2013). In a coordinated effort between the CCMC, NOAA Space Weather Prediction Center (SWPC), and the modeling community, the authors studied the predictive capabilities of models participating in GEM, CEDAR, and SHINE programs for reproducing ground magnetic perturbations, which are strongly related to geomagnetically induced currents. Focusing on selected events and metrics, the authors conducted a validation study which led to one of the participating models being selected for operational usage by NOAA.

In addition to these coordinated community efforts, there have been successful multi-agency validation activities of operational models. From 2017-2020, NOAA SWPC and CCMC, in close collaboration with model developers, started a project to assess improvements in CME arrival time forecasts at Earth using the Air Force Data Assimilative Photospheric Flux Transport (ADAPT) model driven by data from the Global Oscillation Network Group (GONG) ground observatories (Mays et al., 2025). These outputs were then fed into the coupled WSA-ENLIL model (see Appendix A.28 and A.6, respectively) and then compared to the current operational version of WSA-ENLIL (without ADAPT). CCMC performed and validated over 1200 simulations, and with some caveats about nuances, and statistical uncertainties, this study provided SWPC with a quantitative answer to their question of whether an upgrade to the initial operational WSA-ENLIL, would result in a substantially improved representation of heliospheric dynamics. Another example is a comparative bias analysis, conducted by SWPC, between the NOAA WSA–ENLIL model and the operational solar wind model run by the UK Met Office (UKMO). By evaluating both models over identical time intervals and against the same observational references, SWPC identified consistent patterns in model-dependent biases and forecast behavior. Such cross-center analyses help ensure that operational agencies maintain a fair and transparent understanding of model performance, complementing community-wide campaigns, and supporting the development of harmonized validation practices.

Beyond studies focusing on selected historical events, Scoreboard projects facilitated by the CCMC show how a similar community approach can improve space weather forecasting. Scoreboards are built through community leadership, for instance, through collaborations in ISWAT teams. Each Scoreboard has an online platform that displays an ensemble of forecasts from different participating research and operational space weather models worldwide. These forecasts are submitted before an event occurs, and their forecast performance is displayed publicly. A key objective of the Scoreboard projects is to provide the space weather community with a quantified assessment to identify models with potential for operational use, and to demonstrate their forecasting skill to operational agencies and decision-makers. An overview of Scoreboard projects is provided in Section 6 in Kuznetsova et al. (2026b).

The next step beyond community efforts is to translate these collective insights into automated validation infrastructure that can host growing numbers of models, events, and metrics from different domains.

### 5.3 Infrastructure and Tools for Open Validation

The knowledge gained through community collaboration, including agreements on events, time periods, metrics, file formats, and metadata, can be used to build validation infrastructure. This infrastructure can streamline model validation, examine model capabilities, and track modeling progress over time. Besides being flexible and open to the community, this infrastructure should be automated. Automation is crucial because validation is labor-intensive, and new modeling developments happen rapidly, requiring updating and maintenance (MacNeice, 2018). Ideally, such an infrastructure should have standardized formats for submitted (or directly uploaded) model solutions and automated validation reports.

Instead of building individual validation systems, modeling teams can use a shared validation infrastructure already developed and agreed upon by the community. An example of such an infrastructure for open validation is the Comprehensive Assessment of Models and Events using Library tools (CAMEL; Rastätter et al., 2019; https://ccmc.gsfc.nasa.gov/tools/CAMEL/) developed by the CCMC. CAMEL consists of two components, a backend and a web-based frontend. The CAMEL backend uses established CCMC services and infrastructure such as the ROR system and the CCMC Metadata Registry. In contrast, the CAMEL frontend is a web-based platform that displays validation results in the form of numerical and graphical analysis. This analysis can be conducted flexibly by users by selecting a combination of models, model runs, time periods, observational data, and study parameters of interest. The CAMEL framework is designed for on-the-fly metrics calculation with minimal pre-calculation caching.

Building on this shared infrastructure, SWPC has collaborated with the CCMC to contribute historical WSA–ENLIL model outputs reformatted to the CAMEL-compliant metadata structure. This effort ensures that operational model results produced at SWPC can be analyzed within the same validation framework used by research models worldwide. In parallel, SWPC is developing a new ambient solar wind validation platform that incorporates CAMEL metrics (point-to-point, binary, and event-based metrics; Kuznetsova et al., 2026b) to support consistent, automated assessment of long-term model performance. By aligning both historical and real-time validation workflows with community-standard tools and formats, SWPC aims to strengthen interoperability between operational and research validation systems, enabling future integration of CME arrival-time and geomagnetic index validation capabilities. The joint development of an open repository of metrics and visual diagnostics for assessing modeling capabilities is an important aspect of such collaboration. Ideally, this repository can serve as the backend for any validation platform, strengthening collaboration, and reducing duplicated effort. With an open repository, everyone in the community can contribute to the metrics codebase, ensuring it stays up to date with evolving user needs.

Another example of a validation infrastructure at the CCMC is the Ionosphere-Thermosphere Model Assessment and validation Platform (ITMAP, https://ccmc.gsfc.nasa.gov/itmap/), an interactive tool designed to visualize the ionospheric model validation results for the historic geomagnetic storm events (https://kauai.ccmc.gsfc.nasa.gov/CMR/TimeInterval/viewAllTI). ITMAP employs a vast range of observational datasets from ground-based global navigation networks and satellites (such as FORMOSAT-7/COSMIC-2, CHAMP, GOCE, GRACE, GRACE-FO, and Swarm). The current ITMAP ionosphere/thermosphere validation projects focus on key parameters, including total electron content, critical frequency of the F2 layer, height of the F2 layer peak, and neutral density. By using advanced visualization and statistical analyses, ITMAP enables comprehensive model-data comparisons. This process is crucial for pinpointing model strengths and weaknesses, which drives improvements in our ability to forecast the near-Earth space environment.

Making the validation process open and accessible through validation infrastructure promotes transparency, fairness, and trust in model validation. We note that an extensive discussion of validation infrastructure and tools such as NSF National Center for Atmospheric Research (NCAR) METplus and ESA's Space Weather Service Network Online Validation Platform can be found in Section 3.6 in Kuznetsova et al. (2026b) and in Henley et al. (2026).

## 6 Open Development

The goal of open development is to advance the state of the art in modeling by building upon existing capabilities. This includes enabling model development with the fewest impediments to potential contributors. Open source is a necessary but not sufficient element of open development. Properly supported open model development can lead to broader community involvement and more rapid progress. The danger is that unfunded mandates will lead to pro forma compliance without meaningful changes in practice. The concern is that funding agencies think that putting the model code on a publicly accessible code repository is the end goal, instead of one of the many steps in the open science process.

Models are only vibrant and alive to the extent that development continues. A fundamental reality behind all open science considerations must be the recognition that heliophysics modelers belong to a relatively small community with limited financial and human capital resources. Building on the prior work of everyone in the community allows to maximize the potential of all who work in the field. The future health of heliophysics requires the continuous development of new cutting-edge modeling capabilities. The key questions to address are: how do we do this in a manner that is consistent with an open science imperative, and how can an open science focus make this goal achievable more effectively and efficiently?

This section addresses different topics related to the development process: how to train and retain technical figures needed to create and maintain numerical models (Section 6.1); how to incorporate best practices from the software industry (Section 6.2); what are the challenges of adopting an open-source approach for developing operational models and preserving legacy models (Sections 6.3 and 6.5); how funding agencies can help with licensing issues and with setting flexible open science requirements in funding opportunities (Sections 6.4 and 6.6); and, finally, how various forms of collaborative development, inspired by other disciplines and software industry, can help in building the next-generation real-time space weather models and what's the role of funding agencies and simulation service providers in this process (Section 6.7).

## 6.1 Workforce Challenges in the Open Development Era

The world of modeling has changed. 40 years ago, a domain scientist could just develop the modeling tool they needed to address a particular science interest. Today, model development is a specialist activity that demands its own unique training and skill set. It is not uncommon for large modeling groups, as described in Section 3 and Appendix A, to include software engineers in a variety of roles, from code optimization to enhancement of model user friendliness, from infrastructure support to implementation of best software development practices. However, not every modeler or modeling group has access to the expertise of software engineers. For these reasons, Bard et al. (2023), Haiducek (2023), and Ringuette et al. (2023) advocate for increased education and training in software development for scientists and for a tighter transdisciplinary collaboration between researchers and software engineers to develop models that can take advantage of next generation computing capabilities.

Successful open development requires managing and fostering a talent pool. One key aspect is proper credits and recognition of technical figures, including software engineers and scientists deciding to focus more on code development than science analysis. Career advancement in the current academic environment is mostly based on peer-reviewed publications and funds procurement. While in recent times code development and implementation of innovative software engineering solutions have gained recognition at the community level, there is a need for more structured reward systems at the academic institutions and funding agencies level. This is particularly important to entice software engineers to join research groups, rather than commercial companies, and retain them over time periods longer than a typical funded grant project, e.g., more than three years (Ringuette et al., 2023). Code developers should be credited not only for the publications and scientific results that they produced themselves, but also for community achievements based on using the models they developed. This requires a community-wide commitment to a different cultural paradigm from the one adopted so far, which can be built through education of future generations and financial investment by stakeholder funding agencies.

Some of these needs were already recognized in 2018 in a U.S. National Academies of Sciences report for NASA: "NASA Science Mission Directorate should foster career credit for scientific software development by encouraging publications, citations, and other recognition of software created as part of NASA-funded research" (NASEM, 2018, p. 4). Similar recommendations were put forward by the U.S. Decadal Survey: "All the above requires sustained infrastructure for training, hiring, and retaining a new type of interdisciplinary workforce with overlapping expertise ranging from HPC [High Performance Computing]-specialized research software engineers to experienced applied mathematicians, domain scientists, and program managers. Such a sustained, professional workforce is paramount to meeting the nascent modeling challenges in solar and space physics but is not being achieved within the current hierarchy of the funding programs." (NASEM, 2025, p. 158). On the community side, multiple national associations of research software engineers have been established (e.g., Germany: http://www.de-rse.org, the Netherlands: https://nl-rse.org/, UK: https://society-rse.org/, USA: https://us-rse.org), with the goals of changing the academic culture, facilitating recognition of software developers (be them scientists or software engineers) by universities and research institutions, and promoting collaboration between researchers and software experts.

### 6.2 Model Development Best Practices

The pace of change in the computational world continues to steadily accelerate, leading to inflation of the required skill sets. This might hamper the involvement of new contributors, unless best practices from the software industry are adopted. For example, models could be more modular to reduce the areas of expertise that an individual needs to acquire. This strategy has been adopted by some large modeling frameworks, such as SWMF (see Appendix A.24) and MAGE (see Appendix A.15), where different numerical solvers or sub-models are developed as separate modules to be applied in specific domains. Similarly, the numerical solution core of a model could be separated from its interface to the computing environment (parallelization, debugging, graphics, etc.) to allow scientists to focus on the physics implementation, leaving more technical aspects to software engineers. In addition, recent advances in AI, such as large language models, might be leveraged to enable automated generation of code from higher-level descriptions provided directly from users or parsed from literature.

There is reasonable agreement among modelers (see also Section 3 and Appendix A) on the importance of regression and unit testing (including tracking and archiving test results) to ensure reproducibility of model output after code changes and to verify individual pieces of code, especially among large development teams with a long history of code development. This approach is also enabled by the presence of software engineers in the team, which brings expertise from different disciplines and exposes scientists to best practices from the software industry at large. These include: use of version control and code repositories to facilitate shared development within a modeling group; forks and branches to develop and test new features before merging code changes in the main project; atomic commits to isolate code changes and minimize merge conflicts; and code of conduct for external contributions, defining rules and expectations for proposed code modifications to be included in the main project. This cross-fertilization is particularly important for more advanced models, which require new computing infrastructure, such as Amazon Web Services (AWS) cloud computing or graphics processing unit (GPU) clusters, and development approaches (e.g., CI/CD: continuous integration/continuous deployment) to facilitate and accelerate development of complex graphical user interfaces (GUIs) and modeling frameworks, especially in relation to maximizing utilization of high-performance computing clusters.

### 6.3 Open-Source Challenges

Many modelers recognize that publishing their code as open source allows for a wider and more varied user base. The advantages of this approach usually outweigh various concerns (e.g., compromising their funding competitiveness, maintaining control over code correctness, getting credit for the development work, ensuring correct representation of model capabilities and performance) and time/labor resources needed to release the code. Nevertheless, these limiting factors should be considered by both the modeling community and funding agencies when deciding if and how to push for public release of a model code.

For example, one of the most common concerns among modelers is that, once the code is publicly available, everyone can modify it without permission. In theory, this concern can be easily dismissed because all major hosting platforms offer fine controls on who is allowed to contribute code changes to the original code base: external collaborators/users/developers can propose modifications (also known as pull or merge requests) that first need to be approved by the repository owners and administrators. However, this doesn't fully alleviate the real underlying issue: some modelers are worried that people modify their version of the code without communicating with the original model developers, thereby introducing the possibility of bugs and decreased model performance that would reflect badly on the original model once published. This worry is not without merit, as indeed this behavior has occurred. As one modeler put it, the more accessible a model is to use, the more users, especially novice users, will abuse it. Technical solutions (like branch protection and pull/merge requests) cannot solve this problem, since they require external contributors to follow the netiquette of how to properly modify model codes. This requires education at all levels (from students to researchers to users in the private sector) and collaborations from journals and funding agencies, by setting requirements for code changes tracking and for credit recognition and funding acknowledgment for the original model developers.

Though accepting contributions from external developers can be an efficient way to enhance model capabilities (see, e.g., Section 6.7), it can also introduce security vulnerabilities. This challenge is especially acute for operational centers such as SWPC, where contributed code must be thoroughly vetted before it can even be considered for integration. Such vetting takes time and resources.

### 6.4 Funding and Legal Support

As outlined above, the expectations of what open science practices should achieve are disconnected from the implementation of open science requirements at the level of funding and regulatory agencies, with many aspects of open use and open development being currently unfunded. Most of the time, modelers provide the additional products described above (accompanying tools, documentation, user support) for free. This, of course, impacts the quality of the services provided and relies on the goodwill of individual modelers. While NASA offers some focused funding opportunities for developing supporting tools and increasing open source adoption (https://science.nasa.gov/open-science/nasa-open-science-funding-opportunities/), these are usually limited to one-time opportunities, lacking the continuous support needed to maintain large codebases with the corresponding documentation and user training. The U.S. Decadal Survey recognizes the importance of this point and "[...] supports the efforts of the funding agencies to provide opportunities for theory and modeling projects at a range of individual project sizes [...] from pure theory (requiring support for individual researchers, graduate students, or postdocs) to large-scale model development (requiring support for large groups of scientists and other professionals for example, research software developers)" (NASEM, 2025, p. 194). Funding programs that cover all aspects of model developments are also particularly relevant for supporting innovative out-of-the-box solutions that require long development times. The current funding model usually focuses on specific science investigations, with fewer incentives for developing algorithms beyond the state-of-the-art that can solve new problems.

At the same time, requirements to transition existing codebases to open source focus mostly on the funding side, neglecting legal support for navigating the complexities and challenges that multi-institution and large/old legacy code maintainers face. For example, if multiple developers contributed over the course of the years to a closed-source code, publishing it requires getting permission from each developer. Sometimes this is not straightforward, as developers are bound by their institutions to respect patent and intellectual property rights. This can easily turn into an excruciating endeavor involving multiple legal offices. The choice of the license to be used for publishing the code is also an important factor, as not all software licenses meet the open science requirements, nor can all open source licenses be adopted by legacy-code right owners, depending on the license requirements on attribution, use, and future code contributions.

### 6.5 Preservation of Legacy Models

Computers capable of supporting serious modeling have existed for about 40 years, which is about the length of a typical science career. Not surprisingly, we are now at the point where many of the generation that began the modeling enterprise are reaching retirement age, and the models that are their legacy need to be transitioned to a new generation. Many of their models have evolved into group efforts, providing generational continuity. However, some, including some of the community's operational workhorse models, remain as single developer models.

Keeping the development of these models active throughout the next decade presents a prime open development opportunity. Some important codes are already progressing successfully through this generational transitioning process, such as GAMERA from LFM, LaRe3D, or ARMS (Appendix A.2). The quality of code documentation, readability of the code, and clarity of the code's internal structure seem to be critical to their successful transition to the next generation. To achieve this, modeling groups need dedicated funding to hire qualified computational physicists.

### 6.6 Influence of Model Type on Open Development Considerations

Models come in different levels of sophistication and adherence to first-principles physics, from simple attempts to empirically capture the behavior of very specific quantities to very complex model frameworks that strive to capture global evolution of all the relevant physics. These differences lead to different realities in how these models are developed. In general, the more complex the model, the longer it takes to develop. Simple models are more likely to be developed by individuals, with limited investment in both time and money, and are more easily replaced/superseded with time. The more complex the model, the greater the community's investment in both human and financial capital is. These models tend to be developed as a group effort, have a much more sustained presence, and are continually updated and modified to keep abreast of new physical understandings.

Currently, the community relies on self-organization to manage the challenges listed above, but the way forward should include institutional support. Ideas would include workshops, schools, and conferences focusing on the challenges of model development in addition to physical domain science. At the same time, both recommended best practices and guidelines on how to properly develop models (e.g., inclusion of comprehensive testing suites, preparation of clear documentation) should be scaled depending on the modeling group and model user-base size. This is particularly relevant for requirements imposed by funding agencies. The idea is to create a tiered approach to open science requirements such that innovative solutions and ideas from single modelers are not stifled by the burden of too many requirements, while at the same time larger modeling efforts are supported with enough resources.

### 6.7 Community Modeling

The U.S. Decadal Survey recommends that “[i]n addition to maintaining the current range of grant opportunities for theory and modeling, [NASA] should establish a flagship-level heliophysics community science modeling program capable of addressing heliophysics problems that have broad community interest and that require complex community models” (NASEM, 2025, p. 158). The implementation details of this program are left to NASA and the community to decide, however the U.S. Decadal suggests looking at examples existing in other disciplines, such as the Community Earth System Model (CESM, https://www.cesm.ucar.edu/). CESM is a project led by the NSF National Center for Atmospheric Research (NCAR) in collaboration with scientists from multiple U.S. universities and research institutions, aimed at developing a fully coupled global Earth system model for state-of-the-art simulations of the Earth on various time scales. The governance and funding structure ensure continuous support for maintaining and improving the model development, which is split among working groups dedicated to specific topics and sub-domains. Other examples of similar community modeling efforts are the NOAA's Unified Forecast System (https://ufs.epic.noaa.gov/) and the Environmental Protection Agency's Community Modeling and Analysis System (https://www.cmascenter.org/). The strength of these projects is to rely on a large community of contributors, with agreed-upon development timelines and participation rules set by the community itself, managed by a centralized governance. This approach ensures continuity in focusing on the most relevant goals recognized by the community, without restraining innovation and participation, effectively using key open science concepts to advance scientific and operational use of numerical models.

Collaborative development principles followed by large community models can also be adopted for less ambitious, but operationally relevant, projects. Community modeling efforts can focus on only one component of the Sun-to-impact modeling chain, e.g., on pre-processing tools. An example is the Python/Javascript CME Analysis Tool (PyCAT, Milward et al., 2024) co-developed by the NOAA Space Weather Prediction Center (SWPC) and the UK Met Office (UKMO). PyCAT is used by space weather forecasters to interactively fit multi-perspective coronagraph observations to an idealized coronal mass ejection (CME) Cone model for assimilation into the WSA-ENLIL coupled model (see Appendix A.28 and A.6, respectively) to follow CME propagation. PyCAT became operational at UKMO in April 2025 and is currently being transitioned to operational use at SWPC. After the initial operational deployment, the intention is to release PyCAT as open-source code, encouraging community members to help enhance its capabilities by forking the GitHub repository and issuing upstream pull requests. This promises to be an efficient way to promote innovation in an operational product by leveraging community contributions.

In the last decade, innovative forms of collaboration have spread from software engineering and data science communities to academic and research communities. Hackathons (intense collaborative events focused on developing a specific application or functionality) and machine learning (ML) Kaggle competitions (competitive programming events focused on producing the best ML model for a specific dataset) are forms of community open development that take advantage of multiple participants focused on a single goal to accelerate finding a solution to a problem. Hackathons can be considered the synchronous version of open-source development: code repositories allow developers to share their progress over time, but the development is mostly asynchronous, with contributors usually focusing on different aspects of the software, coding and testing modifications locally and then merging them in the shared repository. Hackathons instead compress development into the time frame of up to a few days, putting contributors in the same place to work on the same task. Similarly, ML Kaggle competitions can be likened to a faster approach to open model validation. In these competitions, the host prepares data and metrics to be used to train and tune ML models and provides a platform where results can be uploaded and visualized. Participants continuously submit improved model versions which are then evaluated against an undisclosed dataset, to avoid bias and overfitting. Model performance is visualized on a leaderboard, which is used by participants to guide their development efforts. This is reminiscent of the Scoreboards used to openly validate forecasting models, as discussed in Section 5. These new approaches can be brought to the space weather modeling community for specific use cases and become part of the community modeling program.

A space weather community modeling program requires a coordinated collaborative approach aiming to build real-time modeling systems that address specific forecasting problems, e.g., neutral density and satellite drag modeling, geomagnetic environment modeling to improve geomagnetically induced currents forecasting, ionospheric variability modeling to improve ionosphere specification for navigation and communications, near Earth radiation and plasma environment modeling, and deep space SEP modeling. Such a program would involve the use of community open validation platforms (see Section 5) to demonstrate the potential of modeling system modifications to improve operational forecasting capabilities and participation in open collaboration initiatives, such as the COSPAR ISWAT Initiative (see Section 7.1) to enhance the exchange of ideas and capitalize on a diverse set of expertise from the whole community. The open development component could be built on the efforts and accomplishments of NASA's Space Weather Centers of Excellence, DRIVE centers, LWS Strategic Capabilities, and other space weather modeling programs. Simulation service providers can play an important role in providing technical expertise and support for integrating models and complex simulation pipelines into real-time space weather forecasting workflows, and in acting as a bridge between the research community and end-users (e.g., operational agencies, private sector), effectively enabling the R2O2R process.

## 7 Open Collaboration

Collaboration is central to all scientific research. Collaborations bring together contributors with a broad range of skills and perspectives (such as theorists, modelers, software engineers, mission scientists, data analysts, and space weather forecasters) to work towards a shared goal. Collaboration enables researchers interested in the same topics to join forces and to work more effectively. Collaboration involves sharing research results that are generated throughout the lifetime of a project. Successful collaboration requires trusting working relations (including proper internal credit recognition), shared collaborative environments, and accomplishments demonstrating the value of collaborative efforts. Collaborative efforts are also necessary for development, improvement, and evaluation of space weather predictive capabilities.

Section 7.1 describes the COSPAR ISWAT Initiative as an example of international collaboration with a bottom-up approach, while the other subsections addresses the advantages of an open collaboration between users of simulation services (Section 7.2), simulation service providers and observational data centers (Section 7.3), and operational centers (Section 7.4).

### 7.1. ISWAT — A Hub for Open Self-organized Topical Collaborations

The COSPAR ISWAT (Committee on Space Research International Space Weather Action Teams, https://iswat-cospar.org) Initiative was established to enable joining forces to address challenges across the field of space weather. ISWAT is open to all motivated research groups and individuals committed to active participation. Scientists working on a focused topic can create an ISWAT team and open it for others interested in the same topic to join. Application to join a team includes a short description of how a new member is planning to contribute to team efforts. Requirements for active participation (including time limits for no-activity) are defined by team leads. Anyone can open a new team and/or join established teams. The ISWAT bottom-up initiative opened the doors for early career scientists and researchers who are new to the topic to join a team of experts. ISWAT is an example of an effective platform for open collaborations that removes barriers for becoming a member of an existing team or creating a new team. More details on ISWAT can be found on the ISWAT website and in two recently submitted papers to the COSPAR Space Weather Roadmap Advances in Space Research Special Issue (Bisi et al., 2026; Kuznetsova et al., 2026a).

### 7.2 Open Collaboration Between Users of Simulation Services.

All simulation outputs produced at or delivered to the Community Coordinated Modeling Center (CCMC, https://ccmc.gsfc.nasa.gov) are available for analysis by the entire community. Runs requested or produced by one user can be utilized in research by other users as soon as run results are published. This sets the scene for collaboration between users working with the same runs. The CCMC should develop an infrastructure that will enable users to share their findings and advise other users what simulation settings should be used to produce features indicative of physical phenomena of interest. User contributions to run metadata that link runs with publications, physical phenomena, and science questions are critical.

### 7.3 Open Collaboration Between Simulation Service Providers and Observational Data Centers

Model-data comparisons include preparation of timelines of quantities derived from in-situ observations and from model outputs at specific locations (e.g., orbits or ground stations). Essential ingredients for interconnecting simulation and observational data archives include an agreement on variable naming conventions and the existence of application programming interface (API) calls to retrieve a quantity at specified locations and time periods for both observational and simulation resources. Desired steps toward such APIs include the creation of a naming table that identifies a given physical quantity (e.g., in-situ magnetic field component) across different models, missions, and instrument types, and the provision of detailed documentation for computations needed to get to the returned quantity from the lowest-level data. There may be cases where parameters used to obtain higher level observations could be provided self-consistently by the model itself. For example, Zeeman-split observations from the EZIE spacecraft yield quite directly the absolute magnitude of magnetic fields at a certain altitude in the ionosphere. However, temperature profiles in the upper atmosphere/ionosphere are needed to convert EZIE observations into magnetic field vector data and, subsequently, into 2D current density maps showing electrojet location and structure. If a first-principle model was to be compared to these maps, the model's temperature profiles in the ionosphere (if available) could be used instead of a climatological model such as MSIS or IRI to reprocess low-level EZIE data into the observed electrojet current maps for a more self-consistent comparison.

### 7.4 Open Collaboration Between Operational Centers

Operational centers for space weather forecasting all have the same objective: to provide timely, accurate, and actionable information to their customers. This is a fertile ground for promoting collaboration. Operational centers have a vested interest in promoting open validation platforms (Section 5) to identify the most promising forecast models. When identifying a mutual need, operational centers can work together to address that need. A recent example is the PyCAT application co-developed by the NOAA Space Weather Prediction Center (SWPC) and the UK Met Office (UKMO) (see Section 6.7). Making such applications available as open-source software presents further opportunities for enhancing their capabilities by leveraging community contributions.

## 8 Heliophysics Open Modeling Environment – A HOME for the Modeling Community and Community Modeling

An important outcome of the 2024 Open Science Workshop and the discussion that followed was the creation of a Heliophysics Open Modeling Environment (HOME) for the modeling community across all physical domains in heliophysics.

HOME aims to promote the vitality of the modeling community, guide open-science and open-source mandates tailored to the community needs, increase appreciation for the importance of numerical simulations among colleagues and decision makers at funding agencies, and enable coordination and collaboration on all aspects of open science.

The objective of HOME is to unite and give voice to all scientists and software engineers with primary expertise in modeling and model-related software, as well as those using models and simulation results in their research. Our HOME:

- builds consensus and speaks with a unified voice when talking with funding agencies;
- reviews and proposes open-science policies and codes-of-conduct related to models;
- facilitates more modeling representation on the decision-making committees for open science;
- coordinates efforts to define standards (naming conventions, metadata, licensing, ownership, recognition, modern software engineering techniques, software project management methods, open-science best practices);
- serves as a platform for collaborative model development, frequently referred to as community modeling;

- coordinates with observers to convince agencies to sponsor critical measurements;
- coordinates public outreach activities related to models.

HOME is not a formal organization. It is a bottom-up movement facilitating interconnection of the modeling community across heliophysics domains and funding sources. The HOME movement encompasses all open science themes (use, validation, development, and collaboration).

To facilitate interconnections with broader international heliophysics and space weather communities and to maintain a cross-domain HOME forum, the HOME is set as an overarching activity of the COSPAR ISWAT Initiative (https://iswat-cospar.org/O5; see also Section 7.1 and Kuznetsova et al., 2026a ). Starting from the ISWAT 2025 Working Meeting, the HOME sessions are organized during ISWAT working meetings. HOME sessions will continue at CCMC biennial workshops. Regular HOME forums could be organized in coordination with established domain-focused workshops such as NSF's CEDAR (Coupling, Energetics, and Dynamics of Atmospheric Regions, https://cedarscience.org), GEM (Geospace Environment Modeling, https://gemworkshop.org), and SHINE (Solar Heliospheric and INterplanetary Environment, https://helioshine.org). Maintaining a cross-domain HOME forum for the modeling community will not only facilitate open communication and discussion between model developers, but can also provide a platform for users of the same models to exchange ideas, to share their experience, and to communicate consolidated wish lists to model developers.

HOME is open to partnership with the Earth sciences, (exo)planetary, and astrophysics communities to address common challenges in the implementation of open science in modeling. HOME welcomes balanced partnerships with observational data analysis communities, including DASH (Data, Analysis, and Software in Heliophysics, https://dash.heliophysics.net) and IHDEA (International Heliophysics Data Environment Alliance, https://ihdea.net). HOME can also take advantage of already existing experiences, such as the PETSc (Portable, Extensible Toolkit for Scientific Computation, https://petsc.org) project. PETSc has an active community of contributors, users, supporters, as well as people generally interested in PETSc, its use, or its future. The PETSc governance model employs a community-wide consensus-based decision- making process intended "to ensure that the people who are most affected by and involved in any given change can contribute their knowledge in the confidence that their voices will be heard" (https://petsc.org/release/community/governance/). Specifically, community members need not contribute to the PETSc code base in order to participate in discussions about the library. Promoting contributions – not necessarily in the form of code or documentation – from anyone in the community supports PETSc's form of consensus-based democracy (Fogel, 2023), which provides to all members the option to request a review and formal vote by the PETSc Council, rather than waiting for explicit agreement among all community members regarding a particular change to code base (Adams et al., 2022).

This paper and the companion executive summary by Reiss et al. (2026) are the first outcomes of the HOME initiative. Next steps will include:

- establishing action teams focused on specific models or groups of models to provide a platform for users to exchange ideas and share their experience;
- establishing action teams to build community real-time modeling systems that address specific forecasting problems;
- establishing action teams to address recommendations outlined in Section 9 targeting developers, users, and simulation service providers;
- communicating recommendations targeting funding agencies through white papers.

Moving forward, HOME will be a collaborative effort to keep insights on practices in the modeling community, as presented in this paper, up to date. To achieve this, we will keep the survey presented in Section 3 publicly available, continually expand the content with timely questions, and invite more model developers to participate. In this way, HOME will provide the community with up-to-date insights into the successes and challenges of open science in heliophysics and space weather modeling. These insights will be vital for the community to coordinate steps toward realizing the potential of open science in advancing heliophysics modeling together.

## 9 Summary and Recommendations

We present a community-wide effort to develop a strategy and action plan to advance heliophysics and space weather modeling through open science. This effort includes the establishment of the HOME initiative, as described in the previous Section. Our recommendations are based on feedback from model developers, researchers, and users worldwide, collected through a living open-science survey, discussion sessions at the Open Science Workshop in College Park, Maryland, USA, in 2024, and at the COSPAR ISWAT Initiative Working Meeting in Cape Canaveral, Florida, USA, in 2025.

Models and coupled modeling systems serve our community as tools for numerical experiments that complement observational experiments conducted in space and on the ground. Although programs that fund model development operate with much smaller budgets than space missions and ground-based facilities, their return on investment can be equally significant. Open science shows promise for making progress in modeling more effectively and efficiently, but open science policies and practices originally tailored for observations and data analysis software cannot be extended to modeling through oversimplified analogies; instead, this process must be guided by leadership from the modeling community.

Many heliophysics and space weather models are currently available for open use through simulation service providers, such as the Community Coordinated Modeling Center (CCMC, https://ccmc.gsfc.nasa.gov/) and Virtual Space Weather Modelling Centre (VSWMC, https://spaceweather.hpc.kuleuven.be), and a growing number of models are being prepared for open-source release. Survey results show that most modeling teams are positive toward an open science approach and acknowledge that making a model available for open use through simulation service providers makes their model more broadly accessible to the community than simply releasing the code as open source. Nevertheless, there are differing opinions on how easy or difficult it is to embrace open-science principles. Concerns remain about the possible misuse of models or simulation outputs and the lack of financial and legal support from funding agencies, especially regarding sustained code maintenance, documentation, development, and the transition from closed or legacy codes to open-source codes. Additionally, there are concerns about insufficient recognition and incentives for model developers in the current open science approach, especially a lack of acknowledgment when their models enable community achievements. Insufficient computational resources is also a concern, especially for the more large and complex models.

Progress in modeling through an open science approach is as much a challenge of community work culture as it is a technical one. Following best practices and codes of conduct is important for all actors in the modeling community, including modelers, service providers, and users. Achieving this work culture requires a community-wide commitment to a collaborative mindset, which can be developed through education of future generations and financial investment from stakeholders and funding agencies.

Below, we summarize the key recommendations gathered from the workshop discussions, the modeling team survey, and lessons learned from simulation service providers. Recommendations are organized according to four overlapping open-science themes centered on actions that advance programmatic goals (open use, open development, open validation, and open collaboration; see Section 1.5 for a brief description of each theme), rather than on the open-science policies and practices described in Section 1.2. An example of how this shift in focus affects policy implications for funding agencies is the recognition that open sourcing the code is just one of many steps required for a proper open use of a model by the community. Each recommendation targets one or more groups, including developers (both scientists and software engineers), model users (from research, education, operations, and the private sector), simulation service providers (such as the CCMC and VSWMC), publishers, and funding agencies. For each theme, recommendations are organized moving from the higher level (e.g., funding agencies, broad scope) to the lower level (e.g., users, limited scope). An executive summary of the present material, including high-level recommendations, can be found in Reiss et al. (2026).

## 9.1 Recommendations for the Open Use of Models

Enabling open use of advanced models by the broader community goes far beyond a source code release. Table 5 summarizes the main recommendations on open use of models discussed in Section 4.1. To advance modeling, understanding, and forecasting, state-of-the-art models should be available for open use by the community with as few impediments as possible. Technical expertise, time, and computational resources are all barriers that should be considered. At present, hosting models at simulation service providers significantly minimizes these barriers. In addition to making a model available for open use by the community, it is important to facilitate user training through test cases, run examples, and tutorials. Preparation of such educational material should therefore be supported by funding agencies and acknowledged as a critical activity by modelers to ensure that the community can make the best use of a model. While most of the tasks necessary for enabling open use of models falls on model developers and simulation service providers, model users share the responsibility to ensure that models are used correctly, applied to the intended physical scenarios, and not run outside models' validity regimes. This shared responsibility is essential to ensure that the open use of modeling resources leads to high-quality heliophysics and space weather research.

**Table 5**. *Recommendations for Open Use of Models.*

| Recommendation | Target | Relevant sections |
|---|---|---|
| **Support and Fund Modelers to Enable Open Use of Models**<br>Fund developers to create code documentation, example runs, and tutorials required for open use of models by the heliophysics and space weather community. | Funding agencies | 4 |
| **Sustain Simulation Service Providers for Open Use of Models**<br>Sustain simulation service providers, such as the CCMC and VSWMC, as essential infrastructure for open use of models and simulation results, and model validation. Provide sufficient computational resources to support the open use of state-of-the-art, computationally intensive models. | Funding agencies | 4 |
| **Support Model Onboarding at Simulation Service Providers**<br>Support onboarding and implementation of models in collaborative shared computing environments at simulation service providers for making the newest modeling capabilities available for open use by the community. Modelers' time devoted to onboarding should be funded and recognized. Collaborative environments need to be maintained. | Funding agencies | 4.3 |
| **Best Practices for Simulation Service Providers**<br>Whenever possible, model onboarding should be done in collaborative shared computing environments, to facilitate the involvement of model developers in the installation and upgrades on the simulation service provider's computing infrastructure. | Simulation service providers | 4.3 |
| **Best Practice for Modelers**<br>Provide clear documentation for installing, configuring, compiling, and running the model on different systems. Clearly separate configuration files, inputs, outputs, logs, and other file types. Ideally, configuration files should expose all possible simulation settings to avoid the need for model recompilation. Include run health flags in log files that monitoring software can capture. | Developers | 4.1 |
| **Best Practices for Users**<br>Encourage beginners to use web-based simulation services to benefit from expert guidance when selecting settings for simulating specific physical phenomena. Expert users running models on their own systems should read the documentation to make the best use of the model. Experience using untested simulation settings should be reported back to model developers and simulation service providers. | Users, Run producers | 4.8 |

| Recommendation | Target | Relevant sections |
|---|---|---|
| **Model Running Automation: API Support**<br>Support a common application programming interface (API) for starting and monitoring simulation runs at simulation service providers. This enables interoperability between different simulation service providers and enhances open use through scripting and Jupyter notebooks. | Simulation service providers | 4.1, 4.4, 4.5 |
| **Model Installation Automation**<br>If feasible, formalize the model build process, allowing for automatic installation of the model via containers or package/environment management tools. | Developers | 4.1, 4.3, 4.4 |

### 9.2 Recommendations for the Open Use of Simulation Outputs

The open use of simulation run outputs by the community is as important as the open use of models, because making simulation outputs publicly available allows the community to build directly on the results from prior numerical experiments. Open simulation outputs are especially important for resource-intense simulations, where it is most cost-effective for simulation service providers to work with model developers and the community to produce and curate series of interesting runs for open analysis by all interested parties.

Table 6 summarizes our main recommendations. The essential requirement for open use of simulation outputs is a comprehensive, accessible, and searchable metadata description of each simulation run, that should include unique run ID and information on: 1) who produced the run, where the run was executed, system attributes, where the simulation outputs are stored, and guidelines on attribution; 2) external drivers, including data sources used for data assimilation; 3) parameter settings, including the version number of modeling frameworks and individual modeling components; 4) documentation, including information in configuration files and default model settings; 5) information on the characteristics of the simulated physical phenomena, key findings, and persistent identifiers (DOIs) of publications, presentations, and other public material where run outputs are used. Additionally, open tools for post-processing, visualization, and analysis are critical for making use of simulation outputs, regardless of source-code availability. Current metadata solutions adopted from observational datasets are not practical. They are missing key elements to properly describe complex multi-component, multi-stage simulation runs, from pre-processing raw observational data to producing multi-dimensional outputs in different domains, as well as derived timelines and space weather products.

**Table 6.** *Recommendations for Open Use of Simulation Outputs.*

| Recommendation | Target | Relevant Sections |
|---|---|---|
| **Open Use of Simulation Outputs** | Simulation service | 4.2.3, 4.8 |

| Recommendation | Target | Relevant Sections |
|---|---|---|
| Simulation outputs should have comprehensive and searchable metadata with unique run identifiers. Metadata should consist of all configuration files, input files, internal settings, and essential information, including characteristics of the simulated physical phenomena. If required, DOIs should be created for long-term use and citation. Journals should require citing run IDs and DOIs for runs that produce any simulation outputs used in publications. Run ID should be included in all output file headers produced by the run and in metadata for all derived products. | providers, Developers, Run producers, Publishers | |
| **Efficient Storage**<br>Explore opportunities to store and post-process large simulation outputs near the HPC resource where they are produced (e.g., NASA HECC) to reduce data-transfer bottlenecks. Data reduction techniques should be explored to manage large simulation outputs without losing essential information. Long-term storage solutions should consider different types of storage and cost of storage versus re-running simulations. | Simulation service providers, Funding agencies | 4.2.1 |
| **Live Run Metadata**<br>Simulation service providers should support infrastructure that enables users to keep run metadata up to date by including, for instance, new findings and references to new publications and community projects using output data. This advances search options, discovery, and scientific reuse. | Simulation service providers, Run producers, Users, Developers | 4.2.3 |
| **Supportive Code and Tools for Simulation Outputs**<br>Open tools for post-processing, visualization, and analysis are essential for broad community use of simulation outputs, regardless of source-code availability of the model. Developers should provide scripts to read/convert data into file formats compatible with commonly used software (e.g., Python, ParaView, VisIt) and tools to interpolate output values at any location of interest. | Developers, Simulation service providers | 4.2.1, 4.2.2 |
| **Best Practices for Run Producers and Simulation Service Providers**<br>Whoever generates simulation outputs for publication (e.g., in journal or presentations), including users and developers, has the responsibility to attach run metadata to the output, generate DOIs if needed, and develop a storage policy for accessibility and preservation. Simulation outputs should be made available for download, either in bulk or as selectable individual files. Partnerships between run producers and simulation service providers are encouraged to help implement these requirements. | Simulation service providers, Run producers | 4.2.3, 4.3, 4.6, 4.8 |
| **Best Practices for Users**<br>Request service providers to generate DOIs for runs intended for long storage, provide links to presentations/preprints (with generated DOIs), provide references to papers, include findings and other results/products into live run metadata. | Users | 4.8 |

### 9.3 Recommendations for Open Validation

Open Validation aims to strengthen confidence in models, increase transparency in the R2O2R pipeline, and advance space weather forecasting capabilities. Table 7 shows our recommendations for improving openness in model validation and verification. As discussed in Section 5, users should understand the limitations and uncertainties of the model they are using. Ideally, model developers should conduct tests with ensembles of model runs that systematically explore the model validity range and uncertainties due to inputs, resolution, and model settings. This information should be provided as supplementary material along with the installation at a simulation service provider. When limitations and uncertainties are shared openly, trust in the model increases and the risk of misuse due to wrong expectations decreases.

To enable meaningful model validation activities, it is important to create a shared vision for model validation in the community. This vision includes agreement on essential physical quantities, metrics, time periods or events, metadata, and naming conventions. Streamlining validation procedures requires continued community collaboration throughout the R2O2R pipeline. Once a shared vision for validation is created, the next step is to use this knowledge to automate model validation as much as possible with open tools to facilitate community validation efforts. Automated tools increase transparency, fairness, and trust in model validation and selection of models for operational services. A priority should be placed on forecasting community endeavors such as CCMC Scoreboard projects, where models are run and tested in a realistic quasi-operational space weather environment.

**Table 7.** *Recommendations for Open Validation.*

| Recommendation | Target | Relevant Sections |
|---|---|---|
| **Model Capabilities and Limitations**<br>Provide better information to users on possible model limitations, so they understand which physical phenomena the model (or its specific simulation settings) can or cannot resolve. This information may be obvious to developers, but not necessarily to general users. Make explanatory materials accessible to users. Seek funding opportunities for model developers to continue improving their model capabilities to better meet the growing needs of their users. | Developers, Funding agencies | 5.1 |
| **Standardize Model Validation Procedures**<br>Prioritize the standardization of validation procedures in the heliophysics and space weather communities to support the identification/transition of promising research models to operations. Establish community agreement on events or time intervals, metrics, and essential physical quantities to enable consistent open validation across models. | Developers, Users, Simulation service providers | 5.2 |
| **Validation Infrastructure**<br>Use lessons learned from community efforts to develop open validation tools that promote transparency, fairness, and trust in model validation and model | Developers, Users, Simulation | 5.3 |

| Recommendation | Target | Relevant Sections |
|---|---|---|
| selection for operational services. Take advantage of existing infrastructure and tools at the CCMC. | service providers | |
| **Evaluate space weather forecasting models in a quasi-operational environment**<br>Evaluate space weather forecasting models (a subset of physics-based models) in a realistic, quasi-operational setting to support R2O2R. Support unbiased community-wide projects like Scoreboards at the CCMC, and ensure modelers have adequate resources to participate. Support regular reports summarising the status of forecasting capabilities. | Developers, Users, Simulation service providers | 5.3 |
| **Reference Problems and Sensitivity Assessment**<br>Reference problems with expected outputs should be open to users as a baseline. Furthermore, tests that systematically explore model sensitivity (e.g., inputs, resolution, parameter settings, and more) should become standard practice, and results should be made available to the community. | Developers, Simulation service providers | 5.1 |
| **Model Verification and Stress Testing**<br>Encourage model developers to make their verification tests public and reproducible. Encourage model developers to stress-test their numerical models under extreme physical conditions, such as superstorms, to enhance model robustness/reliability. Simulation service providers could support this process by offering guidance on best practices. | Developers, Simulation service providers | 5.1 |
| **Naming Conventions**<br>Establish naming conventions for physical parameters and derived quantities across models and observational datasets to support validation efforts. Explore solutions used in other disciplines, such as Earth sciences and astronomy. | Simulation and Data Service providers | 4.2.2 |
| **Open Metrics Repositories**<br>Jointly develop and maintain code repositories of metrics and visual diagnostics for assessing modeling capabilities. A shared source code repository that can support different validation platform front ends will strengthen collaboration while reducing duplicated effort. | Simulation service providers, Operational Centers | 5.3 |

### 9.4 Recommendations for Open Development

Open development of models can accelerate the pace at which new modeling capabilities are developed and transitioned to operations. Table 8 summarizes several recommendations. Model development together with the community works best when modelers make their source code publicly available in a repository as early as possible, to increase opportunities for collaboration. Moreover, developers should follow best practices from the software industry (e.g., version control, automated testing, and documentation generation). Critical to the success of open development is appropriate credit and recognition to modeling teams whenever their models are used and modified. Finally, new collaborative approaches to model development (e.g., hackathons, Kaggle competitions, tighter integration between modeling groups and simulation service providers) offer opportunities to solve focused problems and develop source-to-impact modeling capabilities.

**Table 8.** *Recommendations for Open Development.*

| Recommendation | Target | Relevant Sections |
|---|---|---|
| **Developers' Recognition and Credit**<br>Create incentives for open development by ensuring that model developers receive credit, citation, and acknowledgement when their models (or codes) are used by the community. Publishers should require citing modeling papers, referencing model websites, code repositories, simulation service provider websites, and acknowledge model funding. | Funding agencies, Universities, Publishers | 6.1, 6.3 |
| **Create Incentives for Innovation**<br>Create incentives, including sufficient computational resources, for innovation in modeling so that open science supports not only faster progress in existing models but also the emergence of innovative modeling approaches. | Funding Agencies | 6.1, 6.4 |
| **License and Legal Support**<br>Set clear guidelines for selecting open-source licenses and provide recommendations on handling multi-institution or legacy code licensing, patenting, and intellectual property rights. Clarify NOSA licensing issues, such as potential conflicts with other open-source license requirements. | Funding agencies | 6.4 |
| **Source-to-Impact Space Weather Community Modeling**<br>Develop source-to-impact modeling capabilities by leveraging initiatives and programs such as COSPAR ISWAT, NASA's Centers of Excellence, DRIVE Centers, and LWS Strategic Capabilities. Build upon already existing capabilities at the CCMC, such as collaborative modeling environments, real-time modeling systems, and open validation tools. | Funding agencies, Users, Developers, Simulation service providers | 6.7 |
| **Modern Approaches for Community Modeling**<br>Explore modern open model development approaches, including collaborative formats like hackathons, which enable focused, short-term teamwork on key heliophysics and space weather challenges. Also consider competitive formats such as Kaggle challenges, which provide open and fair competition with agreed-upon measures of success for new developments. Both activities provide students and early-career scientists with hands-on development experience. | Developers, Funding agencies, Simulation service providers | 6.7 |
| **Open-Source Code Release**<br>Release model source code as early as possible to enable community involvement and collaborative model improvement. Include a code of conduct in repositories for external contributions, with rules and expectations, to support collaborations with the rest of the community. | Developers | 6.2 |

| Recommendation | Target | Relevant Sections |
|---|---|---|
| Potential vulnerabilities of sensitive/operational code and information require special consideration when participating in open development. | | |
| **Code Development Best Practices**<br>Follow best practices such as version tracking, frequent commits, build automation, automated testing, continuous integration, and automated documentation generation. Testing results should be archived and tracked. | Developers | 6.2 |

### 9.5 Recommendations for Open Collaboration

Open Collaboration aims to accelerate progress in research and forecasting by strengthening partnerships across the modeling community, simulation service providers, data service providers, and operational forecasting centers. Successful collaborations require trustworthy working relationships that can be facilitated by defining expectations and credit mechanisms. Open collaboration also should be inviting to new participants and new ideas. Table 9 summarizes the key recommendations on open collaboration across the heliophysics and space weather community.

**Table 9.** *Recommendations for Open Collaboration.*

| Recommendation | Target | Relevant Sections |
|---|---|---|
| **Open Collaborative Initiatives**<br>Support and participate in open collaboration initiatives like the COSPAR ISWAT Initiative to address focused science and space weather forecasting challenges. | Funding agencies, Users, Developers, Run producers, Simulation service providers | 7.1 |
| **Collaboration between Simulation and Data Service Providers**<br>Foster collaboration between simulation service providers and data service providers. For example, work together to harmonize naming conventions between simulation and observational data archives and develop shared API calls. This would simplify retrieving physical quantities for model validation and data assimilation. | Funding agencies, Simulation and data service providers | 7.3 |
| **Collaboration between Operational Space Weather Forecasting Centers and the Modeling Community**<br>Operational space weather forecasting centers and the modeling community should foster collaboration by developing and sharing forecasting tools, participating in open validation projects to identify promising forecasting models, and designing displays and dashboards tailored to end-users' needs. | Funding agencies, Operational centers, Developers, Simulation service providers | 7.4 |

| Recommendation | Target | Relevant Sections |
|---|---|---|
| **Collaboration on Simulation Outputs**<br>Develop and maintain infrastructure that enables users to share simulation runs, model settings, event information, and key findings. This infrastructure should also enable community contributions to run metadata. | Simulation service providers | 7.2 |
| **Collaboration between Simulation Service Providers**<br>Simulation service providers should work more closely together, for instance, by adopting common interoperability standards to enable the exchange of outputs, related or derived products, data, and metadata between their archives. These standards can improve search, discovery, and compatibility with third-party tools. | Simulation service providers | 4.3, 4.4 |
| **Cross-Domain Collaboration**<br>The heliophysics and space weather community should foster collaborations between heliophysics domains and remain open to working with other scientific fields, such as astronomy and terrestrial weather. | Funding agencies, Users, Developers, Run producers, Simulation service providers | 4.1.3.1 |
| **Advance Modeling through Open Science**<br>Sustain HOME as a platform for modelers and model users to work together, facilitate community modeling, maximize the return on investment in modeling, and advance modeling, understanding, and forecasting in heliophysics and space weather. | Funding agencies, Users, Developers, Run producers, Simulation service providers, Operational centers | 8 |

A key outcome of this community effort is the recognition of the need for ongoing collaboration and exchange on open science in modeling. In Section 8, we introduced a new cross-domain initiative called Heliophysics Open Modeling Environment (HOME), which will serve as a platform for model developers, service providers, and users to work together, facilitate community modeling, enhance the scientific return on modeling investments, and advance modeling, understanding, and forecasting in heliophysics and space weather.

## Appendix A: Model Description Highlights and Modelers' Perspectives

This Appendix includes the model description and highlights provided by the modeling groups participating in the survey presented in Section 3, along with perspectives on successes and challenges with open use, open source, and open science practices.

### A.1 AMPS (Valeriy Tenishev: NASA MSFC, USA; Yinsi Shou: University of Michigan, USA)

The Adaptive Mesh Particle Simulator (AMPS, Tenishev et al., 2021; Shou et al., 2021) is a 3D kinetic modeling framework for space environment problems where particle transport, sources, and interactions matter. It simulates neutral and plasma populations, dust, and energetic particles for comets/planetary exospheres, solar wind–body interactions, and solar energetic particle (SEP) transport/precipitation (Glass et al., 2021; Tenishev et al., 2005; Tenishev et al., 2024). Numerically it uses Direct Simulation Monte Carlo and Monte Carlo test-particle methods with detailed source/sink physics (photo-/electron-impact ionization, charge exchange, sputtering, chemistry) on an adaptive Cartesian mesh. Particles move in prescribed or coupled electromagnetic fields; coupling to magnetohydrodynamic (MHD) and/or empirical field models is supported. The code is high performance computing (HPC)-ready, such as MPI/OpenMP and optional graphics processing unit (GPU) acceleration, and has been validated with spacecraft data (e.g., Rosetta, MAVEN). AMPS is research-grade today and is available at the CCMC for public use (https://ccmc.gsfc.nasa.gov/models/AMPS~2016/).

Open Science Perspectives

A core design principle was to ensure that the code could be adapted for problems not envisioned at the time of development. To achieve this, AMPS implements a layered architecture that supports three distinct levels of user interaction. At the developer level, users can introduce new general-purpose physical models, such as particle collisions, turbulence, sputtering, or energy exchange processes. At the expert user level, users can supply problem-specific routines that, for example, calculate rates of physical processes, branching ratios, or sample velocity distributions resulting from specific interactions. Finally, at the user level, AMPS behavior can be configured entirely through parameters defined in input files without modifying the code.

Testing is a central part of AMPS development. The code includes a robust regression testing framework to ensure the reproducibility of the model results and to verify that new code modifications do not introduce unintended changes. Additionally, unit tests validate individual physical models and functions in isolation from the rest of the codebase.

A.2 ARMS (Rick DeVore: retired independent consultant, CUA, and NASA GSFC, USA; Lars K.S. Daldorff: CUA and NASA GSFC, USA)

The Adaptively Refined MHD Solver (ARMS, DeVore & Antiochos, 2008) is a 3D MHD code in both Cartesian and spherical coordinates designed to simulate a range of solar phenomena. ARMS employs the widely used PARAMESH toolkit (MacNeice et al., 2000) to manage the parallel processing and adaptive meshing aspects of the solver, and well-established Flux Corrected Transport algorithms (DeVore, 1991) to advance the solutions in time. ARMS solves the ideal MHD equations with divergence-free conditions built into the solver, and it conserves helicity exceptionally well. Reconnection is enabled by numerical resistivity introduced by its minimally diffusive, monotone algorithms. Due to the placement of the finest resolution at current sheets, ARMS simulations can attain Lundquist numbers of order $10^4$ or more, sufficient to permit plasmoid instability and other high-Lundquist number behavior, such as explosive eruption onset, expected under solar coronal conditions.

ARMS is currently undergoing extensive revision to modernize the Fortran code, organize and modularize the code according to current standards, improve flexibility and ease of use, modify the input/output approaches, and add essential physics efficiently such as field-aligned thermal conduction. Full documentation is in progress and will be made available to users.

A.3 COCONUT (Barbara Perri, Stefaan Poedts: KU Leuven, Belgium)

The 3D MHD COolfluid COroNa UnsTructured (COCONUT) model (Perri et al., 2022; Brchnelova et al., 2022), simulating solar corona phenomena, has been implemented within the COOLFluiD (Computational Object-Oriented Libraries for Fluid Dynamics) architecture (Lani et al., 2013, 2014). COCONUT uses an unstructured grid and an implicit solver avoiding the Courant–Friedrichs–Lewy time step limitation and speeding up convergence considerably. The initial polytropic MHD model was recently updated to a full MHD model including thermal conductivity, radiative losses, and ad-hoc heating terms (Baratashvili et al., 2024). The model was recently made time accurate (Wang et al., 2025) and dynamically coupled to both EUHFORIA (see Appendix A.8) and Icarus (see Appendix A.12), which enables launching a CME at the bottom of the corona (instead of only at 0.1 AU), and simulating its evolution up to 1 AU and beyond in a dynamic background solar wind (Linan et al., 2023; Guo et al., 2024; Linan et al., 2025).

COCONUT is integrated in COOLFluiD (https://github.com/andrealani/COOLFluiD/wiki). It has also been integrated into the VSWMC test server and will be available on the operational server soon.

A.4 DTM (Sean Bruinsma: CNES, France)

The Drag Temperature Model (DTM) is a semi-empirical model describing the temperature, density, and composition of the Earth's thermosphere in the altitude range 120 to 1,500 km. The first DTM78 (Barlier et al., 1978), was developed in the seventies. The most recent version is DTM2020 (Bruinsma & Boniface, 2021). DTM predicts temperature, concentrations of the main constituents (O, $O_2$, N, He, H), and total density pointwise as a function of location (altitude, latitude, longitude, local solar time), solar and geomagnetic activities, and day-of-year. This type of model is mainly used to compute the atmospheric drag force in satellite orbit determination because of their computational efficiency, robustness, ease of implementation and use, and relatively good precision, but at low resolution.

The solar and geomagnetic proxies limit the temporal resolution of these models implicitly to 1 day and 1 or 3 hours, respectively. As a result, small scale and high-frequency density perturbations, which are present mostly at high latitudes, contribute to the prediction uncertainty because they cannot be modeled. The operational, intermediate, and research versions of the DTM2020 model use different combinations of drivers: F10.7 and Kp, F30 (Dudok de Wit et al., 2014) and Kp, and F30 and Hp60 (Yamazaki et al., 2022), respectively. The solar radio flux at 30 cm is significantly more representative of solar EUV and UV emissions than F10.7, but it is not (yet) an operational index.

The DTM2020 models are publicly available at https://github.com/swami-h2020-eu/mcm/tree/main/src/dtm2020. The operational version of

DTM2020, as well as the previous release DTM2013, are available at the CCMC (https://ccmc.gsfc.nasa.gov/models/DTM~2020/).

A.5 E-CHAIM (David Themens, Sean Elvidge: University of Birmingham, UK)

The Empirical Canadian High Arctic Ionospheric Model (E-CHAIM) is an empirical climatological model of the high latitude ionosphere designed specifically to support ionospheric specification for high-frequency propagation applications (Themens et al., 2017, 2018, 2019; Watson et al., 2021). The model uses a single semi-Epstein layer with height varying scale thickness to represent electron density. The model includes a geomagnetic forcing parameterization in its specification of the peak density, bottomside thickness, and topside thickness, allowing it to capture 15-25% of high latitude ionospheric variability at sub-monthly time scales, mainly associated with negative ionospheric storm responses (Themens et al., 2020). The model furthermore includes a climatological representation of low-energy auroral precipitation (Watson et al., 2021) and a probabilistic model of high latitude sporadic-E (Themens et al., 2024).

Open Science Perspectives

Since E-CHAIM's inception, the model development team actively maintains native Interactive Digital Language (IDL), Matlab, and C versions of the model code to ensure broad availability and ease of use (https://e-chaim.chain-project.net/). The model source code is furthermore freely available with open license and Python wrappers for the C version of the model have been developed and made available by users. To further facilitate ease of use, the model drivers can be updated automatically by the E-CHAIM code itself, where a database of drivers is actively maintained by the model team. Through the provision of multiple versions across three commonly-used languages in industry and academia and the provision of detailed examples and user instructions, we ensure that our code is not only open source, but also open use, where there are little to no barriers to even novice users running the model. This principle of open use has ensured broad adoption of our model not only in the academic community but, perhaps even more broadly, in the commercial and operational sector. The model is now used at nearly 100 institutions across 25 countries and acts as the background model for operational near-real-time data assimilation systems actively applied to support high-frequency communications and remote sensing systems in the Arctic (Reid et al., 2023, 2024).

Because of maintaining an actively updating database of solar wind observations, geomagnetic indices, and solar activity indices, the model team also actively generates and openly publishes daily updates of the index files needed to run the International Reference Ionosphere (IRI, https://irimodel.org). A subsequent effort by the Space Environment and Radio Engineering (SERENE) group at the University of Birmingham also makes daily-updated versions of the interplanetary magnetic field (IMF) and geophysical indices (GPI) files needed for running the Thermosphere-Ionosphere-Electrodynamics Global Circulation Model (TIE-GCM, see Appendix A.25, https://spaceweather.bham.ac.uk/resources/TIE-GCM/), as well as solar driver files for the extended Whole Atmosphere Community Climate Model (WACCM-X, see Appendix

A.27, https://spaceweather.bham.ac.uk/resources/WACCM-X/). These efforts are intended to make these models more accessible and readily usable by members of the community.

Future efforts will see the development of a native Python version of the model and will include new capability for GPU optimization to enable the generation of much larger model ensembles and thereby further support the model's use in real time operational data assimilation systems.

There are concerns regarding open source that go beyond this particular model, that a code base could be used incorrectly to provide bad or incorrect output that then reflects poorly on the model, or that it could be modified such that it then becomes hard to interpret whether subsequent work is indicative of the model or the modifications. The latter can likely be addressed through better community practices of properly documenting and versioning software, even forks, and better journal standards for the treatment of that versioning and documentation.

A.6 ENLIL (Dusan Odstrcil: George Mason University, USA)

ENLIL (http://helioweather.net/, Odstrcil, 2023) is a time-dependent 3D MHD model of the heliosphere that solves equations for plasma mass, momentum, energy density, and magnetic field, simulating the global solar wind and transient disturbances. It uses a modified Lax-Friedrichs (Rusanov) scheme with local Lax-Friedrichs or Harten-Lax-van Leer numerical fluxes and Dedner numerical diffusion. Its inner radial boundary is located beyond the sonic point, typically at 21.5 or 30 solar radii. It can accept boundary condition information from the WSA (see Appendix A.28), MAS (see Appendix A.17), and Heliotomo (see Appendix A.10) models. The outer radial boundary can be adjusted to include planets or spacecraft of interest (e.g., 2 AU to include both Earth and Mars, 5 AU to include Ulysses, 10 AU to include Cassini). It covers 60 degrees north to 60 degrees south in latitude and 360 degrees in longitude.

ENLIL has been used at the CCMC since 2010 (https://ccmc.gsfc.nasa.gov/models/ENLIL~2.8f/, https://ccmc.gsfc.nasa.gov/models/CORHEL-MAS_WSA_ENLIL~5.0/, https://ccmc.gsfc.nasa.gov/models/WSA-Enlil-at-SWPC~3/), at NOAA/SWPC since 2011 https://www.swpc.noaa.gov/products/wsa-enlil-solar-wind-prediction, and coupled with Heliotomo at University of California San Diego since 2016 https://ips.ucsd.edu/IPS-ENLIL_predictions.

A.7 EPREM (Matthew Young: University of New Hampshire, USA)

The Energetic Particle Radiation Environment Model/Module (EPREM) simulates acceleration and transport of ions throughout the heliosphere by numerically solving the focused transport equation (Skilling, 1971; Ruffolo, 1995; Kóta et al., 2005) on a Lagrangian grid in a frame co-moving with the solar wind plasma. EPREM provides the particle acceleration and transport component of the Earth-Moon-Mars Radiation Environment Module (EMMREM,

Schwadron et al., 2010) and of the Predictions of radiation from REleASE, EMMREM, and Data Incorporating CRaTER, COSTEP, and other SEP measurements (PREDICCS) online tool.

Various versions of EPREM have been one-way coupled to established MHD simulations, including BATS-R-US (see Appendix A.24, Kozarev et al., 2013), ENLIL (see Appendix A.6, Mays et al., 2016), and MAS (see Appendix A.17, Schwadron et al., 2015), for the purpose of providing the MHD quantities needed to solve the focused transport equation. Most recently, EPREM developers at the University of New Hampshire and MAS/CORHEL developers at PSI incorporated MAS-coupled EPREM into the SPE Threat Assessment Tool (STAT; Linker et al., 2019), which produces visualizations of energetic proton flux, fluence, and integral flux throughout the heliosphere based on EPREM simulation runs driven by the output of a MAS simulation run.

Open Science Perspectives

An effort to make publicly available an uncoupled implementation of EPREM is currently underway. The code base is hosted on GitLab (https://gitlab.com/open-eprem/eprem), alongside repositories containing a suite of analysis tools (https://gitlab.com/open-eprem/eprempy) and visualization scripts built on those tools (https://gitlab.com/open-eprem/eprem-analysis). All three repositories are currently under active development. All three repositories are also available under an open-source license and welcome contributions. As of this writing, uncoupled EPREM version 0.13 is the latest stable version and uncoupled EPREM version 0.11 is the latest version available for runs-on-request via the CCMC (https://ccmc.gsfc.nasa.gov/models/EPREM~0/). This uncoupled model uses internally generated values of the magnetic-field components, velocity-field components, and density to solve the focused transport equation. The advantage of such a setup is that it does not require access to output from a particular MHD model. The disadvantage is that the current internal MHD models are highly simplified. Because of this disadvantage, the long-term plan for uncoupled EPREM is to develop functionality to allow it to ingest output from one of many — and ideally, any — MHD simulation with recourse to as little MHD model-specific code as possible.

Some of the challenges that we have faced in developing this open-source code base are:

- Updating documentation, as well as condensing various closed-source manuals and institutional knowledge into unified manuals for users and developers.
- Settling on an appropriate place to host the source code. The initial open-source code base resided on GitHub, but there was a clear benefit to grouping EPREM with the analysis packages, so we decided to move it to a GitLab group. This has been overall beneficial, but it required updating our nascent user community. Furthermore, we may deem it beneficial or even necessary to move the three related repositories to a new group (e.g., to develop an open-source EMMREM code base), in which case we will need to proactively alert users to the updated URLs.

- Extracting the uncoupled (or MHD-agnostic) model from multiple hard-coupled repositories. Here, hard-coupled refers to MHD model-specific functionality explicitly coded into EPREM. Over time, multiple repositories evolved with a high degree of redundancy, despite including similar functionality to ingest output from a specific MHD model (e.g., ENLIL) and often differing only by variable names. This has been a challenge not only in terms of teasing out the core EPREM functionality, but also in terms of making sure the result meets the expectations of users who had become familiar with a particular EPREM implementation.
- Deciding on a useful set of visualization routines. This challenge is arguably more subjective than the others, but the Lagrangian nature of EPREM means that some users may find its output less intuitive than the output of simulations on Eulerian grids. We therefore felt that it would be easier to convince potential users of EPREM's value if we could point them to some friendly visualization routines.

One final comment on converting code bases to open-source repositories or deciding to follow an open-source development model from the start: we have encountered developers who are reluctant to make their model publicly available because they are concerned that it will allow any stranger to mangle their code. Preventing such behavior is, of course, straightforward by protecting the main branch and forcing contributors to follow a pull- or merge-request workflow. However, reluctant developers need to know that those options are available before they can convince themselves of the benefits. A modest amount of community education on this topic could make a big difference.

A.8 EUHFORIA (Jens Pomoell, Stefaan Poedts: KU Leuven, Belgium)

The EUropean Heliospheric FORecasting Information Asset (EUHFORIA, Pomoell & Poedts, 2018) is a heliospheric wind and CME propagation and evolution model. It consists of a simple WSA-like coronal model that provides input boundary conditions at 0.1 AU, the inner boundary of the heliospheric model solving the time-accurate ideal MHD equations out to 2 AU typically, although the outer boundary can be up to 5 AU and further (as needed for SEP simulations with PARADISE, see Appendix A.18). EUHFORIA uses a uniform spherical grid, constrained transport, and heliocentric Earth equator coordinates.

EUHFORIA is available via GitHub upon request. It has also been integrated into the VSWMC operational server. The model is routinely run by the UK Met Office (UKMO) and by the ROB in support of space weather forecasting. The daily UKMO runs are archived and available via the ESA Space Weather Portal (https://swe.ssa.esa.int/ral-hparc-pb-federated).

A.9 GITM/Aether (Aaron Ridley: University of Michigan, USA)

The Global Ionosphere Thermosphere Model (GITM) is an atmospheric model that solves the Navier-Stokes equations for a variety of neutral species and a modified set of equations for many ion species (Ridley et al. 2006). GITM can be run in 1D (mostly for testing) and 3D across the globe and regional simulations. GITM works on a stretched fixed altitude grid that has a block-based domain decomposition in the horizontal (longitude/latitude) direction, allowing it to run across many processors. GITM has been adapted to work on Earth, Mars, Titan, and Venus, but here we describe the features of the Earth-based version. In the vertical direction, GITM typically is run with 50 altitude grid cells that start at 100 km and are spaced roughly 0.3 scale-heights apart, where the temperature and mass profile is taken at the subsolar point at the start of the simulation using mass spectrometer and incoherent scatter radar data (MSIS) in order to determine the spacing. The complete vertical momentum equation for the following neutral species is solved for O, $O_2$, $N_2$, N, NO, and He. This means that GITM can develop non-hydrostatic solutions. In fact, GITM allows gravity to vary as a function of altitude in addition to allowing a variety of forces to drive the vertical winds. While GITM is almost always in a vertically forced-balanced state, it is never a strictly hydrostatic solution, since gravity varies, and there are other constant accelerations acting on the atmosphere, such as centrifugal and geometric accelerations. A mass-weighted average bulk vertical wind is determined from the individual species winds. In the horizontal direction, each of these species is advected with the bulk wind. GITM includes the chemistry (but not advection) of $N(^2D)$, $N(^2P)$, $O(^1D)$. In the neutral thermodynamic equation, collisions with the ions are treated as separate frictional (velocity difference) and heat transfer (temperature difference) terms, which is different from other global ionosphere/thermosphere models. Exothermic chemical reactions are the largest source of energy, driven primarily by solar EUV ionization. 5% of the solar EUV energy deposition is directly transferred to the neutrals, but this is adjustable by the user. GITM solves for a wide variety of chemical reactions using a part-implicit method, allowing dynamic stable solutions that can vary rapidly. GITM can be run with different lower boundary conditions on the neutrals including MSIS/Horizontal Wind Model and Hough Mode Extensions for the tidal fields.

For the ions, $O^+$ and $NO^+$ are advected. The following ions are chemically active, but are not advected: $O^{2+}$, $N^{2+}$, $N^+$, $O^+(^2D)$, $O^+(^2P)$, and $He^+$. Bulk ion drifts along the field lines are driven by gravity, pressure gradients (including electron pressure), and collisions with neutral winds. Across the field lines, electric fields, gravity, pressure gradients (including electron pressure), and collisions with neutral winds are included. The divergence of the ion velocity along the field line is considered in the continuity equation in addition to the advective term. The ion and electron energy equations are solved using semi-implicit schemes. For the ions, collisions with neutrals and electrons are considered, while for the electrons, collisions with ions are considered as well as collisional loss-terms with specific neutral species, but frictional and heat transfer collisions with neutrals are not included. Below 45°, GITM calculates a self-consistent dynamo electric field. Above 45°, the electric potential and aurora are specified by external models, such as the Weimer (2005) potential model and the Feature Tracking of the Aurora (Wu et al., 2021a; Wu & Ridley, 2024) model of the aurora. A wide variety of other models can also be used. The auroral average energy and energy flux are used to calculate 100 monoenergetic beams, which are then used with the Fang et al. (2010, 2013) energy deposition relationships to derive ionization rates as a function of altitude for each horizontal grid position. Using the 100 monoenergetic beams, GITM can simulate Maxwellian and Kappa distributions for electron and ion diffuse auroras as well as discrete and wave-driven auroras for electrons. GITM can be run using the IGRF magnetic field or a dipole that can be aligned with the geographic rotation axis or tilted and offset.

Typical global simulations have resolutions of 1° latitude by 4° longitude (on 100 processors), but GITM can be run with a wide variety of resolutions and has been run with global scales down to 0.5° latitude by 2° longitude resolution (on 800 processors). GITM uses MPI to communicate between blocks, with ghost cells outside of the block domain. This allows each block to be independent, treating ghost cells as boundaries. A variety of Python codes are available to prepare driver files (including solar wind and interplanetary magnetic fields, auroral electrojet indices, a configuration file, and FISM EUV specification). Python codes are also included for post-processing and comparing to a variety of data sources.

GITM is freely available on GitHub ([Ridley et al., 2025]) and at the CCMC via ROR ([https://ccmc.gsfc.nasa.gov/models/GITM~23.01/](https://ccmc.gsfc.nasa.gov/models/GITM~23.01/)).

Aether is an as-yet-to-be-published model of the ionosphere and thermosphere. It is very similar to GITM in its capabilities and underlying physics, but its architecture is quite different. A summary of differences includes:

- Aether is written in C++. To avoid 3D loops for math, the Armadillo library is used, allowing entire data cubes to be manipulated with a single command. Internally, the Armadillo library uses OpenMP to parallelize the loops. The design choice of C++ was made because most students learn to code in C++, while almost no students learn to code in Fortran. While there is most likely a performance penalty for working with C++, the inclusion of many more programmers offsets this penalty.
- There are two underlying grids in Aether, one for the neutrals and one for the ions. The available grid types include: spherical (similar to GITM), cubesphere (take a cube and blow out its 6 sides to spherical shape, getting rid of the pole problem), and dipole (a grid that is aligned with the dipolar magnetic field). The two grids run in parallel with each other, passing different states to each other as the simulation progresses. Technically, each grid is supported for both neutrals and ions, but the dipolar grid is optimized for ions, while the two radially aligned grids are best for the neutrals. The user specifies the number of points in each of the three dimensions for the two grid systems. Aether uses a quadtree-based 2D domain decomposition, so that the user doubles the resolution when they use 4x the processors (i.e., for a spherical grid, the user can ask for 1, 4, 16, 64, 256, etc. processors). Finer tuning can be easily made by adjusting the number of cells in each direction in the input files.
- Chemistry, EUV, planetary characteristics, neutral and ion species, and collision terms are all described with text files, allowing the user to change these without recompiling the code. This allows users who are hesitant to alter code to modify the atmosphere that they desire to simulate.
- Ensembles are natively implemented in Aether, meaning that the user can run multiple simulations of the same event, perturbing different drivers and parameters within the code. All drivers (e.g., F10.7, IMF, solar wind, and AE) can be easily perturbed in different ways. Chemical reaction rates can also be perturbed. Users add the number of ensemble members they would like as well as which variables they would like to perturb in the input file. Post-processing codes create means and standard deviations of all output states.

Aether is freely available at https://github.com/AetherModel.

A.10 Heliotomo (Bernard V. Jackson: University of California San Diego, USA)

The University of California San Diego time-dependent 3D reconstruction determines heliospheric structure and predicts CME and corotating structure arrival at Earth using interplanetary scintillation (IPS) and/or Thomson scattering analyses (e.g., Jackson et al., 2020). This technique is unique in that it derives the location and extent of solar wind structures from heliospheric remote sensing alone and does not use a predetermined shape to derive their structure or speeds. Developed over more than two decades (Jackson et al., 2001, 2003; Jackson & Hick, 2005; Jackson et al., 2006, 2010, 2013, 2023a), the 3D reconstruction proceeds by least-squares fitting a kinematic or MHD solar wind outflow to line-of-sight observations. The line-of-sight weighting of the IPS scintillation level or speed, or Thomson scattering brightness, or the ratio of brightness and polarization brightness plus perspective outflow views provide an iterated fit to the observations. Since 2005, these 3D temporal reconstructions have been employed to forecast solar wind density and speed, and can extract solar surface magnetic field components up to five days into the future (Dunn et al., 2005; Jackson et al., 2023b). Recent heliospheric imager brightness analyses have enabled these 3D spatial and temporal reconstructions to match in-situ observations down to mesoscale sizes (see Jackson et al., 2020).

Heliotomo is available at the CCMC via ROR (https://ccmc.gsfc.nasa.gov/models/HeliosphericTomographyIPS~24/).

A.11 HYPERS (Yuri Omelchenko: Space Science Institute, USA)

HYPERS (Omelchenko et al., 2021; Omelchenko & Karimabadi, 2012, 2023) is a hybrid—particle-in-cell (PIC) ions of multiple species and massless fluid electrons—code that belongs to a newer family of kinetic (PIC) models at the CCMC (https://ccmc.gsfc.nasa.gov/models/HYPERS-Global~2021/). HYPERS is a unique multiscale code, taking advantage of adaptive numerical algorithms in both space and time. For spatial adaptivity, HYPERS employs a spatially adaptive mesh, which is uniform in logical Cartesian coordinates and stretched in configuration (physical) space. Temporal updates of electromagnetic fields and macro-particles in HYPERS are asynchronous and self-adaptive, driven by the Event-driven Multi-Agent Planning System (EMAPS). EMAPS incorporates elements of discrete-event simulation and artificial intelligence (Omelchenko & Karimabadi, 2023). The numerical spatio-temporal adaptivity in HYPERS enables large-scale, massively parallel runs of Earth's magnetosphere and provides superior accuracy and execution speed metrics, as compared to similar hybrid models.

Open Science Perspectives

Physics-driven kinetic models are known for being resource-intensive (CPU, memory, and storage intensive), compared to their well-established fluid counterparts. As discussed below, this naturally creates extra challenges for onboarding 3D kinetic models at a simulation service provider for open use.

- To achieve reasonable accuracy in global 3D simulations of the Earth's magnetosphere, HYPERS simulations typically employ more than 10,000 parallel cores (up to >100,000 cores) and run longer than 10-15 wall-clock hours. Such resource-intense simulations need to be carefully set up to avoid wasting computational resources. This typically requires significant prior experience with hybrid simulations, conducting test (2D and 3D) runs, and possibly holding direct consultations with model developers. Kinetic runs also generate large data for which users would typically have difficulty providing long-term storage. Therefore, instead of focusing on runs-on-request for 3D kinetic models, simulation service providers may find it more practical to work with model developers on constructing databases of interesting runs for open analysis by all interested parties. This effectively means transitioning from runs-on-request to data-on-request for kinetic models. This paradigm is also consistent with modern open-science data requirements for publications.
- Currently, CCMC visualization capabilities mostly revolve around the generation of standard 1D and 2D plots of simulation data using in-house software. Although this kind of output may be sufficient for basic analysis of simulation results, it still lacks the quality of advanced 3D visualization (e.g., Paraview based) tools that some model developers may employ. The CCMC team and end users will benefit from working directly with modelers on making advanced viz tools available for open use.
- The open-science simulation community would greatly benefit from specific funding opportunities for model developers so that they are able to: 1) improve the performance of their models, 2) store high-quality simulation data from benchmark (published) simulations for long-term analysis, and 3) supply simulation service providers with documentation and visualization tools for analysis of model outputs.
- Should simulation service providers decide to harbor multiple kinetic models of similar physical approximation (e.g., multiple hybrid codes), the provider team is encouraged to work closely with their respective developers on establishing a unified input deck for common test problems. This will help both developers and users conduct model comparisons and further improve simulation capabilities.
- Hybrid codes are now used to model many different plasma systems. For instance, the HYPERS code can be used for modeling plasmas ranging from magnetospheres to shock-driven ion acceleration to laboratory experiments. The scientific community would benefit from simulation service providers being able to safely maintain the source code, related scripts, documentation, and input decks for these different applications (e.g., HYPERS-Global, HYPERS-Shock). To summarize, putting more emphasis on the data-on-request (vs runs-on-request) approach for resource-intensive applications, supporting model improvement efforts, collecting user feedback, and improving model visualization capabilities should be considered by the modeling community as key principles for ultimate success of open science.

- In 3D high-resolution simulations, HYPERS and other particle-in-cell codes typically generate terabytes of 6D particle data, periodically saved for post-processing. Storing these data permanently for open-science access may easily become a prohibitive task. Extra work is needed to reduce the amount of stored kinetic information without significantly degrading its physical resolution. For instance, this could be achieved by applying precision-controlled machine learning algorithms (e.g., clustering) for merging (compressing) particle data before saving it for post-processing. The quality of particle merging can be ensured by tolerances on particle positions and momenta. These tolerances will act to balance the accuracy of preserved kinetic information and degree of data reduction. Smaller tolerances will mean smaller kinetic errors and little reduction of particle data. Larger (relaxed) tolerances will result in merging more particles, thus reducing their final numbers saved for permanent storage.

A.12 Icarus (Christine Verbeke, Tinatin Baratashvili, Stefaan Poedts: KU Leuven, Belgium)

The relatively new heliospheric wind and CME evolution model Icarus (Verbeke et al., 2022; Baratashvili et al., 2022) has been implemented in the numerical framework MPI-AMRVAC since version 3.0 (Keppens et al., 2023). It uses finite-volume discretization with a radially stretched grid, and block-adaptive mesh refinement with combined refinement criteria to speed up the simulation runs (Baratashvili et al., 2025). Icarus solves the ideal MHD equations in a corotating spherical finite-volume grid. It was recently upgraded to enable time-dependent boundary conditions at the inner heliospheric boundary, i.e., at 0.1 AU, based on updated photospheric magnetogram information varying during the simulation. This yields more realistic solar wind and CME evolution modeling during solar maxima and extends the applicability of Icarus to studies beyond 1 AU.

Icarus is integrated as one of the test cases in MPI-AMRVAC (https://github.com/amrvac/amrvac/tree/master/tests/mhd/icarus). It has also been integrated in the VSWMC test server and soon it will be available on the VSWMC operational server and on the CCMC ROR service (https://ccmc.gsfc.nasa.gov/models/Icarus~3/).

A.13 IMPTAM (Natalia Ganjushkina: University of Michigan, USA)

The Inner Magnetosphere Particle Transport and Acceleration model (IMPTAM) traces distributions of ions (protons, O+ and He+) and electrons in the drift approximation with arbitrary pitch angles from the plasma sheet to the inner L-shell regions with energies up to hundreds of keVs in time-dependent magnetic and electric fields (Ganushkina, 2023). Relativistic effects for electrons are taken into account in the drift velocities. Liouville's theorem is used to gain information on the entire distribution function, including loss process attenuation. For the obtained distribution, we apply radial diffusion by solving the Fokker-Planck equation (Schulz & Lanzerotti, 1974). IMPTAM is a flexible, module-based model that can use various representations for its main parts: particle transport, boundary conditions, background magnetic and electric fields, and loss processes.

In operational systems, it is often critical to rapidly determine whether spacecraft anomalies are, in fact, due to a dynamically changing environment, from hostile sources, or from incorrect operational procedures. IMPTAM can quickly assess and specify the current and predicted state of the radiation environment at keV energies. For example, the severity of the keV electron radiation environment at medium Earth orbit around the times when worst-case severe environments for surface charging were detected by LANL satellites at geosynchronous orbit was investigated with IMPTAM (Ganushkina et al., 2024).

IMPTAM is driven by (1) solar wind number density, (2) dynamic pressure, (3) velocity, (4) total strength and Y- and Z-components of IMF, and (5) Dst and (6) Kp indices. If these driving parameters can be forecasted, they can drive IMPTAM to forecast keV electron fluxes in the inner magnetosphere. 1 day of real time modeled in 1 hour of IMPTAM time is the slowest running time estimate. Near-real time comparisons have been made with GOES 13 and 15 MAGED data, the only keV electrons dataset available in near real time, until the end of GOES 13 and 15 operations (Ganushkina et al., 2019).

IMPTAM has been operational online (https://imptam.fmi.fi, http://imptam.engin.umich.edu) since February 2013 (Ganushkina et al., 2015). IMPTAM is currently being onboarded at the CCMC for the ROR service (https://ccmc.gsfc.nasa.gov/models/IMPTAM~1/).

A.14 iPIC3D (Fabio Bacchini, Pranab Deka, Paul Wilhelm, Nicolas Moens: KU Leuven, Belgium; Stefano Markidis: KTH, Sweden)

iPIC3D (https://github.com/iPIC3D/) is designed for large-scale kinetic simulations of space plasmas and planetary magnetospheres (Markidis et al., 2010; Peng et al., 2015). It solves the fully kinetic Vlasov–Maxwell system using a semi-implicit particle-in-cell (PIC) method. Electrons and ions are modeled as macro-particles that follow the (relativistic) equations of motion, while charge, current, and pressure tensor fields are interpolated from the particles to a 3D Cartesian grid. The adopted implicit scheme and strategies allow larger time steps and coarser spatial grids than explicit PIC codes, reducing 3D resolution needs by more than four orders of magnitude.

iPIC3D targets heterogeneous supercomputers, using a hybrid MPI + OpenMP/CUDA(+HIP) parallel approach which enables exascale-class performance on 32,768 AMD APUs on the El Capitan supercomputer at Lawrence Livermore National Laboratory. To manage output size, iPIC3D includes an in-situ, lossy compressor based on a Gaussian mixture model which achieves compression ratios over 1000:1, while preserving beam features, heating, and non-thermal tails. These capabilities enable iPIC3D to simulate realistic solar wind–magnetosphere interactions, revealing the impact of electron-scale dynamics on global magnetosphere dynamics. In its most recent version, iPIC3D also implements the exactly energy-conserving method (ECSIM, Lapenta, 2017) and its relativistic counterpart (RelSIM, Bacchini, 2023).

Open Science Perspectives

iPIC3D has been open source since 2012, supporting broad community adoption. Development is led by the team at the KTH Royal Institute of Technology, with roots at Los Alamos National Laboratory, the University of Illinois at Urbana-Champaign, and KU Leuven. Active users and contributors include research groups from KU Leuven, the University of Colorado Boulder, and NASA. The code has been a core component in European Commission-funded projects such as EPiGRAM (Markidis et al., 2016) and SPACE (Shukla et al., 2025; https://www.space-coe.eu/), which promote open-source tools for exascale computing. The European Commission's emphasis on open research infrastructures continues to guide the project's commitment to openness and accessibility.

A.15 MAGE (Viacheslav Merkin: Johns Hopkins University, USA)

The Multiscale-Atmosphere Geospace Environment (MAGE, http://cgs.jhuapl.edu/MAGE/) model is being developed by the NASA DRIVE Science Center for Geospace Storms (CGS, https://cgs.jhuapl.edu). MAGE spans the domains of geospace, from the lower atmosphere to the thermosphere-ionosphere, to the different regions of the magnetosphere. It resolves global dynamics and mesoscale processes throughout geospace with highly precise numerical techniques.

MAGE 1.0 is the latest version currently available at the CCMC (https://ccmc.gsfc.nasa.gov/models/MAGE~1.0/), consisting of the global magnetosphere Grid Agnostic MHD with Extended Research Applications (GAMERA) model (Zhang et al., 2019; Sorathia et al., 2020) coupled with the Rice Convection Model (RCM, Toffoletto et al., 2003) and the Redeveloped Magnetosphere-Ionosphere Coupler/Solver (REMIX) code (Merkin & Lyon, 2010). MAGE 1.0, coupling with the TIE-GCM (see Appendix A.25, Qian et al., 2014), has been in science production for a few years (Lin et al., 2021, 2022b; Pham et al., 2022; Bao et al., 2023) and is currently being onboarded at the CCMC.

Open Science Perspectives

MAGE development does not just consist of coupling existing codes together, but it largely involves writing new models from scratch or rewriting legacy code. This is done primarily for scientific reasons but is also at times necessary to enable the model to comply with open science requirements. For instance, the MAGE development team is currently rewriting RCM to enable much higher spatial and energy resolution and a more efficient coupling and deeper integration with GAMERA and the rest of MAGE. The Ionosphere/generalized Polar Wind model (IPWM, Varney et al., 2016a,b) is being redeveloped as the High-latitude Ionosphere Dynamics for Research Applications (HIDRA) model (Albarran et al., 2024) which uses GAMERA transport algorithms to enable much higher resolving power. The SAMI3 (see Appendix A.20, Huba &

Krall, 2013) ionosphere/plasmasphere model is also undergoing significant upgrades to enable efficient coupling with the rest of MAGE.

MAGE 0.75 has been available at the CCMC since April 2024 (https://ccmc.gsfc.nasa.gov/news/mage/) and has already accumulated more than 400 runs on request. This success has been enabled by a few factors. First, MAGE development follows the best practices of team software development. The model runs a hierarchy of nightly and weekly tests on the NCAR Derecho HPC system, checking both code updates and system changes. All code development is done in a version-controlled environment, all code updates are done via pull requests, and both test results and code updates are tracked in a dedicated Slack channel. Updates to the main branch are approved by a small group of developers, including scientists and software engineers, who have admin privileges. The team maintains a Slack user help channel where users ask questions that can be answered by other users or by developers. Second, we found it critical to be able to support a dedicated software engineering effort on the modeling team side that interfaces directly with the CCMC staff. This is important both during the onboarding process and during subsequent maintenance. CCMC staff are included on the team Slack space and receive code updates via git tags on a dedicated branch in the repository. Finally, MAGE is accompanied by a Python ecosystem, called Kaipy, that facilitates pre- and post-processing. Pre-processing software includes tools generating a grid or obtaining the necessary data to initiate the model. Additional Python scripts within the MAGE repository build upon the Kaipy package to facilitate the automation of runs on modern HPC systems such as NASA HECC platform Pleiades or NSF NCAR's Derecho (https://www.cisl.ucar.edu/capabilities/derecho). Visualization and analysis tools in Kaipy allow users to make default visualization or bring the data into Python for customization. Kaipy resides in a separate repository (Wiltberger et al., 2025) with its own online documentation (https://kaipy-docs.readthedocs.io/en/latest/). It is available under the permissive BSD 3-Clause license (https://spdx.org/licenses/BSD-3-Clause.html).

MAGE itself is currently being prepared for an open-source release. We find that one of the primary challenges that hinders more expeditious release of research software is the abundance of poorly documented and unlicensed software in legacy codes that is inevitably included in any scientific model. The MAGE development team is taking a judicious approach by systematically eliminating unlicensed or limited-use legacy code. In some cases, it leads to rewriting significant parts of the code which inevitably and significantly delays the process. Finally, an issue not to be overlooked, especially in a large multi-institutional team, is that releasing a code under an open-source license requires coordination with, review by, and permission from legal and technology transfer offices at multiple institutions, which can add up to a sometimes-lengthy process. Anyone who is embarking on a path of developing a large open-source community model should be prepared for a long road ahead.

A.16 MARBLE (Chris Bard: NASA GSFC, USA)

The Magnetosphere Aurora Reconnection Boundary Layer Explorer (MARBLE) project is a planned implementation of collisionless Hall MHD (or kinetic Hall MHD) for planetary magnetospheres that self-consistently models the propagation of electrons and field-aligned currents from magnetic reconnection sites to the ionosphere.

As a first step for this project, we created the AGATE simulation code (Bard & Dorelli, 2026), a Python-based framework developed primarily for solving MHD equations while maintaining adaptability to other equation sets. The code employs a modular, object-oriented architecture that separates interface specifications from numerical implementations, allowing users to customize numerical methods and physics models. Built on a Godunov-type finite-volume scheme, AGATE currently supports the ideal, Hall, and Chew-Goldberger-Low MHD equations, with multiple acceleration options ranging from NumPy to GPU-enabled computation via NVIDIA CUDA. AGATE is available open source (https://git.smce.nasa.gov/marble/agate-open-source/) under the NASA Open Source Agreement (NOSA, https://spdx.org/licenses/NASA-1.3.html).

Open Science Perspectives

Recent open-source requirements for all NASA-funded scientific software represent a significant shift from the closed-development ecosystem traditionally seen in heliophysics modeling. For the MARBLE project, this presents a unique opportunity to develop a large-scale, production code right from the outset with open science principles at its core.

AGATE aims to serve as a foundation for developing a community-driven, open-source global magnetosphere code. To be effective, such a code must be robust, user-friendly, comprehensible, and integrate smoothly with scientists' existing research methods. We anticipate that this open-source community approach will function as an educational tool, an experimental platform, and a mechanism for incorporating advanced capabilities from proprietary global codes.

To achieve this vision, we chose to write AGATE in Python. Python enables faster development cycles and works seamlessly with open-source community libraries like Kamodo, PlasmaPy, SpacePy, and other PyHC ecosystem tools. Although Python isn't inherently as fast as compiled languages, AGATE compensates by incorporating performance-enhancing technologies such as Numba and Cupy to boost computational speed. The Python framework also supports a modular design philosophy, making it easier to incorporate, replace, or develop new features, thereby creating a versatile resource for the scientific community.

AGATE is not intended to replace any existing codes. Having multiple independent codes is essential for validating results and techniques. However, we believe an open science modeling community should prioritize continuous improvement of our collective capabilities. This requires making cutting-edge features implemented in individual codes universally accessible, whether through complete public release or through user interfaces like CCMC ROR. AGATE offers a promising third approach: public modules to a community code that implements new features as open-source components while allowing proprietary code bases to remain private. Though this approach might seem to contradict open science principles, it provides a practical solution to an ongoing tension between grant funding, which often improves only one codebase, and purely open development, which frequently lacks sufficient funding for code developers. This hybrid model creates a pathway for broader scientific advancement while respecting existing development constraints.

Finally, the primary challenge for AGATE has been navigating the complex open-source release process at NASA, especially understanding how the NOSA license impacts code development and collaboration with the community. It is very unclear how NOSA fits with a modern open-source project approach, including community pull requests, forking, and combining multiple code bases. Currently, AGATE is unable to accept community pull requests in the same manner as a typical public git repository, e.g., Sunpy. These licensing issues must be clarified to create a truly open science and open-source modeling community between universities, government agencies, corporations, and individual researchers.

A.17 MAS/CORHEL/CORHEL-CME (Jon Linker, Ron Caplan: Predictive Science, Inc., USA)

CORona-HELiosphere (CORHEL, Riley et al., 2012) is a suite of coupled models and tools for describing the solar corona and solar wind for specific events and time periods, first developed at Science Applications International Corporation (SAIC) and later at Predictive Science Incorporated (PSI, https://predsci.com). At the heart of CORHEL is the Magnetohydrodynamic Algorithm outside a Sphere (MAS) code (https://predsci.com/mas/, Linker et al., 1999; Lionello et al., 2009; Mikić et al., 2018), which integrates the time-dependent resistive magnetohydrodynamic (MHD) equations in three-dimensional spherical coordinates, and has been used extensively in models of coronal structure (e.g., Mikić & Linker, 1996; Linker et al., 1999; Lionello et al., 2009; Downs et al., 2013; Mikić et al., 2018), coronal dynamics (e.g., Lionello et al., 2006; Linker et al., 2011), and coronal mass ejections (CMEs) out to the Earth (Linker et al., 2003; Lionello et al., 2013; Török et al., 2018). Recently, Downs et al. (2025) demonstrated a time-evolutionary model of the solar corona that ran in real time for 32 days. As part of the Solar Particle Event (SPE) Threat Assessment Tool (STAT), MAS has been coupled to the Energetic Particle Radiation Environment Module (EPREM, see Appendix A.7, Schwadron et al., 2010) in simulations of SPEs (Linker et al., 2019; Young et al., 2021).

Open Science Perspectives

MAS has a long history of development. Originally designed in the 1990s to run efficiently on vector supercomputers, it was almost completely rewritten for massively parallel computers in the 2000s. In the last several years, it has again undergone substantial restructuring for computation on multi-GPU systems (Caplan et al., 2019). Over its 30 plus year history, there have also been substantial algorithmic improvements and significant expansion of the physical model to include anisotropic thermal conduction, radiative losses, coronal heating, and solar wind acceleration, incorporating models of wave turbulence. The MAS code is written in modern Fortran and parallelized for CPUs and GPUs using MPI, OpenACC (https://www.openacc.org), and Fortran's `do concurrent` standard parallelism (Caplan et al., 2023). It solves the MHD equations on a non-uniform, logically rectangular staggered grid using finite differences. The non-uniformity of the grid helps MAS to efficiently resolve small-scale structures such as the transition region and active regions, while allowing for coarser grid points over the global scale. The code employs implicit, semi-implicit, and super-timestepping methods for the efficient computation of longer time evolution (Caplan et al., 2024). The code exhibits parallel strong scaling up to thousands of CPU cores (MPI ranks) on large HPC systems. A 3D decomposition is used to distribute the domain across MPI ranks in a topologically aware communicator. Inter-rank communication is performed asynchronously whenever possible.

The experience at PSI has been that the vast majority of external users would like access to PSI model results for time periods they are studying, but without needing to understand the details of preparation of input data, or setting up, installing, and running simulations. CORHEL was originally developed to facilitate Open Use, Open Validation, and Open Collaboration by allowing non-experts to run MAS for user-selected time periods, abstracting away many of the required data processing details (e.g., selection and processing of magnetic maps from observatories). Originally a series of Bash scripts that link together different codes (primarily Fortran), versions of CORHEL have been delivered to the CCMC (https://ccmc.gsfc.nasa.gov/models/CORHEL-MAS_WSA_ENLIL~5.0/, https://ccmc.gsfc.nasa.gov/models/CORHEL~MAS-TDM~6.0/) and Air Force Research Laboratory, with the expectation of scientific programmers/software engineers installing MAS/CORHEL and executing runs for a variety of users. A focus was creating a GUI that allows users to tailor specifics of data input and parameters. Several validation studies have been performed using CCMC runs on request (e.g., Gressl et al., 2014; Reginald et al., 2014; Jian et al., 2015, 2016). PSI also posts standard solar/heliospheric solutions on the PSI website for time periods dating back to the 1970s up until the present.

CORHEL-CME (Linker et al., 2024), recently made available at the CCMC (https://ccmc.gsfc.nasa.gov/models/CORHEL-CME~1/), has furthered this approach by providing a GUI that takes users through the steps of modeling CMEs in a realistic background for specific events or time periods. Utilizing CORHEL-CME, users can perform state-of-the-art models of pre-eruptive configurations and CME initiation in solar active regions and obtain Sun-to-Earth simulations with quick-look diagnostic images and movies provided after run completion. This effort has pioneered the use of NASA Amazon Web Services (AWS, https://aws.amazon.com/) for runs on request. The GUI is hosted and the runs are performed on AWS systems. Significant effort has gone into making runs executable in a couple of days on modest resources (a single 8-GPU system). A continuous integration/continuous deployment (CI/CD) paradigm has been employed. New versions of the interface are uploaded to a development node, where they are tested (including scanning for compliance with NASA security requirements) prior to implementation on the production node.

Recently, MAS was made open-source and is now freely available at https://github.com/predsci/mas. In addition to MAS, additional components of CORHEL are available on GitHub as part of the Solar Wind Generator (SWiG) package (Caplan & Stulajter, 2025). Further elements will be released as part of its successor, DYNAmically evolving Model of CMEs and SEPs (DYNAMCS).

A.18 PARADISE (Nicolas Wijsen, Stefaan Poedts: KU Leuven, Belgium)

The Particle Radiation Asset Directed at Interplanetary Space Exploration (PARADISE) models the transport and acceleration of solar energetic particles (SEPs) through the solar wind by solving the focused transport equation under realistic solar wind conditions. It integrates the equivalent set of Itô stochastic differential equations forward in time for many pseudo-particles, while assuming a solar wind configuration obtained from a magnetohydrodynamic model, like EUHFORIA (see Appendix A.8) or Icarus (see Appendix A.12). PARADISE can model the acceleration evolution of energetic particle distributions in corotating interaction region shocks or CME shocks in a dynamic solar wind containing a large-scale transient structure (Wijsen et al., 2019; Wijsen, 2020). PARADISE is written in C++. PARADISE has also been coupled to COCONUT (see Appendix A.3, Husidic et al., 2024a, 2024b).

PARADISE has been integrated into the VSWMC test server and will be available on the operational server soon.

A.19 RAM-SCB (Vania Jordanova: Los Alamos National Laboratory, USA)

The Ring current–Atmosphere interactions Model with Self-Consistent magnetic field (RAM-SCB) combines a kinetic model of ring current ions ($H^+$, $He^+$, $N^+$, and $O^+$) and electrons with a three-dimensional (3D) force-balanced model of the terrestrial magnetic field. The expansion of the code to include a self-consistent calculation of the electric field is underway. The model has grown from a research-grade code with limited options and static magnetic field (Jordanova et al., 1994) to a rich, highly configurable research and operations tool with a multitude of new physics and output products (e.g., Engel et al., 2019; Jordanova et al., 2023). A significant development of RAM-SCB was achieved during the SHIELDS project (Jordanova et al., 2018) when the models were modernized and made more user friendly and coupled to the SWMF (see Appendix A.24, Welling et al., 2018).

Open Science Perspectives

To facilitate the work in a large team, and to maintain and grow external collaborations worldwide, it was decided to make the model available as open source (Jordanova et al., 2025). Open-source tools to work with the model outputs are provided by Morley et al. (2024). RAM-SCB can be built from source following the instructions detailed in its user manual, but developers' help is usually required. To use RAM-SCB, understanding of the physics included in the model and familiarity with Unix-like environments are strongly recommended. The introduction of students enrolled in the Los Alamos Space Weather Summer School to RAM-SCB was proven to be highly efficient and beneficial to all, resulting in new code development, research studies, and peer-reviewed publications.

The model has been used extensively over the past >30 years to study various aspects of inner magnetosphere dynamics, including application to satellite surface-charging hazards (Yu et al., 2019). As shown by Morley et al. (2023), RAM-SCB is also available now to the scientific community through the CCMC ROR service (https://ccmc.gsfc.nasa.gov/models/RAM-SCB~v.2.2/). The development of RAM-SCB throughout the years was supported with NASA, NSF, and DOE research grants. More funding opportunities and support for model developers are needed to ensure further maintenance and progress.

A.20 SAMI3 (Joe Huba: Naval Research Laboratory, USA)

SAMI3 (Sami3 is Also a Model of the Ionosphere, Huba & Joyce, 2010) is a global physics-based 3D model of the Earth's ionosphere/plasmasphere system. It is based on SAMI2 (Sami2 is Another Model of the Ionosphere, Huba et al., 2000), a two-dimensional model of the ionosphere. SAMI3 models the plasma and chemical evolution of seven ion species (H+, He+, $N^+$, $O^+$, $N_2^+$, $NO^+$, and $O_2^+$) as well as metal ions ($Fe^+$, $Mg^+$). It uses a semi-implicit transport scheme for parallel dynamics and the partial donor cell method for E×B transport perpendicular to B, which reduces numerical diffusion. Ion inertia is included in the ion momentum equation for motion along the geomagnetic field. This is important in modeling the topside ionosphere where the plasma transitions from collisional to collisionless (Huba & Krall, 2013). A significant capability of SAMI3 is that it models the development of equatorial plasma bubbles, which impact communication and navigation systems, and plasmaspheric ducts when coupled to a first-principles thermosphere model, such as WACCM-X (see Appendix A.27) and HIAMCM (Huba et al., 2017).

Open Science Perspectives

The predecessor to SAMI3, SAMI2, has been open source since 2000 (Huba & Richardson, 2021; https://github.com/NRL-Plasma-Physics-Division/SAMI2) and has been downloaded by more than 1000 researchers and used in many ionospheric studies and Ph.D. theses. An advantage to SAMI2 is that it is relatively simple to use and can be modified easily for specific applications. One example is the development of the KIPM model, which is based on SAMI2, used to predict ionospheric conditions over Korea (Kim et al., 2022). Currently, SAMI3 is available upon request from J.D. Huba (jdhuba@gmail.com) or from the CCMC (https://ccmc.gsfc.nasa.gov/models/SAMI3~3.22/). Similar to SAMI2, SAMI3 has been made available to several researchers and has been used in several research studies. A recent example is the inclusion of data assimilation into SAMI3 to improve its forecasting capabilities (Ma et al., 2024). A minor issue with the release of SAMI2 and SAMI3 source code is that, in some cases, the codes have been modified and results published without the authors notifying the main developer (J.D. Huba) for feedback or comments.

### A.21 SEPMOD (Janet Luhmann: University of California Berkeley, USA)

SEPMOD is a physics-based model designed to produce time profiles of solar energetic particle (SEP) fluxes based on interplanetary shock and observer magnetic connectivity information derived from WSA-ENLIL-Cone CME event simulations (see Appendix A.28 and A.6, respectively). It is currently available for use at the CCMC (https://ccmc.gsfc.nasa.gov/models/SEPMOD~v2.20221228/). It currently lacks information inside ENLIL's inner boundary at 20 solar radii, making its results most applicable for extended SEP time profiles rather than rapid onset timing. It is also a contributor to the NASA SEP Intensity Scoreboard, where it is set up to automatically run using near-real-time WSA-ENLIL-Cone model results.

### A.22 SPS (Oleksandr Koshkarov, Oleksandr Chapurin, Vadim Roytershteyn, Gian Luca Delzanno: Los Alamos National Laboratory)

The Spectral Plasma Solver (SPS) solves the kinetic Vlasov-Maxwell equations. It uses a spectral expansion of the distribution function to obtain a set of three-dimensional partial differential equations for the coefficients of the expansion, akin to the well-known moment expansion. The uniqueness of the approach is that, with suitable basis functions, the low-order coefficients of the expansion provide a fluid description of the plasma (with a particular closure), while kinetic physics can be retained by adding more terms to the expansion. We refer to this property as fluid-kinetic coupling. Thus, by adjusting the number of spectral coefficients locally in space and time, one can optimize the number of degrees of freedom and perform large-scale modeling that accurately accounts for the feedback of microscopic processes.

The standard version of SPS is based on asymmetrically-weighted Hermite basis functions in velocity space, since these functions satisfy fluid-kinetic coupling (Delzanno, 2015; Vencels et al., 2016; Koshkarov et al., 2021). SPS has been in development for the past 15 years. A lot of SPS method formulation and algorithmic development was performed on the 1D1V (one spatial dimension, one velocity dimension) Vlasov-Poisson model, including adjusting the number of spectral moments in time (Vencels et al., 2015), comparisons between the spectral approach and particle-in-cell techniques (Camporeale et al., 2016), changing the shift and scaling of the spectral basis function in time (Pagliantini et al., 2023a), developing energy-conserving time integrators (Pagliantini et al., 2023b), combining the spectral approach with a particle-in-cell approach (Chapurin et al., 2024) and studying the effect of different collisional operators (Issan et al., 2025). The latter is important to mitigate filamentation in velocity space and helps with numerical stability. Note that, although not yet available in a 3D3V production code, Legendre polynomials also satisfy the fluid-kinetic coupling property and they have been successfully applied to the 1D1V Vlasov-Poisson equations (Manzini et al., 2016).

The standard SPS approach for the 3D3V Vlasov-Maxwell equations uses a discontinuous-Galerkin discretization on uniform meshes in physical space (Koshkarov et al., 2021), while a version based on Fourier spectral discretization of physical space has also been developed (Roytershteyn & Delzanno, 2018). The former is our preferred and most widely used approach as it maintains data locality and can therefore be massively parallelized. A variety of high-order explicit and implicit time integrators are available. SPS is based on the Portable, Extensible Toolkit for Scientific Computation (PETSc, https://petsc.org), which naturally handles domain decomposition and MPI parallelism. SPS is written in C with ~15,000 lines of code and uses HDF5 or binary files for output. We often run SPS with ~103 spectral terms in velocity space. These runs are computationally expensive, easily requiring many millions of core-hours, large memory and TBs of storage space. SPS example applications include studies of the solar wind turbulence cascade (Roytershteyn et al., 2019) and radiation from pulsed electron beams (Delzanno & Roytershteyn, 2019).

Open Science Perspectives

SPS follows traditional collaborative coding practices. The code is hosted on the Los Alamos National Laboratory internal repositories for version control, where unit testing, continuous integration, and containerization can be applied. However, being developed by only a handful of individuals and typically driven by applications, one ongoing challenge has been to maintain adequate documentation. Another challenge, particularly in the early stages of the project, was the lack of software engineers in the development team.

Open science practices are normally followed by (1) documenting the input parameters necessary to replicate a given problem and sharing the SPS output, typically in association with publications in the open literature; (2) sharing the simulation outputs either upon request or through specific collaborative projects.

Given its kinetic nature, SPS can require significant compute, memory, and storage resources. As these resources might not be available everywhere, or not at the level required to tackle the most difficult problems, an open use paradigm would be very valuable. For example, simulation outputs could be shared and stored at CCMC, which could also provide on-demand analysis and visualization services. National laboratories could also be important resources, as typically they are at the forefront of high-performance computing and have significant computational resources.

SPS is currently not open source. We share the belief that the open source paradigm, particularly in a broader collaborative framework, offers distinct advantages for developing more robust, accessible codes that can be widely adopted. However, the development of high-quality, validated scientific software must also be supported by adequate, stable and long-term funding, essentially to provide a stable *home* for model developers. At the moment, such dedicated support is lacking and model development is often funded only intermittently, in conjunction with broader science investigations.

A.23 SWASTi (Prateek Mayank: Indian Institute of Technology Indore, India, now at UCAR, USA)

The Space Weather Adaptive Simulation framework (SWASTi; Mayank et al., 2022, 2023, 2024) is a 3D physics-based framework for solar wind and CME modeling. It couples a semi-empirical coronal domain (PFSS with optional Schatten current sheet, delivering WSA-type speed maps from synoptic magnetograms) with an inner-heliosphere MHD model (PLUTO, https://plutocode.ph.unito.it/) to evolve plasma from 0.1 to 2.1 AU (the corona spans 1 solar radius–0.1 AU). The system forecasts background solar wind structure and has been validated against L1 observations over several Carrington rotations. The CME module includes (i) a non-magnetized elliptic cone and (ii) a magnetized flux rope (FRi3D) inserted at 0.1 AU. SWASTi is a research/pre-operational framework under active development.

Open Science Perspectives

To build on these efforts, the community-facing infrastructure is being expanded to streamline adoption and collaboration. Interactive Jupyter tutorials and hands-on workshop workflows have already seeded initial contributions. SWASTi is now being packaged into Docker containers that bundle all dependencies for turnkey local installation, and a continuous-deployment pipeline is under development to automate end-to-end testing, building, and release. Next steps include deployment on the CCMC ROR platform and launch of a public website offering near-real-time forecasts and runs-on-request capabilities. In parallel, integration of GPU-accelerated PLUTO kernels is underway to reduce full-domain MHD runtimes.

A.24 SWMF (Gabor Toth: University of Michigan, USA)

The Space Weather Modeling Framework (SWMF, https://clasp.engin.umich.edu/research/theory-computational-methods/space-weather-modeling-framework/, Toth et al., 2005, 2012; Gombosi et al., 2021) comprises several models that can simulate space physics phenomena from the chromosphere of the Sun to the upper atmosphere of Earth as well as other planets, moons, and comets. These models include BATS-R-US for the solar corona, inner heliosphere, outer heliosphere, and magnetosphere; RCM, CIMI, RAM-SCB (see Appendix A.19), and HEIDI for the inner magnetosphere (De Zeeuw et al., 2004); RIM for ionosphere electrodynamics (Ridley et al., 2004); PWOM for polar wind outflow (Glocer et al., 2009); FLEKS for particle-in-cell modeling and neutral particles; and SOFIE/M-FLAMPA for solar energetic particles (Zhao et al., 2024; Liu et al., 2025). The SWMF can also model the interaction of the solar wind with the local interstellar medium forming the outer heliosphere.

The SWMF has more than a million lines of source code, about 70% in Fortran 90, the rest in C++. It uses the MPI, OpenMP, and OpenACC libraries for parallel execution. The SWMF can run on a single core of a laptop as well as on tens of thousands of cores of a supercomputer and/or on hundreds of GPUs. The full source code has been available for registered users since the first release of the SWMF in 2005. In 2020, most of the SWMF source code was released under a non-commercial open-source license, and in 2024 the full SWMF became open-source (https://github.com/SWMFsoftware) under the Apache 2 license. Various configurations of the SWMF have been available through the CCMC ROR service for more than two decades (https://ccmc.gsfc.nasa.gov/models/SWMF~20140611/, https://ccmc.gsfc.nasa.gov/models/SWMF~2023/, https://ccmc.gsfc.nasa.gov/models/SWMF~AWSoM_R~1.0/). The geospace configuration of the SWMF has been running operationally at the Space Weather Prediction Center (SWPC) of the National Oceanic and Atmospheric Administration (NOAA) since 2016 and was statistically compared to other geospace simulation results at the CCMC in Ridley et al., 2016.

Open Science Perspectives

The philosophy behind making the SWMF publicly available is that this brings much more benefit than the occasional negative impacts. Exposing the source code to a broad community makes the developers work hard to create a reliable, well-documented and well-written code base. The SWMF is tested nightly on several computer platforms, including CPUs and GPUs, with different compilers. The reliability of the code is achieved by maintaining a high level of success rate for hundreds of unit and functionality tests. The nightly test results have been available and archived online since 2009 (https://csem.engin.umich.edu/SWMFTESTS/). The SWMF documentation is created together with and checked against the nightly tests. Maintaining the SWMF requires constant monitoring of the tests, fixing issues, and updating documentation. The large variety of applications and the user base motivate a lot of requests for new features that require code development. Neither the maintenance nor the code development are funded by grants in general, with a few exceptions. The SWMF is maintained and developed as a side activity for science-focused NASA and NSF projects. The biennial SWMF user meetings hosted at the University of Michigan have provided an opportunity for users and developers to interact with each other since 2014.

A.25 TIE-GCM (Wenbin Wang: UCAR, USA)

The NSF NCAR Thermosphere Ionosphere Electrodynamics – General Circulation Model (TIE-GCM) is a three-dimensional model of the coupled thermosphere-ionosphere system that self-consistently solves the neutral continuity, momentum, and energy equations, the ion continuity equation, the electron and ion energy equation, and the neutral wind dynamo (Roble et al., 1988; Richmond et al., 1992; Qian et al., 2014). TIE-GCM is currently the thermosphere-ionosphere component model of a fully coupled whole geospace model, MAGE (see Appendix A.15). The recently released TIE-GCM 3.0 can run at multiple resolution in longitude and latitude (5°, 2.5°, 1.25°, and 0.625°) when an azimuthal averaging–reconstruction technique has been introduced to increase the model horizontal resolution (Dang et al., 2021). The model upper boundary is also extended upward to ~700-800 km in solar minimum and greater than 1000 km in solar maximum to allow the model to simulate the dynamic changes of neutral mass density and drag for low Earth orbit satellites, with helium added as a major species (Sutton et al., 2015; Cai et al., 2022), and to investigate ionospheric interhemispheric transport at low latitudes. For high-latitude magnetospheric inputs, the model uses the empirical Weimer convection pattern (Weimer, 2005) driven with solar wind and interplanetary magnetic field data, or Heelis convection pattern specified with the Kp index (Heelis et al., 1982), but can also use results from the Assimilative Mapping of Ionosphere Electrodynamics (AMIE, Richmond et al., 1998) procedure or from coupling with magnetospheric MHD models as in MAGE (Pham et al., 2022; Lin et al., 2021, 2022a). The lower boundary condition (atmospheric tides) is given by the Global Scale Wave Model (GSWM, Hagan et al., 2001), or by nudging with external data such as those from the WACCM-X (see Appendix A.27), to facilitate studies of wave coupling across different layers of the atmosphere. TIE-GCM 3.0 also includes parameterization schemes for some specific scientific processes, such as subauroral polarization streams (Wang et al., 2012), electrojet turbulent heating (Liu et al., 2016), and field-aligned ion drag in the neutral momentum equation (Lei et al., 2012), as well as solar flares and eclipses. The model can also output diagnostic analysis terms of the equations solved to facilitate understanding of the physical processes that are responsible for the modeled thermosphere-ionosphere structures and their variability.

Open Science Perspectives

An early version of the TIE-GCM (2.0) was released to the public more than a decade ago (https://www.hao.ucar.edu/modeling/tgcm/). The model has been frequently downloaded and widely used in scientific research. With the wide community usage for both space science research and space weather application, the model has been debugged, improved, validated, and it has contributed to science advances: all of this would not have been possible without the public release of model source code. The new TIE-GCM 3.0 version is publicly available and updated at https://github.com/NCAR/tiegcm with detailed documentation of the necessary input parameters for the model run. A new scripting system has also been developed to guide the users in setting up the compiling and running environment on their platform. Quick look and post-processing graphic tools are also available for users to diagnose model outputs in the repository. TIE-GCM is also available at the CCMC (https://ccmc.gsfc.nasa.gov/models/TIE-GCM~3/).

A.26 Vlasiator (Minna Palmroth: University of Helsinki, Finland)

The Vlasiator code (https://www.helsinki.fi/en/researchgroups/vlasiator, Palmroth et al., 2018; Ganse et al., 2023) simulates the global near-Earth space as driven by the solar wind and bounded by the electrostatic solution of the ionosphere (Ganse et al., 2025). Every grid cell in the 3D position space (r-space) includes a 3D velocity space, which contains the ion velocity distribution functions. Vlasiator describes ion-kinetic physics self-consistently without numerical noise. The velocity distribution functions are used to calculate moments (e.g., density, temperature, and pressure), which are then used to update the electromagnetic field in the r-space. Electrons are a massless charge-neutralizing fluid described using Ohm's law, which includes the Hall and electron pressure terms, and which closes the set of equations in r-space. Vlasiator’s inner boundary is currently at 4.7 Earth radii. The r-space simulation box spans approximately a million kilometers in each dimension, with dynamically adaptive resolution (Palmroth et al., 2023, Kotipalo et al., 2024). Vlasiator is fully validated and reproduces satellite measurements in detail (Palmroth et al., 2018, Turc et al., 2023, Palmroth et al., 2023).

Vlasiator performance is continually improved and optimized to run on the world’s largest supercomputers and to fully use heterogeneous computing, including GPUs (Palmroth, 2022, Papadakis et al., 2024, Battarbee et al., 2025).

A.27 WACCM-X (Hanli Liu: UCAR, USA)

The Whole Atmosphere Community Climate Model with thermosphere/ionosphere extension (WACCM-X) is part of the NCAR Community Earth System Model (CESM, Danabasoglu et al., 2020). It aims to capture the coupling of different atmosphere regions through dynamical, radiative transfer, photochemical, and electrodynamical processes (Liu et al., 2018). Examples include effects of atmospheric waves from the weather system on the thermospheric circulation and composition, and on ionospheric irregularities. A species-dependent spectral-element method has recently been adapted to replace the finite-volume method to solve atmospheric dynamics. For radiative transfer, the Rapid Radiative Transfer Model for General circulation models (RRTMG) is used from the troposphere to the lower mesosphere, while a nonlocal thermal equilibrium calculation is used at higher altitudes (Gettelman et al., 2019). A versatile and robust solver, MOZART, is used for chemistry modeling (Emmons et al., 2020). The ionospheric electrodynamics, transport, and energetics solvers are adapted from TIE-GCM. Currently, WACCM-X is not used as an operational model.

Open Science Perspectives

CESM and its predecessor Community Climate System Model have been open-source community models since its initial development to support the broad research community, for example by comparative analysis and interpretation of observations, exploring atmospheric coupling processes, hypothesis testing, and mission planning. As one of the atmosphere components, WACCM-X has also been open source. WACCM-X v1 was publicly released in 2010 as part of CESM1 (Liu et al., 2010), and WACCM v2 was released in 2018 as part of CESM2 (Liu et al., 2018). The upcoming release of CESM3 is currently scheduled for the Summer of 2026. Moreover, WACCM-X model output histories have also been made publicly available, including WACCM-X Specified Dynamics runs and high-resolution simulations. WACCM-X has been available for ROR at the CCMC since 2022 (https://ccmc.gsfc.nasa.gov/models/WACCMX~2.2/).

A.28 WSA (Charles Nick Arge: NASA GSFC, USA)

The Wang-Sheeley-Arge (WSA) model (Arge & Pizzo, 2000; Arge et al., 2003, 2004, 2024; McGregor et al., 2008; Wallace et al., 2019) is a combined empirical and physics-based model of the corona and solar wind. Global photospheric magnetic maps are used as input to the coronal portion of WSA, which comprises two potential-field-type models. The inner one is a traditional magnetostatic potential field source surface (PFSS) model (Schatten et al., 1969; Altschuler & Newkirk, 1969; Wang & Sheeley, 1992), which solves Laplace's equation using a spherical harmonic expansion to determine the coronal field out to 2.51 solar radii (Rs). The output of the PFSS model serves as input to the outer coronal model, the Schatten Current Sheet model (Schatten, 1971), which provides a more realistic magnetic field topology of the upper corona. An empirical relationship between the magnetic field expansion factor and the solar wind speed (e.g., Arge et al., 2003, 2004; Samara et al., 2024) is then used to assign the speed at this outer boundary. The model provides the radial magnetic field and solar wind speed at the outer coronal boundary surface, which is then fed into a simple, quick running 1D kinematic solar wind model (Arge & Pizzo, 2000; Arge et al., 2004) or advanced 3D MHD solar-wind propagation models such as MS-FLUKSS (Kim et al., 2020; Singh et al., 2020; Manoharan et al., 2015), LFM-Helio (Pahud et al., 2012; Merkin et al., 2016), Gamera (Zhang et al., 2019), and ENLIL (see Appendix A.6, Odstrcil et al., 2005; Lee et. al, 2013, 2015; McGregor et al., 2011). Densities and temperatures, which are not provided by WSA and required by MHD models, are deduced by assuming, for example, mass flux conservation and pressure balance (e.g., Kim et al., 2020; Merkin et al., 2016).

WSA has been running at the CCMC since 2005, both in real-time and as Runs-on-Request (https://ccmc.gsfc.nasa.gov/models/WSA~5.4/, https://ccmc.gsfc.nasa.gov/models/WSA~6/). The WSA Dashboard is also available at the CCMC (https://ccmc.gsfc.nasa.gov/wsa-dashboard/).

**Appendix B: S3 Storage and Scientific Computing**

Amazon Simple Storage Service (S3, https://aws.amazon.com/pm/serv-s3/), is a highly scalable and durable object storage service, fundamentally different from traditional filesystems such as hard drives or cluster storage. It operates over web protocols (HTTP) and is not based on the familiar POSIX system of files and directories.

This unique architecture gives S3 compelling features such as virtually unlimited capacity, limited only by cost, and extreme redundancy, but it also means it is not a low-latency service. S3 is ill-suited for tightly coupled High-Performance Computing (HPC) tasks, such as physical simulations that rely on MPI and require millisecond access to shared storage. S3 excels in loosely coupled workflows where tasks are independent, such as in many ML/AI training jobs. The recommended architecture for HPC on AWS is a hybrid approach: use a high-performance, POSIX-compliant filesystem (like Amazon FSx for Lustre) for the actual computation, and use S3 for long-term storage, staging data to and from the filesystem at the beginning and end of the job. Ultimately, while S3's capabilities are improving, its object-store nature means that tightly coupled scientific software must be significantly re-architected to use it directly for computation.

Table B1 summarizes the CCMC storage infrastructure components, comparing cost, latency, usage limitations, and use cases.

**Table B1.** *CCMC Storage Infrastructure Components, Comparisons, and Best Use.*

| **Infrastructure component** | **Annual cost for 1 Pb** | **Latency (time to first byte)** | **Use case and other comments** |
|---|---|---|---|
| AWS Cloud **S3** (**object** storage)<br><br>Current usage: 180 TB | $260k | 50 – 250 ms | Use case: temporary storage before transitioning to EFS, EBS, or on-premises.<br><br>Files must be downloaded to local or network storage before they can be used for computations, visualization, and analysis.<br><br>Complex storage management and security challenges.<br><br>Migration from traditional file systems requires significant software rewriting. |
| AWS Cloud **EFS** (**network** file system)<br><br>Current usage: 8 TB | $3,600k | 20 ms | Use case: shared collaborative environments.<br><br>Can be shared between instances. |
| AWS Cloud **EBS** (**local** storage)<br><br>Current usage: 56 TB | $960k | 2.5 ms (fastest) | Use case: local hard drives for each computing instance.<br><br>Cannot be used with another server at the same time, cannot be shared between compute instances. |
| On-premises Dell **PowerScale**<br><br>Current capacity: 2.5 PB | $150k | 5 ms | Use case: most cost-effective solution for high-performance large long-term storage, web visualization, and analysis. |
| On-premises **BlackPearl** modern tape archive<br><br>Current capacity: 4.5 PB | $25k | Retrieval and copy to high performance storage: minutes to hours | Use case: long-term preservation cost-effective solution for less frequently used run outputs. |

| Infrastructure component | Annual cost for 1 Pb | Latency (time to first byte) | Use case and other comments |
|---|---|---|---|
| | | | Not suitable for computations, visualization, and/or web applications that require direct access. |

Latency in Table B1 does not include the impact of network bandwidth for files transferred between locations. Download speed measurement can be influenced by many factors, many outside of anyone's control (network congestion or link failures, operating system load). To illustrate the impact of remote access on download speed, we performed an experiment with retrieving the same data files (3 files of identical size, 3.6 GB) from different locations to a target AWS instance located in Virginia using a network card capable of up to 20 Gbps network bandwidth. Results are shown in Table B2.

**Table B2.** *Results of Data Retrieving Experiment from Different Locations to AWS Instance in Virginia.*

| | CCMC website via server-side file query | Direct HTTPS from CCMC PowerScale filesystem | S3 bucket in Virginia | S3 bucket in the West coast |
|---|---|---|---|---|
| **Bandwidth (average)** | 9 MB/s | 174 MB/s | 278 MB/s | 84 MB/s |
| **Wait time per file** | 7 min | 21 sec | 13 sec | Wide variation: slowest was 7900 seconds (132 minutes) |
| **Note** | File query can access files located on various file systems | This method bypasses the server-side file query | This is an S3 bucket in the *same region* as the target | This highlights the impact of network congestion and unpredictability |

**Appendix C: Acronyms and Definitions**

Acronyms are grouped by category: Computing (both hardware and software), General, Instruments (including spacecraft and ground-based observational infrastructures), Models (including supporting tools and services), and Organizations.

### C.1 Computing

**AI**: Artificial Intelligence

**API**: Application Programming Interface

**APU**: Accelerated Processing Unit

**AWS**: Amazon Web Services

**BP5**: Binary Pack, version 5 (file format for the ADaptable Input/Output System, ADIOS, framework)

**BSD**: Berkeley Software Distribution

**CD**: Continuous Deployment

**CDF**: Common Data Format (NASA file format for heliophysics data)

**CF**: Climate and Forecast metadata standard

**CI**: Continuous Integration

**CPU**: Central Processing Unit

**CUDA**: NVIDIA Compute Unified Device Architecture

**DOI**: Digital Object Identifier

**EBS**: Amazon Elastic Block Store

**EFS**: Amazon Elastic File System

**FAIR**: Findable, Accessible, Interoperable, Reusable

**FITS**: Flexible Image Transport System (file format)

**FSx**: Amazon file system manager

**GPL**: General Public License

**GPU**: Graphics Processing Unit

**GUI**: Graphical User Interface

**HAPI**: Heliophysics data Application Programmer's Interface

**HDF**: Hierarchical Data Format (file format; commonly used versions are 4, HDF4, and 5, HDF5)

**HIP**: Heterogeneous-computing Interface for Portability

**HPC**: High-Performance Computing/Computer

**HTTP(S)**: Hyper Text Transfer Protocol (Secure)

**I-ADOPT**: InteroperAble Descriptions of Observable Property Terminology (metadata standard)

**ID**: Identifier

**IDL**: Interactive Data Language

**IP**: Internet Protocol

**JSON**: JavaScript Object Notation (file format)

**ML**: Machine Learning

**MPI**: Message Passing Interface (communication protocol for parallel computing)

**NetCDF**: Network Common Data Form (file format)

**NOSA**: NASA Open Source Agreement software license

**NPZ**: NumPy Zip (file format)

**OpenACC**: Open ACCelerator (API for parallel computing)

**OpenMP**: Open Multi-Processing (API for parallel computing)

**POSIX**: Portable Operating System Interface

**RAM**: Random Access Memory

**REST**: REpresentational State Transfer (software architectural style)

**S3**: Amazon Simple Storage Service

**SPASE**: Space Physics Archive Search and Extract (metadata standard for accessing heliophysics resources)

**TAP**: Table Access Protocol (IVOA communication protocol for accessing astrophysics and heliophysics data)

**URL**: Uniform Resource Locator

**UWS**: Universal Worker Service (IVOA API for managing asynchronous execution)

**VOUnits**: Virtual Observatory Units format (IVOA metadata standard)

**VOTable**: Virtual Observatory Table format (IVOA data exchange format)

**VTK**: Visualization ToolKit

**WCS**: FITS World Coordinate System (metadata standard)

### C.2 General

**AE**: Aurora Electrojet index (quantification of auroral zone magnetic activity)

**CME**: Coronal Mass Ejection

**EUV**: Extreme UltraViolet

**F10.7**, **F30**: 10.7 and 30 cm solar radio Flux indices

**GPI**: GeoPhysical Indices

**Hp60**: Hourly planetary open-ended geomagnetic index

**IMF**: Interplanetary Magnetic Field

**IPS**: InterPlanetary Scintillation

**Kp**: planetary K-index (quantification of disturbances in the horizontal component of the Earth's magnetic field)

**MHD**: MagnetoHydroDynamic

**MSIS**: Mass Spectrometer and Incoherent Scatter radar data

**R2O2R:** Research-to-Operations-to-Research

**ROI**: Region Of Interest

**SEP**: Solar Energetic Particle

**SPE**: Solar Particle Event

**UV**: UltraViolet

### C.3 Instruments/Missions

**CHAMP**: CHAllenging Minisatellite Project (spacecraft measuring Earth's gravitational potential and magnetic field)

**COSMIC-2**: Constellation Observing System for Meteorology, Ionosphere, and Climate (second mission, also known as FORMOSAT-7)

**COSTEP**: COmprehensive SupraThermal and Energetic Particle (a detector on board the Solar and Heliospheric Observatory, SOHO, spacecraft, in orbit around the Earth-Sun Lagrangian point 1, L1)

**CRaTER**: Cosmic Ray Telescope for the Effects of Radiation (a detector on board the Lunar Reconnaissance Orbiter, LRO, spacecraft, in orbit around the Moon)

**EZIE**: Electrojet Zeeman Imaging Explorer (constellation of three CubeSats observing the aurora electroject)

**FORMOSAT-7**: FORMOsa SATellite (see COSMIC-2)

**GOCE**: Gravity Field and Steady-State Ocean Circulation Explorer (spacecraft mapping Earth's gravity field)

**GOES**: Geostationary Operational Environmental Satellite (series of satellites monitoring terrestrial and space weather)

**GRACE**: Gravity Recovery and Climate Experiment (a pair of spacecraft measuring Earth's gravity field anomalies)

**GRACE-FO**: GRACE Follow-On (successor of GRACE)

**MAGED**: MAGnetospheric Electron Detector (onboard GOES-13 and successors)

**MAVEN**: Mars Atmosphere and Volatile Evolution (spacecraft in orbit around Mars)

**STEREO**: Solar TErrestrial RElations Observatory (pair of spacecraft in orbit around the Sun at 1 AU, observing the Sun and measuring solar wind plasma, radio, and energetic particles)

### C.4 Models

**ADAPT**: Air Force Data Assimilative Photospheric flux Transport

**AGATE**: Accelerated Grid-Agnostic TEchniques (backronym)

**AMIE**: Assimilative Mapping of Ionosphere Electrodynamics

**AMPS**: Adaptive Mesh Particle Simulator

**ARMS**: Adaptively Refined Magnetohydrodynamic Solver

**BATS-R-US**: Block Adaptive Tree Solar-wind Roe Upwind Scheme

**CAMEL**: CCMC Comprehensive Assessment of Models using Library tools

**CAMELS**: Cosmology and Astrophysics with MachinE Learning Simulations

**CAM-NET**: Compressible Atmospheric Model-NETwork

**CESM**: Community Earth System Model

**CIMI**: Comprehensive Inner-Magnetosphere Ionosphere

**COCONUT**: COolfluid COroNa UnsTructured

**COOLFluiD**: Computational Object-Oriented Libraries for Fluid Dynamics

**CORHEL**: CORona HELiosphere model

**CORHEL-CME**: CORHEL with Coronal Mass Ejection

**DTM**: Drag Temperature Model

**DYNAMCS**: DYNAmically evolving Model of CMEs and SEPs

**E-CHAIM**: Empirical Canadian High Arctic Ionospheric Model

**ECSIM**: Energy Conserving Semi-Implicit Method

**EMAPS**: Event-driven Multi-Agent Planning System

**EMMREM**: Earth-Moon-Mars Radiation Environment Module

**ENLIL**: ancient Sumerian god of wind, air, earth, and storms

**EPREM**: Energetic Particle Radiation Environment Module

**EPiGRAM**: Exascale ProGRAmming Model

**EUHFORIA**: EUropean Heliospheric FORecasting Information Asset

**FISM**: Flare Irradiance Spectral Model

**FLAMINGO**: Full-hydro Large-scale structure simulations with All-sky Mapping for the Interpretation of Next Generation Observations

**FLEKS**: FLexible Exascale Kinetic Simulator

**FRi3D**: Flux Rope in 3D (CME magnetic topology model)

**GAMERA**: Grid Agnostic MHD with Extended Research Applications

**GSWM**: Global Scale Wave Model

**GITM**: Global Ionosphere Thermosphere Model

**HEIDI**: Hot Electron Ion Drift Integrator

**HIAMCM**: HIgh Altitude Mechanistic general Circulation Model

**HIDRA**: High-latitude Ionosphere Dynamics for Research Applications

**HYPERS**: HYbrid Particle Event-Resolved Simulator

**IGRF**: International Geomagnetic Reference Field

**IMPTAM**: Inner Magnetosphere Particle Transport and Acceleration Model

**iPIC3D**: implicit Particle-In-Cell in 3D model

**IPWM**: Ionosphere/generalized Polar Wind Model

**IRI**: International Reference Ionosphere

**ITMAP**: CCMC Ionosphere-Thermosphere Model Assessment and validation Platform

**KIPM**: Korean Ionosphere physics-based Prediction Model

**LaRe3D**: Lagrangian–Eulerian Remap code for 3D MHD

**LFM**: Lyon-Fedder-Mobarry global MHD code

**MAGE**: Multiscale-Atmosphere Geospace Environment

**MARBLE**: Magnetosphere Aurora Reconnection Boundary Layer Explorer

**MAS**: Magnetohydrodynamic Algorithm outside a Sphere

**METplus**: Model Evaluation Tools verification system

**M-FLAMPA**: Multiple Field-Line-Advection Model for Particle Acceleration

**MOZART**: Model for OZone and Related chemical Tracers

**MPI-AMRVAC**: MPI Adaptive Mesh Refinement Versatile Advection Code

**MS-FLUKSS**: Multi-Scale FLUid-Kinetic Simulation Suite

**OpenGGCM**: Open Geospace General Circulation Model

**PARADISE**: Particle Radiation Asset Directed at Interplanetary Space Exploration

**PFSS**: Potential-Field Source-Surface model

**PIC**: Particle-In-Cell method

**PREDICCS**: Predictions of radiation from REleASE, EMMREM, and Data Incorporating CRaTER, COSTEP, and other SEP measurements

**PWOM**: Polar Wind Outflow Model

**PyCAT**: Python/javascript CME Analysis Tool

**RAM-SCB**: Ring current–Atmosphere interactions Model with Self-Consistent magnetic (B) field

**RCM**: Rice Convection Model

**REMIX**: Redeveloped Magnetosphere-Ionosphere Coupler/Solver

**REleASE**: Relativistic Electron Alert System for Exploration

**RelSIM**: Relativistic Semi-Implicit Method

**RIM**: Regional Ionosphere Model

**ROR**: CCMC Runs-on-Request simulation service

**RRTMG**: Rapid Radiative Transfer Model for General circulation models

**SAMI2**: Sami2 is Another Model of the Ionosphere

**SAMI3**: Sami3 is Also a Model of the Ionosphere

**SEPMOD**: Solar Energetic Particle MODel

**SHIELDS**: Space Hazards Induced near Earth by Large Dynamic Storms

**SOFIE**: SOlar wind with FIeld lines and Energetic particles

**SPS**: Spectral Plasma Solver

**STAT**: SPE Threat Assessment Tool

**SWASTi**: Space Weather Adaptive Simulation framework

**SWMF**: Space Weather Modeling Framework

**SWiG**: Solar Wind Generator

**TIE-GCM**: Thermosphere-Ionosphere-Electrodynamics Global Circulation Model

**TIE-GCM ROPE**: TIE-GCM Reduced-Order Probabilistic Emulator

**WACCM**: Whole Atmosphere Community Climate Model

**WACCM-X**: WACCM with thermosphere/ionosphere eXtension

**WSA**: Wang-Sheeley-Arge (solar wind model)

### C.5 Organizations/Projects/Initiatives/Movements

**CCMC**: Community Coordinated Modeling Center

**CEDAR**: Coupling, Energetics, and Dynamics of Atmospheric Regions (NSF program)

**CGS**: Center for Geospace Storms (NASA DRIVE center)

**COSPAR**: COmmittee on SPAce Research

**DASH**: Data, Analysis, and Software in Heliophysics

**DOE**: U.S. Department of Energy

**DRIVE**: NASA Diversify, Realize, Integrate, Venture, Educate centers

**EO**: ESA Earth Observation directorate

**ESA**: European Space Agency

**GEM**: Geospace Environment Modeling (NSF program)

**HECC**: NASA High-End Computing Capability project

**HOME**: Heliophysics Open Modeling Environment

**IHDEA**: International Heliophysics Data Environment Alliance

**IMPEx**: Integrated Medium for Planetary Exploration project

**ISTP(Next)**: (next) International Solar Terrestrial Physics program

**ISWAT**: COSPAR International Space Weather Action Teams

**IVOA**: International Virtual Observatory Alliance

**LANL**: Los Alamos National Laboratory

**LWS**: Living With a Star (NASA program)

**MIT**: Massachusetts Institute of Technology

**NASA**: National Aeronautics and Space Administration

**NASEM**: National Academies of Sciences, Engineering, and Medicine

**NCAR**: National Center for Atmospheric Research

**NOAA**: National Oceanic and Atmospheric Administration

**NSF**: National Science Foundation

**PETSc:** Portable, Extensible Toolkit for Scientific Computation project

**PSI**: Predictive Science Incorporated

**PyHC**: Python in Heliophysics Community

**RDA**: Research Data Alliance

**SAIC**: Science Applications International Corporation

**SERENE**: University of Birmingham Space Environment & Radio Engineering group

**SHINE**: Solar Heliospheric and Interplanetary Environment (NSF program)

**SMD**: NASA Scientific Mission Directorate

**SPACE**: Scalable Parallel Astrophysical Codes for Exascale (European Union center of excellence)

**SWPC**: NOAA Space Weather Prediction Center

**SWx TREC**: University of Colorado Space Weather Technology, Research, and Education Center

**UCAR**: University Corporation for Atmospheric Research

**UKMO**: United Kingdom Met Office

**UNESCO**: United Nations Educational, Scientific, and Cultural Organization

**VSWMC**: Virtual Space Weather Modelling Centre

**WDCC**: World Data Center for Climate

**Acknowledgments**

The 11th CCMC Community Workshop, College Park, MD, June 3-7, 2024, was focused on Open Science in Modeling and was sponsored by NSF Award 2423907 Innovations in Open Science Planning Workshop. We thank the then NSF program officer, Dr. Tai-Yin Huang, for her insightful discussions.

C. Corti, M.M. Kuznetsova, M.A. Reiss, J. Yue, M.-Y. Chou, D. De Zeeuw, C. Didigu, M. Lesko, P. MacNeice, M.L. Mays, M. Petrenko, L. Rastätter, J. Topper, T. Tsui, J. Wang, C. Wiegand, and Y. Zheng acknowledge support of CCMC from NASA Heliophysics and the NSF.

M.M. Kuznetsova acknowledges encouragement from Antti Pulkkinen to establish a forum for the modeling community.

E. Samara was supported in part by the NASA Solar Orbiter project and NASA Heliophysics Division Space Weather.

B. Cecconi acknowledges support from CNRS, Observatoire de Paris and CNES.

M. M. Bisi acknowledges support from the Natural Environment Research Council of UK Research and Innovation (Grant number NE/X019004/1 [RISER]) and via the RAL Space In House Research programme funded by the Science and Technology Facilities Council of UK Research and Innovation (award ST/M001083/1).

S. Poedts is funded by the European Union (ERC-AdG agreement No 101141362, Open SESAME). Views and opinions expressed are, however, those of the author(s) only and do not necessarily reflect those of the European Union or the European Research Council. Neither the European Union nor the granting authority can be held responsible for them.

S. Poedts also acknowledges support from the projects C16/24/010 (C1 project Internal Funds KU Leuven), G0B5823N and G002523N (WEAVE) (FWO-Vlaanderen), and 4000145223 (SIDC Data Exploitation (SIDEX2), ESA Prodex).

## Open Research

No experimental data nor simulation data were used in the preparation of this manuscript.

## Conflict of Interest Statement

The authors have no conflicts of interest to disclose.

## Author Contributions

Conceptualization, paper coordination, and Sections 1, 2, and 9: C. Corti, M.M. Kuznetsova, M.A. Reiss, J. Yue, and J. Karpen.

Contributions to modelers' perspectives (Section 3 and Appendix A): C.N. Arge, F. Bacchini, C. Bard, S. Bruinsma, R.M. Caplan, L.K.S. Daldorff, P.J. Deka, C.R. DeVore, S. Elvidge, N. Ganushkina, J. Huba, B.V. Jackson, V. Jordanova, J.A. Linker, J.G. Luhmann, H. Liu, S. Markidis, P. Mayank, V. Merkin, N. Moens, D. Odstrcil, Y.A. Omelchenko, M. Palmroth, S. Poedts, A.J. Ridley, Y. Shou, V. Tenishev, D.R. Themens, G. Toth, W. Wang, R.-P. Wilhelm, M.A. Young, C. Corti, M.A. Reiss, M.M. Kuznetsova, and J. Karpen.

Contributions to Sections 4-8 and Appendices B, C: M.M. Kuznetsova, M.A. Reiss, C. Corti, J. Karpen, J. Yue, C.N. Arge, B. Cecconi, M.-Y. Chou, D. De Zeeuw, G.L. Delzanno, C. Didigu, M. El Alaoui, S. Fung, K. Garcia-Sage, J. Green, Z. Huang, L.K. Jian, T. Kodikara, D. Kondrashov, M. Lesko, J.A. Linker, J.G. Luhmann, P. MacNeice, A. Masson, M.L. Mays, P. M. Mehta, V. Merkin, Y. Omelchenko, E. Palmerio, C. Pandey, M. Petrenko, S. Poedts, E. Provornikova, L. Rastätter, L. Rusaitis, N. Sachdeva, E. Samara, C. Shi, D. Sur, A. Taktakishvili, D. Themens, J. Topper, G. Toth, T. Tsui, C. Verbeke, J. Wang, K. Whitman, C. Wiegand, M. Wiltberger, M. Young, Y. Zheng, I. Zakharenkova, and K. Zhang.

Review and editing: J. Karpen, C. Corti, M.M. Kuznetsova, M.A. Reiss, J. Yue, M.M. Bisi, M.K. Georgoulis, T. Kodikara, T. Pulkkinen, A. Chartier, D. da Silva, A. Faturahman, V.E. Ledvina, W. Liu, E. Resnick, and R.S. Weigel.